\documentclass[a4paper,11pt]{article}
\usepackage{jheppub}
\usepackage[T1]{fontenc}
\usepackage{mathtools}
\usepackage{mathrsfs}
\usepackage{booktabs}
\usepackage{array}
\usepackage{tabularx}
\usepackage{flafter}
\usepackage{placeins}
\usepackage{subcaption}

\newcommand{\dd}{\mathrm{d}}
\newcommand{\ii}{\mathrm{i}}
\newcommand{\ee}{\mathrm{e}}
\newcommand{\cJ}{\mathcal J}
\newcommand{\cH}{\mathcal H}
\newcommand{\arcosh}{\operatorname{arcosh}}
\newcommand{\OO}{\mathcal O}
\newcommand{\Vmin}{V_{\min}}
\newcommand{\Vtop}{V_{\mathrm{top}}}
\newcommand{\K}{\mathsf K}
\newcommand{\Eell}{\mathsf E}
\newcommand{\Pell}{\mathsf \Pi}
\newcommand{\Ham}{\hat H}
\newcommand{\RR}{\mathbb R}
\newcolumntype{Y}{>{\raggedright\arraybackslash}X}

\title{Power-law tunneling in the Razavy hyperbolic double well: Heun connection formulas and their resonances}

\author{Jie Ren}
\author{and Xuanzhuang Wu}
\affiliation{School of Physics, Sun Yat-sen University, Guangzhou, 510275, China}

\emailAdd{renjie7@mail.sysu.edu.cn}
\emailAdd{wuxzh23@mail2.sysu.edu.cn}

\hypersetup{
  pdftitle={Power-law tunneling in the Razavy hyperbolic double well: Heun connection formulas and their resonances},
  pdfauthor={Jie Ren and Xuanzhuang Wu},
  pdfkeywords={Nonperturbative Effects; Solitons Monopoles and Instantons}
}

\abstract{
We study nonperturbative level splitting in the symmetric Razavy
hyperbolic double well, relating its double-confluent Heun connection
problem to exact WKB and quasi-exact solvability.  In the weak-coupling
limit the wells separate logarithmically and approach two Morse
systems.  Reflection factorizes the two-ended spectral condition into
separate even and odd quantization conditions.  For each normalizable
Morse level with noninteger decay exponent, the irregular-block
connection formula yields the leading splitting with its full
gamma-function coefficient, two analytic fluctuation corrections, and
the leading logarithmic three-traversal contribution.  The noninteger
power of the coupling follows from the logarithmic growth of the barrier
action.  We express the classical periods in terms of complete elliptic
integrals and derive the first quantum-period correction.  At integer
exponent, a convergent reduction onto the two resonant Floquet modes
resolves the singular fixed-order expansion into two local spectral
branches and reproduces the known finite-spin tunneling gap.
}


\begin{document}

\maketitle

\section{Introduction}
\label{sec:introduction}

The level splitting of a symmetric double well is the simplest
nonperturbative effect in quantum mechanics.  Perturbation theory about
either minimum gives the same series for the even and the odd member of
each doublet, so their difference vanishes to all orders in $\hbar$.
The splitting is controlled instead by tunneling through the barrier
and is exponentially small.  For this reason the double well has become
the standard testing ground for the ideas that connect perturbative and
nonperturbative physics: large-order behavior, multi-instanton
expansions, resurgent transseries, and exact WKB analysis
\cite{Bender:1969,Bender:1973,Voros:1983,Delabaere:1997,
ZinnJustin:2004,ZinnJustin:2004b,Dunne:2014,Sueishi:2020,
vanSpaendonck:2024}.  The potential studied in most detail is the
quartic one.

In this paper we study a double well of a different kind, the
hyperbolic potential introduced by Razavy \cite{Razavy:1980},
\begin{equation}
  \Ham(\hbar;\kappa,\beta)
  =-\hbar^2\partial_x^2
   +\frac{\kappa^2}{2}\cosh 2x-\beta\kappa\cosh x,
  \qquad \kappa>0,\quad \beta>2\kappa.
  \label{eq:intro_razavy_hamiltonian}
\end{equation}
All variables are dimensionless.  Up to an additive constant the
potential is $\kappa^2(\cosh x-\beta/2\kappa)^2$: it is confining and
has two minima, at $x=\pm\arcosh(\beta/2\kappa)$, separated by a barrier
of height $(\beta/2-\kappa)^2$.  The parameter $\kappa$ controls the
coupling between the wells, which move apart as $\kappa\to0^+$, and the
ratio $\beta/\hbar$ controls the number of levels below the barrier.
The model has three features that make it a useful complement to the
quartic well.

First, its Schr\"odinger equation is of Heun type.  In the variable
$z=\kappa\ee^x$ it becomes a double-confluent Heun equation (DCHE),
whose two irregular singular points of rank one, $z=0$ and $z=\infty$,
correspond to the two ends of the real line.  A bound state is a
solution that is recessive at both points, so that the spectrum is
determined by a two-point connection problem
\cite{LaySlavyanov:1998,LayBaySlavyanov:1998,AbadGomezSesma:2008}.  The
quartic double well leads instead to a triconfluent Heun equation
\cite{Lay:1997Quartic,Batic:2015}, with a single irregular singular
point of rank three, and its bound states connect different Stokes
sectors of that one singularity \cite{Osherov:2011}.  Within the Heun
class these are two particularly simple double wells, since neither
equation has an additional regular singular point
\cite{Ronveaux:1995,Slavyanov:2000}.

Second, the model is quasi-exactly solvable (QES).  When
$\beta/\hbar=M$ is a positive integer, Razavy's terminating solutions
\cite{Razavy:1980} span an $M$-dimensional invariant space, and the
lowest $M$ levels are the eigenvalues of an $M\times M$ matrix.  Up to
an additive constant, this matrix is the Hamiltonian
$H_{\rm spin}=-S_z^2-BS_x$ of a spin $S=(M-1)/2$ with easy-axis
anisotropy in a transverse field, with $B=2\kappa$ in units $\hbar=1$.
The spin--coordinate correspondence behind this identification was
developed in
\cite{ZaslavskiiUlyanovTsukernik:1983,ZaslavskiiUlyanov:1984,
ZaslavskiiUlyanov:1987,UlyanovZaslavskii:1992}.  The algebraic sector
fits the general $\mathfrak{sl}(2)$ theory of QES operators and its
formulation in terms of orthogonal polynomials
\cite{Turbiner:1988QES,GonzalezLopezKamranOlver:1993,BenderDunne:1996,
FinkelGonzalezLopezRodriguez:1996}, and its polynomial families and
parity structure are known in detail
\cite{KhareMandal:1998,Finkel:1999,Becker:2017}.  It provides exact
spectra against which general formulas can be tested.  The interplay
between quasi-exact solvability and resurgence has been studied for
related trigonometric and polynomial potentials in \cite{Kozcaz:2018}.

Third, the equation has a gauge-theory and conformal-field-theory
interpretation.  The operator \eqref{eq:intro_razavy_hamiltonian}
coincides with the quantum Seiberg--Witten curve of four-dimensional
$\mathcal N=2$, $SU(2)$ gauge theory with two fundamental
hypermultiplets of equal mass
\cite{Ashok:2016,Aminov:2022,Imaizumi:2021}, which arises in the
Nekrasov--Shatashvili limit of the $\Omega$-background
\cite{NekrasovShatashvili:2009}.  On the conformal-field-theory side of
the AGT correspondence \cite{Alday:2010,Gaiotto:2013}, equations of
Heun type arise as semiclassical limits of null-vector equations in
Liouville theory, and their connection coefficients have been obtained
in closed form in terms of classical Virasoro conformal blocks
\cite{Bonelli:2023}.  These connection formulas have found applications in black-hole perturbation
theory and in holography
\cite{Aminov:2022,Bonelli:2022Kerr,Dodelson:2023}.  For the DCHE the
relevant object is a rank-one irregular block: a power series in
$\kappa^2$ that depends on the Floquet exponent $\nu$ of the equation.
The Razavy well is a bound-state problem with real parameters to which
this formula applies directly, and in which its content can be compared
both with exact algebraic spectra and with the instanton picture of
tunneling.

The expansion in $\kappa^2$ that is natural from this point of view is
not the semiclassical expansion.  At fixed $(\kappa,\beta)$, the limit
$\hbar\to0^+$ gives the usual splitting of order $\ee^{-S_B/\hbar}$,
with a finite barrier action $S_B$.  In this paper we instead take
$\kappa\to0^+$ at fixed $\beta$ and $\hbar$, and we set $\hbar=1$
unless stated otherwise.  The minima then move apart as
$x\simeq\pm\log(\beta/\kappa)$, and each well approaches a Morse
potential \cite{UlyanovZaslavskii:1992,Konwent:1998,Becker:2017} with
finitely many bound states,
\begin{equation}
  E_n^{(0)}=-\frac{r_n^2}{4},
  \qquad r_n=\beta-1-2n>0,
  \qquad n=0,1,\ldots.
  \label{eq:intro_morse_levels}
\end{equation}
Under the barrier a Morse state decays as $\ee^{-r_nd/2}$ with the
distance $d$ from its well, and the two wells are a distance
$2\log(\beta/\kappa)$ apart.  The tunneling amplitude between them is
therefore a power of the coupling, $\kappa^{r_n}$, with an exponent
that is in general not an integer.  Equivalently, the barrier action
grows logarithmically, $S_B\simeq r_n\log(1/\kappa)$, and the instanton
factor $\ee^{-S_B}$ becomes $\kappa^{r_n}$.  In this limit the spectrum
is a transseries in the coupling: analytic series in $\kappa^2$, which
describe the fluctuations, are multiplied by powers of $\kappa^{r_n}$,
one for each traversal of the barrier, and by powers of $\log\kappa$.
We call $\kappa\to0^+$ the weak-coupling limit because it decouples the
two wells, although $\kappa$ itself is not the tunneling matrix
element.  For
the quartic well the analogous limit separates the minima
algebraically, and the splitting remains exponentially small in the
inverse coupling (see table~\ref{tab:quartic-razavy-comparison} in
section~\ref{sec:two_limits}).

At the QES values the exponent $r_n=M-1-2n$ is an integer, and the
splitting is known from the theory of spin tunneling.  The leading
weak-field gap of $H_{\rm spin}$ is
\begin{equation}
  \Delta E_n^{\rm spin}
  \sim2\left(\frac B2\right)^{2S-2n}
  \frac{(2S-n)!}{n!\,[(2S-2n-1)!]^2},
  \qquad B\to0,
  \label{eq:intro_prior_qes_gap}
\end{equation}
for $n=0,\ldots,\lfloor S-1/2\rfloor$
\cite{KorenblitShender:1978,Garanin:1991,UlyanovZaslavskii:1992}, and
its first relative correction, of order $B^2$, was derived in
\cite{Ulyanov:1999}.  The double-Morse interpretation of the Razavy
well and the extension of the weak-field exponent to noninteger $2S$
were discussed in \cite{Konwent:1998}.

The same potential has also been studied in the semiclassical limit at
fixed $(\kappa,\beta)$.  Garg derived the splitting, prefactor included,
from Herring's formula and evaluated it for the easy-axis spin, which
maps onto this potential \cite{Herring:1962,Garg:2000}.  The
fluctuation determinant of the Razavy instanton was computed with
spectral zeta functions in \cite{AlonsoIzquierdo:2014}.  The power law
in $\kappa$ should not be confused with algebraic instantons
\cite{Patrascioiu:1978,MottolaRouet:1982,Aziz:2026}, which connect
minima at which the potential is flat, $V-\Vmin\propto\eta^{2p}$ with
$p\geq2$ at a distance $\eta$ from a minimum.  Such instantons approach
the vacuum as a power of Euclidean time, and the dilute instanton gas
does not apply to them.  The minima of the Razavy potential remain
harmonic, with frequency $\sqrt{\beta^2-4\kappa^2}\to\beta$ as
$\kappa\to0^+$, and the instanton approaches them exponentially; the
power of $\kappa$ comes from the logarithmic growth of the barrier
action.

Compared with the quartic well, however, the model has received little
attention as a setting for instanton transseries and resurgent spectral
relations.  The purpose of
this paper is to determine the splitting for general $\beta$, including
its coefficient, its corrections, and its behavior at integer $r_n$,
and to show how the Heun connection formula, exact WKB, and the QES sector
describe the same quantity.

\subsection{Summary of results}
\label{sec:intro_summary}

Our starting point is the connection matrix of the DCHE given in
\cite[sec.~4.4.2]{Bonelli:2023}, which we specialize to the two $L^2$
boundary conditions of \eqref{eq:intro_razavy_hamiltonian}.  The
spectral determinant is the Wronskian of the solutions recessive at
$z=0$ and at $z=\infty$.  Reflection $x\mapsto-x$ exchanges the two
singular points, and we use it to factorize this determinant into an
even and an odd factor, \eqref{eq:local_parity_factorisation}.  Let
$\nu$ be the Floquet exponent, defined by
$\psi_\nu(x+2\pi\ii)=\ee^{2\pi\ii\nu}\psi_\nu(x)$ on the branch
$E\to-\nu^2$ as $\kappa\to0$, and let
$\mathcal F(\nu,\beta;\kappa^2)
=-\beta^2\kappa^2/[2(4\nu^2-1)]+\OO(\kappa^4)$ be the classical
irregular block.  The spectrum is then determined by
\begin{equation}
  \kappa^{2\nu}\exp\!\left[-\frac12\partial_\nu\mathcal F\right]
  \frac{\Gamma(-2\nu)\,
        \Gamma\!\left(\frac{1-\beta}{2}+\nu\right)}
       {\Gamma(2\nu)\,
        \Gamma\!\left(\frac{1-\beta}{2}-\nu\right)}
  =\epsilon,
  \qquad
  E=-\nu^2+\kappa^2\partial_{\kappa^2}\mathcal F,
  \label{eq:intro_quantisation}
\end{equation}
with $\epsilon=+1$ for even and $\epsilon=-1$ for odd states.  The
vanishing of the full determinant gives only the square of the first
relation, and the sign $\epsilon$ distinguishes the two members of a
doublet.  The second relation is of Matone type
\cite{Matone:1995,Flume:2004}.  The corresponding characteristic-value
expansion at noninteger Floquet exponent was obtained from the
$N_f=2$ irregular block in \cite{PiatekPietrykowski:2016}.

For noninteger $r_n$ and small $\kappa$, the solutions of
\eqref{eq:intro_quantisation} lie near the poles of
$\Gamma\bigl(\frac{1-\beta}{2}+\nu\bigr)$, which are the Morse levels
$\nu=r_n/2$.  The two signs $\epsilon=\pm1$ displace $\nu$ from the
pole in opposite directions, and the resulting gap is
\begin{equation}
  \Delta E_n\equiv E_{n,\mathrm o}-E_{n,\mathrm e}
  =2\frac{\Gamma(\beta-n)}{n!\,\Gamma(r_n)^2} \kappa^{r_n}
  \left[1-\frac{\beta^2(r_n+2)}{(r_n^2-1)^2}\kappa^2 +\OO(\kappa^4)\right]
  +\OO\!\left(\kappa^{3r_n}|\log\kappa|^2\right).
  \label{eq:intro_generic_gap}
\end{equation}
The tunneling factor $\kappa^{r_n}$ is the factor $\kappa^{2\nu}$ of
the connection formula evaluated at the Morse level.  Its coefficient,
the residue of the gamma functions, continues the factorials of
\eqref{eq:intro_prior_qes_gap} to noninteger $\beta$.
This leading term also follows from an elementary matching argument
(section~\ref{sec:morse_limit}), whereas the corrections require the
connection formula.  The bracket is the fluctuation series of the
one-traversal sector, generated by the irregular block.  We obtain its
$\kappa^2$ and $\kappa^4$ terms, \eqref{eq:generic_splitting_c1} and
\eqref{eq:generic_splitting_c2_explicit}.  The first continues the spin
result of \cite{Ulyanov:1999} to noninteger exponent, and both agree
with the exact QES spectra of appendix~\ref{app:qes_examples} wherever
they remain finite at integer $\beta$.  At noninteger
$\beta$ the expansion reproduces the numerically computed gaps of
figure~\ref{fig:razavy_level_examples}.  The last term in
\eqref{eq:intro_generic_gap} is the three-traversal sector, whose
leading coefficient, a quadratic polynomial in $\log\kappa$, is given
in \eqref{eq:generic_multitraversal_coefficient}.  The two remainders
belong to different sectors of the transseries: for $0<r_n<1$ the
three-traversal term is larger than the $\kappa^2$ correction.  The
expansion holds for each fixed Morse level but is not uniform as
$r_n\to0^+$ or as $r_n$ approaches a positive integer.

The splitting \eqref{eq:intro_generic_gap} can also be understood in
exact WKB.  Below the barrier the four real turning
points define a genus-one curve, as for the quartic double well, and
the exact quantization condition of a symmetric double well
\cite{Delabaere:1997,Sueishi:2020,vanSpaendonck:2024,
MisumiPazarbasi:2025} applies with the Razavy quantum periods,
\eqref{eq:exact_wkb_quantisation}.  We express the classical periods
through complete elliptic integrals and show that their first quantum
correction is generated by a second-order differential operator in the
parameters, \eqref{eq:first_quantum_period_operator_main}.  At fixed
energy $E=-r^2/4$ the action for one traversal of the barrier grows
logarithmically as $\kappa\to0^+$,
\begin{equation}
  S_B=r\log\frac1\kappa+\mathcal C_0(\beta,r)+o(1),
  \label{eq:intro_barrier_action}
\end{equation}
with the finite part $\mathcal C_0$ given in
\eqref{eq:barrier_finite_part}, so that the instanton factor becomes
$\ee^{-S_B}\sim\kappa^r\ee^{-\mathcal C_0}$.  This accounts for the
power law.  The classical periods reproduce the leading Stirling
approximation to the gamma-function coefficient of
\eqref{eq:intro_generic_gap}, and the first quantum correction to the
barrier period reproduces the next term.  For the lowest QES doublets
the classical coefficient is in error by up to $16\%$ and the corrected
one by about $1\%$ (table~\ref{tab:wkb_qes_amplitudes}).  The remaining
difference is due to higher orders in $\hbar$ of the one-traversal
term, since the multi-traversal corrections vanish in the limit.

When $r_n$ is a positive integer, as it is for every doublet of the QES
sector, the coefficients of the block have poles
at $\nu=r_n/2$, of an order that grows with the power of $\kappa^2$,
although the spectrum itself is regular.  We refer to them as Kac
poles, since they are the classical counterparts of the poles of
conformal blocks at degenerate intermediate weights.  Avoided crossings
in Heun problems and the reorganization of such poles into branch
functions have been studied in
\cite{SlavyanovVeshev:1997,Gorsky:2017ndg,Beccaria:2016wop,
Alekseev:2018kcn,Alekseev:2019gkl,Demulder:2016mja}.  In the present
problem the mechanism can be made explicit.  For $\beta=M=r+1+2n$ and
$\nu=r/2+\delta$ with $|\delta|=\OO(\kappa^r)$, a Feshbach--Schur
reduction of the Floquet problem onto its two resonant modes, with a
convergent expansion for the remaining ones, gives the two
characteristic values
\begin{equation}
  E_\pm=E_c
  \pm\sqrt{r^2\delta^2+\mathcal A_{r,n}^2\kappa^{2r}}
  +\OO(\kappa^{r+2}),
  \qquad
  \mathcal A_{r,n}=\frac{(n+r)!}{n!\,[(r-1)!]^2},
  \label{eq:intro_resonant_sheets}
\end{equation}
where $E_c$ is a smooth function of $\kappa$ and $\delta$.  At $\delta=0$ the two
branches are the odd and the even state, and their separation
reproduces \eqref{eq:intro_prior_qes_gap}.  Expanding
\eqref{eq:intro_resonant_sheets} at fixed $\delta\neq0$ gives the
leading Kac pole of the block at every resonant order in closed form,
\eqref{eq:kac_pole_tower}, a result that we also derive directly from
the Floquet recurrence.  Conversely, the sum of these poles
cancels the pole of the gamma functions in
\eqref{eq:intro_quantisation} and leaves $\epsilon=\pm1$ on the two
branches.  Without it, the gamma functions alone would account for
only half of the gap.  This last step uses a continuation of the Matone
relation into the resonant region, which we state as an assumption in
section~\ref{sec:resonant_block}.  We do not construct a
conformal-block representation of the full connection matrix that is
uniform near resonance.

The QES sector also constrains the spectrum away from the integer
values.  If $N_{\rm sub}$ denotes the number of eigenvalues strictly
below the central maximum, then
\begin{equation}
  N_{\rm sub}(\kappa,\beta;\hbar)
  \leq2\left\lceil\frac{\beta/\hbar-1}{2}\right\rceil,
  \qquad 0<\kappa<\frac{\beta}{2},
  \label{eq:intro_subbarrier_bound}
\end{equation}
for all real $\beta/\hbar>0$.  The right-hand side is twice the number
of bound states of each limiting Morse well.  The proof compares the
Hamiltonian with the one at the next odd integer value of
$\beta/\hbar$ by means of the min--max principle, and it extends the
results of \cite{UlyanovZaslavskii:1992,Konwent:1998} for integer
$\beta/\hbar$.  The bound is sharp, since
it is attained for sufficiently small $\kappa/\hbar$, when all Morse
doublets lie below the barrier.

Finally, appendix~\ref{app:recurrence_connection} gives a
representation of the same spectral determinant that does not involve
conformal blocks.  It is obtained from the large-order behavior of the
solution of a four-term recurrence, following
\cite{Wolf:1998,Schaefke:1984}.  In deriving it we correct the
normalization of the published connection coefficient.

The paper is organized as follows.
Section~\ref{sec:spectral_problem} sets up the spectral problem,
contrasts the weak-coupling and semiclassical limits, and derives the
Morse limit.  Section~\ref{sec:heund} formulates the connection problem
for the two parities and derives the splitting at noninteger exponent.
Section~\ref{sec:exact_wkb} develops the exact WKB description: the
quantization condition, the periods and their first quantum
correction, and the comparison with the exact coefficient.
Section~\ref{sec:resonance_qes} treats the QES sector, the resonances
at integer exponent, and the bound on the number of levels below the
barrier.  Section~\ref{sec:sum}
contains our conclusions and a list of open problems.  The appendices
collect the block recursion and the Kac poles
(appendix~\ref{app:floquet_formal}), the estimates behind the two-mode
reduction (appendix~\ref{app:floquet_feshbach}), the period
calculations (appendix~\ref{app:period_calculations}), explicit QES
spectra (appendix~\ref{app:qes_examples}), and the connection formula
based on a recurrence relation
(appendix~\ref{app:recurrence_connection}).

\section{The Razavy spectral problem and the Morse limit}
\label{sec:spectral_problem}

In this section we fix the parameters, distinguish the weak-coupling
limit from the semiclassical limit at fixed potential, describe the
curve defined by the turning points, and derive the two Morse systems
that determine the limiting doublets.

\subsection{Hamiltonian and double-well geometry}
\label{sec:double_well_chamber}

We study the self-adjoint Schr\"odinger operator
\begin{equation}
  \begin{aligned}
    \Ham&=-\hbar^2\frac{\dd^2}{\dd x^2}+V(x),\\
    V(x)&=a\cosh 2x-b\cosh x
      \equiv\frac{\kappa^2}{2}\cosh 2x-\beta\kappa\cosh x,
  \end{aligned}
  \label{eq:razavy_hamiltonian}
\end{equation}
on $L^2(\RR,\dd x)$, with $\hbar>0$, $a>0$, and $b>4a$.
The two parametrizations of the Razavy potential are related by
$a=\kappa^2/2$ and $b=\beta\kappa$, or equivalently
\begin{equation}
  \kappa=\sqrt{2a},\qquad \beta=\frac{b}{\sqrt{2a}}.
  \label{eq:parameter_dictionary}
\end{equation}
We write $\Ham(\hbar;\kappa,\beta)$ when its parameter dependence
is needed.  Dividing the spectral equation by $\hbar^2$ gives the
exact rescaling
\begin{equation}
  \hat \kappa=\frac{\kappa}{\hbar},\qquad
  \hat\beta=\frac{\beta}{\hbar},\qquad
  \hat E=\frac{E}{\hbar^2},\qquad
  \frac{1}{\hbar^2}\Ham(\hbar;\kappa,\beta)
  =\Ham(1;\hat \kappa,\hat\beta).
  \label{eq:hbar_rescaling_dictionary}
\end{equation}
The spectral problem therefore depends on $\hbar$ only through
$(\hat\kappa,\hat\beta)$.  For instance, the QES condition is
$\hat\beta=M\in\mathbb{Z}^+$, which reads $\beta=M$ at $\hbar=1$.  Since
$V(x)\to+\infty$ as $|x|\to\infty$, the operator
\eqref{eq:razavy_hamiltonian} has compact resolvent and a discrete,
simple spectrum, and since $V(-x)=V(x)$ every eigenfunction has
definite parity.

We denote by $E_{n,\mathrm e}$ and $E_{n,\mathrm o}$, with
$n=0,1,\ldots$, the eigenvalues in the even and odd sectors.  They are
the spectra of the half-line problems with $\psi'(0)=0$ and
$\psi(0)=0$, respectively.  The two sequences interlace strictly
\cite[sec.~IV.3]{Becker:2017}, so that the globally ordered spectrum
$\mathcal E_j$ is
\begin{equation}
  \begin{gathered}
    E_{0,\mathrm e}<E_{0,\mathrm o}<E_{1,\mathrm e}<E_{1,\mathrm o}<\cdots,\\
    \mathcal E_{2n}=E_{n,\mathrm e},\qquad
    \mathcal E_{2n+1}=E_{n,\mathrm o},\qquad
    \Delta E_n:=E_{n,\mathrm o}-E_{n,\mathrm e}>0.
  \end{gathered}
\end{equation}
The weak inequalities
$E_{n,\mathrm e}\leq E_{n,\mathrm o}\leq E_{n+1,\mathrm e}$ follow from
the min--max principle, because the Dirichlet condition at the origin
restricts the Neumann form domain by one linear condition.  Equality is
excluded because a solution with $\psi(0)=\psi'(0)=0$ vanishes
identically.  We use this labeling throughout the spectrum, although
$E_{n,\mathrm e}$ and $E_{n,\mathrm o}$ form a tunneling doublet only
when the wells are weakly coupled at the relevant energy.

The completed-square form
\begin{equation}
  V(x)=\kappa^2\left(\cosh x-\frac{\beta}{2\kappa}\right)^2
  -\frac{\beta^2+2\kappa^2}{4}
  \label{eq:completed_square_potential}
\end{equation}
and the derivative
\begin{equation}
  V'(x)=\kappa\sinh x\bigl(2\kappa\cosh x-\beta\bigr)
  \label{eq:stationary_equation}
\end{equation}
show that the potential is a symmetric double well if and only if
\begin{equation}
  \beta>2\kappa.
  \label{eq:double_well_chamber}
\end{equation}
The minima at $x_0=\arcosh(\beta/2\kappa)$, the central maximum, the barrier height, and the local
oscillation frequency are
\begin{equation}
\begin{gathered}
  \Vmin=-\frac{\beta^2}{4}-\frac{\kappa^2}{2},\qquad
  \Vtop=V(0)=\frac{\kappa^2}{2}-\beta \kappa,\\
  \Vtop-\Vmin=\left(\frac{\beta}{2}-\kappa\right)^2,\qquad
  \Omega_0=\sqrt{2V''(x_0)}
  =\sqrt{\beta^2-4\kappa^2}.
    \label{eq:well_geometry}
\end{gathered}
\end{equation}
At $\beta=2\kappa$ the two minima merge at the origin, where the
curvature vanishes, and for $\beta<2\kappa$ the origin is the unique
minimum.  We work throughout in the region
\eqref{eq:double_well_chamber}.

\begin{figure}[p]
  \centering
  \captionsetup{font=small,labelfont=bf,labelsep=period}
  \captionsetup[subfigure]{font=small,labelfont=normalfont,
    justification=centering,singlelinecheck=false,skip=3pt}
  \begin{subfigure}{\textwidth}
    \centering
    \includegraphics[width=\linewidth]{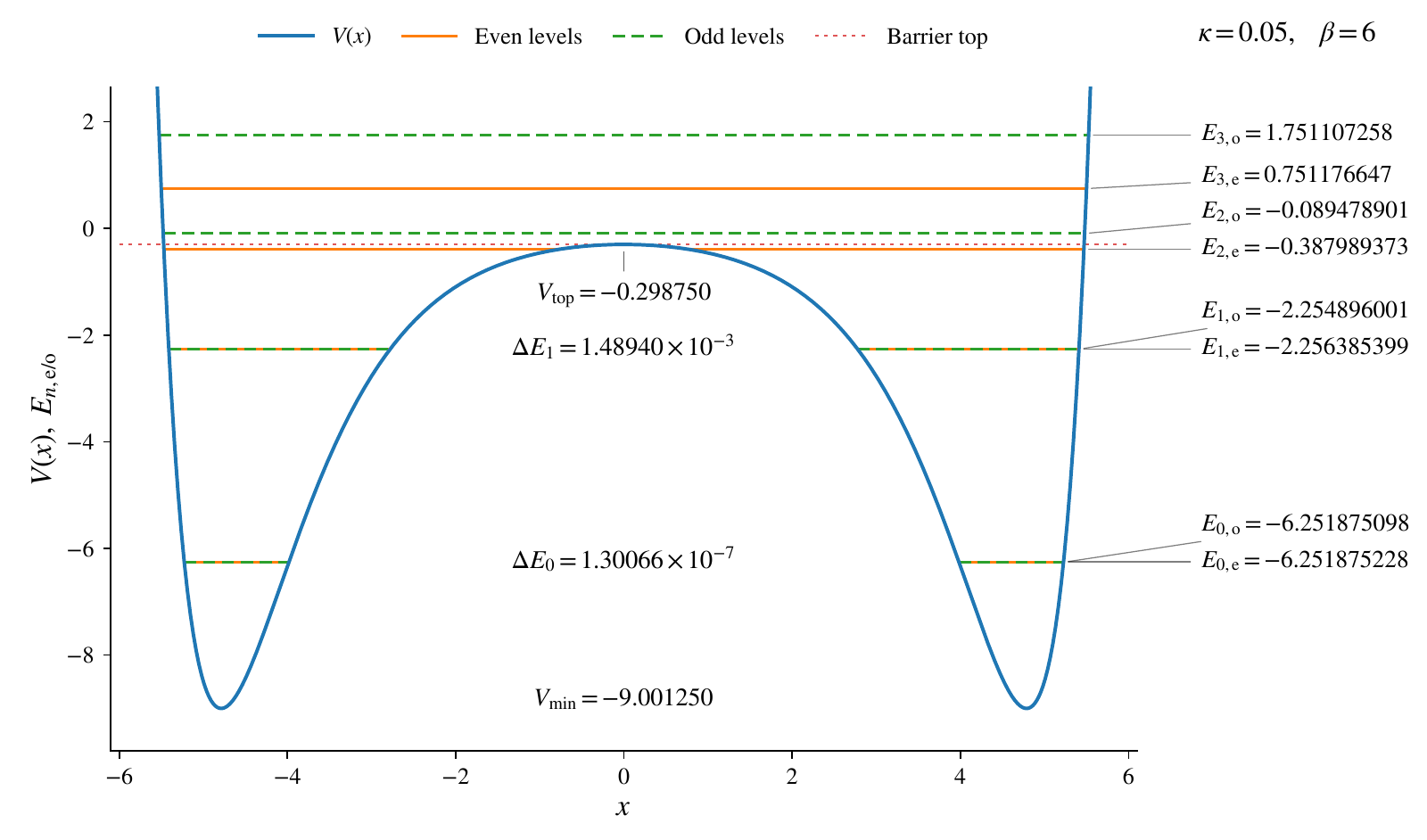}
    \caption{$\beta=6$: five levels below the barrier and three above;
      $E_{2,\mathrm e}<\Vtop<E_{2,\mathrm o}$.}
    \label{fig:razavy_levels_beta6}
  \end{subfigure}

  \medskip

  \begin{subfigure}{\textwidth}
    \centering
    \includegraphics[width=\linewidth]{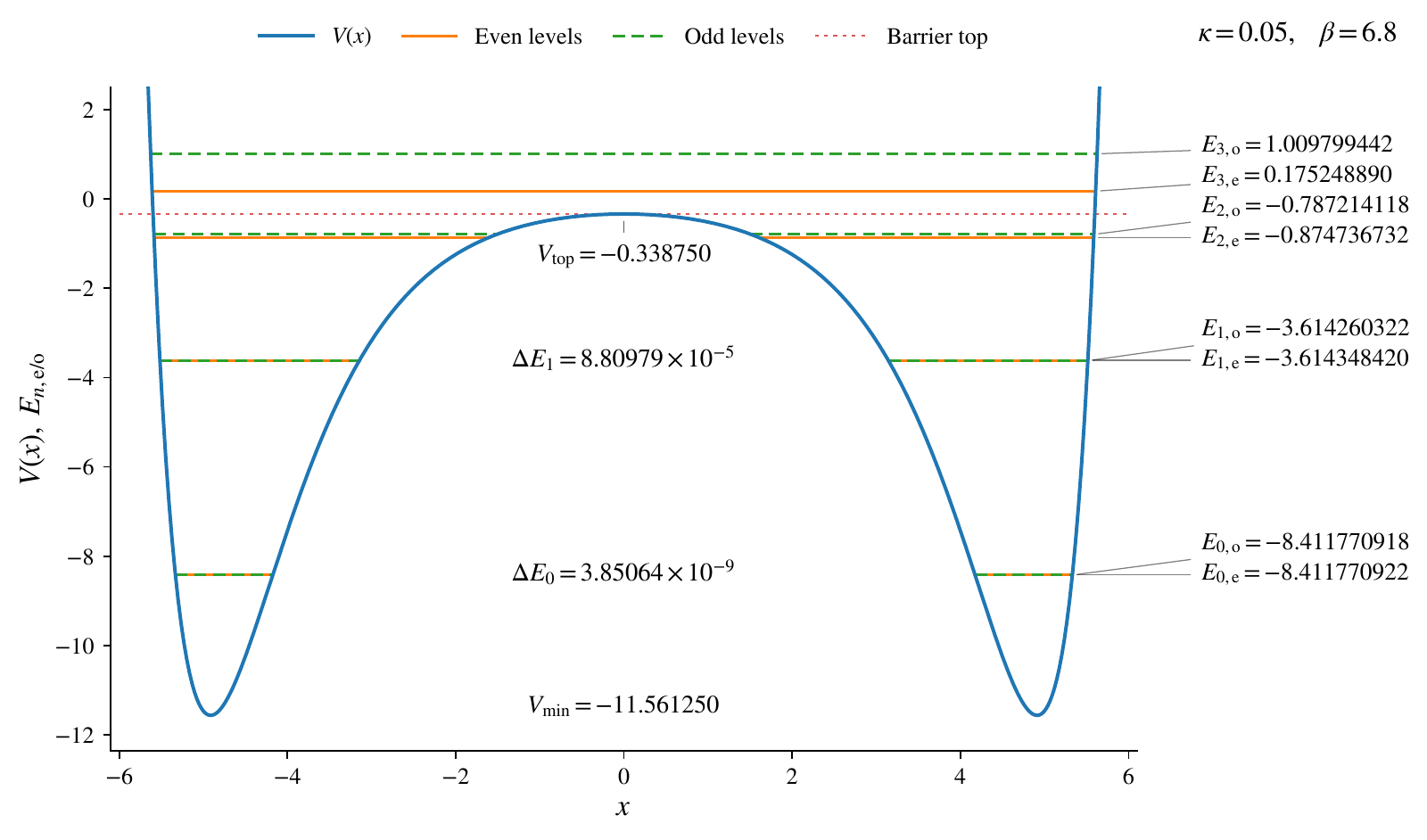}
    \caption{$\beta=6.8$: the $n=0,1,2$ parity pairs lie below the
      barrier; $E_{3,\mathrm e}$ and $E_{3,\mathrm o}$ lie above.}
    \label{fig:razavy_levels_beta68}
  \end{subfigure}
  \caption{The Razavy potential and the lowest eight eigenvalues of
    \eqref{eq:razavy_hamiltonian} at $\kappa=0.05$ and $\hbar=1$,
    computed by shooting in each parity sector.  The blue curve is
    $V(x)$.  Solid orange and dashed green segments show the even and
    odd levels over their classically allowed intervals, and the dotted
    line marks the barrier top.  The annotations give the splittings
    $\Delta E_n=E_{n,\mathrm o}-E_{n,\mathrm e}$, which are not resolved
    on the scale of the plot for the lowest pairs.}
  \label{fig:razavy_level_examples}
\end{figure}

Figure~\ref{fig:razavy_level_examples} shows the lowest eight
eigenvalues at $\kappa=0.05$ for $\beta=6$ and $\beta=6.8$.  The lowest
pairs are nearly degenerate.  For $\beta=6$ the barrier top lies between the two
levels of the $n=2$ pair, $E_{2,\mathrm e}<\Vtop<E_{2,\mathrm o}$, so
that five levels lie below it.  For $\beta=6.8$ both members of that pair lie
below the barrier, which gives six such levels.  The $n=3$ levels lie
above the barrier in both panels and do not form a tunneling doublet.

\subsection{Two distinct small-gap limits}
\label{sec:two_limits}

The splitting can be made small in two inequivalent ways.  In the
semiclassical limit one fixes the potential and sends $\hbar\to0^+$.  A
given level, labeled by $N$ below, then has a splitting proportional to
$\exp(-S/\hbar)$, where $S$ is the finite action for one traversal of
the barrier.  This holds for the Razavy and for the quartic double
well.  For the quartic double well
\begin{equation}
  \Ham_{\rm q}
  =-\hbar^2\partial_x^2+V_{\rm q}(x),
  \qquad V_{\rm q}(x)=-x^2+\lambda x^4,\quad \lambda>0,
  \label{eq:quartic_hamiltonian}
\end{equation}
the minima lie at $x=\pm x_{\rm q}$ with
$x_{\rm q}=(2\lambda)^{-1/2}$, where
$V_{{\rm q},\min}=-1/(4\lambda)$, and the barrier top is at
$V_{{\rm q},\mathrm{top}}=0$.  The barrier action at the bottom of the
well is
\begin{equation}
  S_{\rm q}^{(0)}
  =\int_{-x_{\rm q}}^{x_{\rm q}}
    \sqrt{V_{\rm q}(x)+\frac{1}{4\lambda}}\,\dd x
  =\frac{\sqrt{2}}{3\lambda}.
  \label{eq:quartic_bottom_action}
\end{equation}
For the Razavy potential, \eqref{eq:completed_square_potential} similarly
gives
\begin{equation}
  S_{\rm R}^{(0)}
  =\int_{-x_0}^{x_0}\sqrt{V(x)-\Vmin}\,\dd x
  =\beta\arcosh\frac{\beta}{2\kappa}
   -\sqrt{\beta^2-4\kappa^2}.
  \label{eq:razavy_bottom_action}
\end{equation}
Both actions are finite at fixed parameters, so that the splitting is
exponentially small as $\hbar\to0^+$.

The limit studied in most of this paper is different.  We set
$\hbar=1$, keep $\beta$ and the level index $n$ fixed, and send
$\kappa\to0^+$.  In the variables of
\eqref{eq:hbar_rescaling_dictionary} this is the limit $\hat\kappa\to0$
at fixed $\hat\beta$.  The minima then move apart as
\begin{equation}
  x_0=\log\frac{\beta}{\kappa}-\frac{\kappa^2}{\beta^2}+\OO(\kappa^4),
  \qquad 2x_0\sim2\log\frac{\beta}{\kappa}.
  \label{eq:minimum_separation}
\end{equation}
At a Morse energy $E=-r^2/4$ the barrier action, computed in
section~\ref{sec:wkb_small_kappa}, behaves as
\begin{equation}
  S_B\left(-\frac{r^2}{4};\kappa\right)
  =r\log\frac1\kappa+\mathcal C_0(\beta,r)+o(1),
  \qquad
  \ee^{-S_B}\sim \kappa^r\ee^{-\mathcal C_0(\beta,r)}.
  \label{eq:comparison_logarithmic_action}
\end{equation}
The power of $\kappa$ is thus the instanton factor of a family of
potentials whose barrier widens logarithmically.  We write $N$ for the
level number in the WKB expansion at fixed potential and $n$ for the
index within each parity sector, which labels the Morse levels when
$r_n>0$.  The two indices enumerate the same doublets, and we keep
separate symbols only to distinguish the two asymptotic regimes.

For comparison, the limit $\lambda\to0^+$ of
\eqref{eq:quartic_hamiltonian} at fixed $\hbar$ moves the minima apart
as $\lambda^{-1/2}$ and raises the barrier as $\lambda^{-1}$.  The
rescaling $y=\sqrt{\lambda}\,x$ gives
\begin{equation}
  \lambda\Ham_{\rm q}
  =-(\hbar\lambda)^2\partial_y^2-y^2+y^4,
\end{equation}
so that $\hbar\lambda$ is the effective semiclassical parameter.  Up to
algebraic prefactors the low-lying gaps are of order
$\exp[-\sqrt{2}/(3\hbar\lambda)]$, and each well approaches a harmonic
oscillator.
The weak-coupling limit of the quartic well is therefore a
semiclassical limit, whereas that of the Razavy well is not.
Table~\ref{tab:quartic-razavy-comparison} compares the two models.

\begin{table}[!t]
  \centering
  \caption{Comparison of the quartic and Razavy double wells.}
  \label{tab:quartic-razavy-comparison}
  \setlength{\tabcolsep}{5pt}
  \renewcommand{\arraystretch}{1.16}
  \small
  \begin{tabularx}{\textwidth}{@{}>{\bfseries\raggedright\arraybackslash}p{2.55cm}YY@{}}
    \toprule
    Feature & \textbf{Quartic double well} & \textbf{Razavy double well} \\
    \midrule
    Potential and minima
    & $V_{\rm q}=-x^2+\lambda x^4$, $\lambda>0$; minima at
      $x=\pm(2\lambda)^{-1/2}$ and barrier top $V_{\rm q}(0)=0$.
    & $V_{\rm R}=\frac{\kappa^2}{2}\cosh 2x-\beta\kappa\cosh x$,
      $\beta>2\kappa>0$; minima at
      $x=\pm\arcosh(\beta/2\kappa)$.
      \\
    \addlinespace[0.35em]
    Differential equation
        & Triconfluent Heun equation, with one irregular singular point
      of rank three at infinity.
    & Double-confluent Heun equation, with two irregular singular
      points of rank one at $z=0$ and $z=\infty$, where
      $z=\kappa\ee^x$.
      \\
    \addlinespace[0.35em]
        Bound states
    & Connection between recessive solutions in different Stokes
      sectors of the same singular point.
    & Connection between the recessive solutions at the two singular
      points, with separate conditions for even and odd states.
      \\
    \addlinespace[0.35em]
        Semiclassical limit
    & At fixed $\lambda$, $\hbar\to0^+$ gives a splitting of order
      $\ee^{-S_{\rm q}/\hbar}$ with a finite barrier action.
    & At fixed $(\kappa,\beta)$, $\hbar\to0^+$ gives a splitting of
      order $\ee^{-S_B/\hbar}$ with a finite barrier action.
      \\
    \addlinespace[0.35em]
        Weak-coupling limit ($\hbar=1$)
    & As $\lambda\to0^+$ the minima separate as $\lambda^{-1/2}$, and
      the splitting is of order $\exp[-\sqrt{2}/(3\lambda)]$ up to
      algebraic prefactors.
    & As $\kappa\to0^+$ at fixed $\beta$,
      $x_0\sim\log(\beta/\kappa)$, and the wells become Morse wells
      with $E_n^{(0)}=-r_n^2/4$ and $\Delta E_n\propto\kappa^{r_n}$,
      $r_n=\beta-1-2n>0$.
      \\
    \addlinespace[0.35em]
        Quasi-exact solvability
    & No analogous algebraic sector.
    & At $\beta/\hbar=M\in\mathbb{Z}^+$ the lowest $M$ levels are the
      eigenvalues of a spin Hamiltonian with $S=(M-1)/2$.
      \\
    \bottomrule
  \end{tabularx}
\end{table}

\subsection{Turning points and the genus-one WKB curve}
\label{sec:spectral_curve}

For an energy below the barrier, $\Vmin<E<\Vtop$, we set
\begin{equation}
  u=\cosh x,\qquad
  \xi=\frac{\beta}{2\kappa},\qquad
  \eta=\frac{\sqrt{E-\Vmin}}{\kappa},\qquad
  u_\pm=\xi\pm\eta.
  \label{eq:turning_point_parameters}
\end{equation}
The condition on the energy is equivalent to
\begin{equation}
  0<\eta<\xi-1,
  \qquad 1<u_-<u_+,
  \label{eq:turning_point_ordering}
\end{equation}
and the turning points of the right well are $x_\pm=\arcosh u_\pm$.
The classical momentum and the measure become
\begin{equation}
  E-V(x)=\kappa^2(u_+-u)(u-u_-),
  \qquad
  \dd x=\frac{\dd u}{\sqrt{u^2-1}}.
  \label{eq:classical_momentum_u}
\end{equation}
In the variable $w=\ee^x\in\mathbb C^*$ the four turning points are the
roots of the self-reciprocal quartic
\begin{equation}
  w^2[E-V(x)]
  =-\frac{\kappa^2}{4}w^4+\frac{\beta \kappa}{2}w^3+Ew^2
   +\frac{\beta \kappa}{2}w-\frac{\kappa^2}{4}.
  \label{eq:reciprocal_turning_quartic}
\end{equation}
Setting $v=wp$, where $p=\sqrt{E-V}$ is the classical momentum, turns
\eqref{eq:reciprocal_turning_quartic} into the equation of a genus-one
curve in the variables $(w,v)$.
The pairing of its branch points under $w\mapsto1/w$ expresses the
reflection symmetry $x\mapsto-x$.  In the invariant coordinate
$u=(w+w^{-1})/2$, the measure in \eqref{eq:classical_momentum_u} leads
to the elliptic curve
\begin{equation}
  \Sigma_E:\qquad
  y^2=(u^2-1)(u_+-u)(u-u_-).
  \label{eq:razavy_spectral_curve}
\end{equation}
It has genus one for distinct branch points $-1,1,u_-,u_+$, and
$p\,\dd x$ is a meromorphic differential on it.

The quartic problem \eqref{eq:quartic_hamiltonian} has the curve
\begin{equation}
  p_{\rm q}^2=E+x^2-\lambda x^4.
  \label{eq:quartic_spectral_curve}
\end{equation}
For $-1/(4\lambda)<E<0$ its four turning points are real and distinct,
and the curve again has genus one.  Below the barrier the two models
therefore have the same cycles.  We take the $A$-cycle to encircle the
classically allowed interval of the right well and the $B$-cycle to
encircle the forbidden interval between the wells.  The $A$-period
governs the quantization in a single well, while one traversal of the
barrier corresponds to half of the $B$-cycle.  For this reason the
exact quantization condition of a symmetric double well involves the
square root of the Voros multiplier of the $B$-cycle.  Orientations and
normalizations are fixed in section~\ref{sec:exact_wkb}.

The curve \eqref{eq:razavy_spectral_curve} degenerates in three ways.
As $E\to\Vmin$ the branch points $u_\pm$ coalesce at the minimum.  As
$E\to\Vtop$ the branch point $u_-$ approaches $1$ and the two inner
turning points merge at the barrier top.  Finally, as $\kappa\to0^+$ at
fixed $\beta$ and fixed Morse energy, both $u_\pm$ go to infinity, and
a finite Morse curve remains after a rescaling near either well.

\subsection{The Morse limit and its bound states}
\label{sec:morse_limit}

We introduce separate coordinates for the right and the left well,
\begin{equation}
  z_R=\kappa\ee^x,
  \qquad
  z_L=\kappa\ee^{-x},
  \qquad
  z_R z_L=\kappa^2.
  \label{eq:morse_coordinate}
\end{equation}
Reflection exchanges $z_R$ and $z_L$, and the origin $x=0$ corresponds
to $z_R=z_L=\kappa$.  In terms of $z=z_R$ the potential is exactly
\begin{equation}
  V(x;\kappa,\beta)
  =\frac{z^2}{4}-\frac{\beta z}{2}
   -\frac{\beta \kappa^2}{2z}+\frac{\kappa^4}{4z^2}.
  \label{eq:razavy_morse_expansion}
\end{equation}
At fixed $z$ the limit $\kappa\to0^+$ is the Morse Hamiltonian
\begin{equation}
  H_{\rm M}
  =-\left(z\frac{\dd}{\dd z}\right)^2
   +\frac{z^2}{4}-\frac{\beta z}{2},
  \qquad
  \dd x=\frac{\dd z}{z},
  \label{eq:morse_potential}
\end{equation}
on $L^2((0,\infty),\dd z/z)$, and the coordinate $z_L$ gives an
identical copy.  The two copies interact only near
$z_R\sim z_L\sim\kappa$, a point that moves to $z=0$ in either
coordinate as $\kappa\to0^+$, so that they decouple in the limit.

For $\hbar=1$ the bound states of one copy have the energies
\begin{equation}
  E_n^{(0)}=-\frac{r_n^2}{4},
  \qquad
  r_n=\beta-1-2n>0,
  \qquad n=0,1,\ldots.
  \label{eq:morse_energies}
\end{equation}
The normalized wavefunctions are
\begin{equation}
  \phi_n(z)=\left[\frac{r_n n!}{\Gamma(n+r_n+1)}\right]^{1/2}
       z^{r_n/2}\ee^{-z/2}L_n^{(r_n)}(z),
  \qquad
  \int_0^\infty\frac{\dd z}{z}\,|\phi_n(z)|^2=1.
  \label{eq:normalised_morse_state}
\end{equation}
These are the Morse levels and wavefunctions \cite{Morse:1929} in our
normalization of the kinetic term.  The condition $r_n>0$ selects the
normalizable states.  For each such $n$, both $E_{n,\mathrm e}$ and
$E_{n,\mathrm o}$ tend to $E_n^{(0)}$ as $\kappa\to0^+$.  Levels with
$r_n\leq0$, such as the $n=3$ levels in
figure~\ref{fig:razavy_level_examples}, exist for every $\kappa>0$ but
do not converge to a Morse bound state.

For general $\hbar$ the bound states of each Morse well are labeled by
\begin{equation}
  n=0,\ldots,\Big\lceil\frac{\hat\beta-1}{2}\Big\rceil-1,
  \label{eq:morse_level_count}
\end{equation}
a range that is empty for $0<\hat\beta\leq1$.  In
section~\ref{sec:qes_subbarrier_count} we show that twice the number of
these states is a sharp upper bound, valid for all $\kappa$, on the
number of levels below the barrier.

At small $z$ the state \eqref{eq:normalised_morse_state} behaves as
$z^{r_n/2}$.  Each localized state therefore reaches the origin $x=0$
with an amplitude of order $\kappa^{r_n/2}$, and the matrix element
between the two wells is of order $\kappa^{r_n}$.  The power of $\kappa$ in the splitting is thus
$r_n$, twice the exponent with which the Morse wavefunction decays
under the barrier.  The estimate can be made quantitative with Herring's
formula \cite{Herring:1962,Garg:2000}, which expresses the splitting
through the localized state at the origin,
$\Delta E_n\simeq4\,\phi_n\,\partial_x\phi_n|_{x=0}$ in our
normalization.  Inserting the small-$z$ behavior of
\eqref{eq:normalised_morse_state} gives $2\Gamma(n+r_n+1)\kappa^{r_n}/[n!\,\Gamma(r_n)^2]$, which is the
leading term of the gap derived from the connection formula in
section~\ref{sec:heund_splitting}.  The connection problem is needed
for the corrections to this estimate and for its behavior at integer
$r_n$.

Equation \eqref{eq:razavy_morse_expansion} also shows that
Rayleigh--Schr\"odinger perturbation theory in $\kappa^2$ about a
single Morse well cannot describe the doublet.  The perturbation series
is defined at fixed $z_R$ or at fixed $z_L$, whereas the splitting is
determined by the wavefunction near the origin.  For $r_n>1$ the
first-order shift, which is common to both members of the doublet, is
finite, and the mean energy of the doublet is
\begin{equation}
  E_{n,\mathrm{centroid}}(\kappa)
  =-\frac{r_n^2}{4}
   -\frac{\beta^2}{2(r_n^2-1)}\kappa^2+o(\kappa^2),
  \label{eq:morse_centroid_shift}
\end{equation}
while the splitting starts at order $\kappa^{r_n}$.  The terms omitted
in \eqref{eq:morse_centroid_shift} do not form a power series in
$\kappa^2$, because contributions from two traversals of the barrier
enter at order $\kappa^{2r_n}$ and carry logarithms that originate in
the factor $\kappa^{2\nu}$ of the connection formula.  At $r_n=1$ the
coefficient in \eqref{eq:morse_centroid_shift} has a pole, a first
instance of the resonances studied in
section~\ref{sec:resonance_qes}, and for $0<r_n<1$ the first-order
expectation value diverges.  All of these features are captured by the
connection problem to which we now turn.

\FloatBarrier

\section{The double-confluent Heun connection problem}
\label{sec:heund}

In this section we set $\hbar=1$, which is restored in
section~\ref{sec:exact_wkb}.  With
\begin{equation}
  z=\kappa\ee^x,
  \qquad
  \psi(x)=z^{-1/2}\Phi(z),
  \label{eq:liouville_transformation}
\end{equation}
the Razavy eigenvalue equation becomes
\begin{equation}
  \Phi''(z)+\left[
  -\frac14+\frac{\beta}{2z}+\frac{E+\frac14}{z^2}
  +\frac{\beta \kappa^2}{2z^3}-\frac{\kappa^4}{4z^4}
  \right]\Phi(z)=0.
  \label{eq:razavy_dche}
\end{equation}
This is a double-confluent Heun equation with irregular singularities
of rank one at $z=0$ and $z=\infty$, which correspond to $x=-\infty$
and $x=+\infty$.  In the weak-coupling limit they belong to the two
Morse wells of section~\ref{sec:morse_limit}, described by the
coordinates $z_L=\kappa^2/z$ and $z_R=z$.  A bound state is a solution
of \eqref{eq:razavy_dche} that is recessive at both singularities, so
that the spectral problem is a connection problem between them.

We first fix the normal form of the equation and construct the local
solutions at each singularity.  We then use the reflection symmetry to
factorize the spectral determinant into an even and an odd part, and
finally obtain the splitting and its first corrections from the
connection formula in terms of irregular conformal blocks.  The
derivation assumes a noninteger exponent $r_n$; the integer case is
treated in section~\ref{sec:resonance_qes}.  The recursion for the
block coefficients and the multi-traversal terms are collected in
appendix~\ref{app:floquet_formal}, and an alternative construction of
the connection coefficient from a recurrence relation is given in
appendix~\ref{app:recurrence_connection}.

\subsection{Normal and standard forms}
\label{sec:heund_normal_form}

Following \cite{Bonelli:2023}, we write the double-confluent Heun
equation without a first-derivative term as
\begin{equation}
  \Phi''(z)+\left[-\frac{L_1^2}{4}+\frac{m_1L_1}{z}
  +\frac{u}{z^2}+\frac{m_2L_2}{z^3}
  -\frac{L_2^2}{4z^4}\right]\Phi(z)=0.
  \label{eq:canonical_dche}
\end{equation}
We call \eqref{eq:canonical_dche} the \emph{normal form}.  The Razavy
equation \eqref{eq:razavy_dche} corresponds to
\begin{equation}
  L_1=1,\qquad L_2=\kappa^2,\qquad
  m_1=m_2=\frac{\beta}{2},\qquad
  u=E+\frac14,
  \qquad \lambda\equiv L_1L_2=\kappa^2.
  \label{eq:razavy_dche_dictionary}
\end{equation}
For the construction of the local solutions we use the \emph{standard
form} of Mathematica's
$\mathrm{HeunD}[\bar q,\bar\alpha,\bar\gamma,
\bar\delta,\bar\epsilon,z]$ \cite{Wolfram:HeunD},
\begin{equation}
  Y''(z)+\left(\frac{\bar\gamma}{z^2}
       +\frac{\bar\delta}{z}+\bar\epsilon\right)Y'(z)
       +\frac{\bar\alpha z-\bar q}{z^2}Y(z)=0.
  \label{eq:dche_standard_form}
\end{equation}
We assume $\bar\gamma\bar\epsilon\ne0$, so that both $z=0$ and
$z=\infty$ are irregular singularities of rank one.  Mathematica's
function is normalized at the ordinary point $z=1$, with $Y(1)=1$ and
$Y'(1)=0$.  The connection problem requires instead a solution
normalized at a singular point, which we denote by $\mathrm{HeunD}_0$
and define in section~\ref{sec:local-solutions}.

The normal form \eqref{eq:canonical_dche} and the standard form
\eqref{eq:dche_standard_form} are related by
\begin{align}
  \Phi(z)&=G(z)Y(z),\qquad
  G(z)=z^{1+m_2}\exp\!\left(\frac{L_1z}{2}+\frac{L_2}{2z}\right)
      =z^{\bar\delta/2}
       \exp\!\left(\frac{\bar\epsilon z-\bar\gamma/z}{2}\right),
  \label{eq:dche_standard_gauge}\\
  \bar q&=\frac{\lambda}{2}-u-m_2(m_2+1),\qquad
  \bar\alpha=L_1(1+m_1+m_2),\qquad
  \bar\gamma=-L_2,\nonumber\\
  \bar\delta&=2+2m_2,\qquad \bar\epsilon=L_1.
  \label{eq:dche_standard_general_dictionary}
\end{align}
For the Razavy equation this gives
\begin{equation}
  \bar q=\frac{\kappa^2}{2}-E-\frac{(\beta+1)^2}{4},\qquad \bar\alpha=\beta+1,\qquad
  \bar\gamma=-\kappa^2,\qquad \bar\delta=\beta+2,\qquad \bar\epsilon=1.
  \label{eq:dche_standard_dictionary}
\end{equation}
Note that $\bar\gamma=-\kappa^2$ is negative, whereas
$L_2=\kappa^2>0$.

The form in DLMF eq.~(31.12.2) \cite{DLMF:Heun} differs in two
respects: the constant term in the coefficient of $Y'$ is normalized to
one, and the symbols $\gamma$ and $\delta$ are exchanged.  With
$w=\bar\epsilon z$ and $U(w)=Y(w/\bar\epsilon)$ one finds
\begin{equation}
  U''+\left(\frac{\delta_{\rm D}}{w^2}
           +\frac{\gamma_{\rm D}}w+1\right)U'
     +\frac{\alpha_{\rm D}w-q_{\rm D}}{w^2}U=0,
  \qquad
  \begin{gathered}
    q_{\rm D}=\bar q,\quad \alpha_{\rm D}=\bar\alpha/\bar\epsilon,\\
    \gamma_{\rm D}=\bar\delta,\quad
    \delta_{\rm D}=\bar\gamma\bar\epsilon.
  \end{gathered}
  \label{eq:dche_dlmf_conversion}
\end{equation}
For the Razavy equation, where $\bar\epsilon=1$, only the exchange of
$\gamma$ and $\delta$ is needed.

\subsection{Local solutions and the connection problem}
\label{sec:local-solutions}

The connection problem relates bases of solutions defined at the two
irregular singularities.  We construct both bases from a single
function $\mathrm{HeunD}_0$, first as formal series and then by Borel
summation.  Throughout, $z^\rho=\exp(\rho\operatorname{Log}z)$ with
$-\pi<\arg z<\pi$, and continuation beyond this branch is understood on
the Riemann surface of the logarithm.

\paragraph{Formal solutions of the standard equation.}
The standard equation \eqref{eq:dche_standard_form} has a unique formal
power-series solution with constant term one,
\begin{equation}
  \mathrm{HeunD}_0
    [\bar q,\bar\alpha,\bar\gamma,\bar\delta,\bar\epsilon,z]
  =\sum_{n=0}^{\infty}c_n z^n,
  \qquad c_{-1}=0,\quad c_0=1.
  \label{eq:heund0_formal_series}
\end{equation}
Substitution into
$z^2Y''+(\bar\gamma+\bar\delta z+\bar\epsilon z^2)Y'
 +(\bar\alpha z-\bar q)Y=0$ gives the three-term recurrence
\begin{equation}
  \bar\gamma(n+1)c_{n+1}
   +\bigl[n(n+\bar\delta-1)-\bar q\bigr]c_n
   +\bigl[\bar\alpha+(n-1)\bar\epsilon\bigr]c_{n-1}=0,
  \qquad n\ge0.
  \label{eq:heund0_recurrence}
\end{equation}
In particular,
\begin{equation}
  c_1=\frac{\bar q}{\bar\gamma},\qquad
  c_2=\frac{\bar q(\bar q-\bar\delta)
                 -\bar\alpha\bar\gamma}{2\bar\gamma^2}.
  \label{eq:heund0_initial_coefficients}
\end{equation}
The coefficients generically grow factorially, so that
\eqref{eq:heund0_formal_series} is an asymptotic series.  For
exceptional parameter values it converges or terminates; for example,
it terminates at degree $N$ if $c_{N+1}=0$ and
$\bar\alpha+N\bar\epsilon=0$.

Both formal solutions at $z=0$ can be written in terms of this series,
\begin{align}
  \widetilde Y_{01}(z)
   &=\mathrm{HeunD}_0
       [\bar q,\bar\alpha,\bar\gamma,\bar\delta,\bar\epsilon,z],
       \label{eq:heund0_zero_first}\\
  \widetilde Y_{02}(z)
   &=\ee^{\bar\gamma/z-\bar\epsilon z}z^{2-\bar\delta}\,
      \mathrm{HeunD}_0
       [\bar q+\bar\delta-2,\bar\alpha-2\bar\epsilon,
        -\bar\gamma,4-\bar\delta,-\bar\epsilon,z].
       \label{eq:heund0_zero_second}
\end{align}
The second expression follows by substituting
$Y=\ee^{\bar\gamma/z-\bar\epsilon z}z^{2-\bar\delta}H(z)$
into the standard equation.  The leading behaviors are $1$ and
$\ee^{\bar\gamma/z}z^{2-\bar\delta}$, respectively.  If $Y_{01}$ and
$Y_{02}$ denote Borel sums of the two series in a common sector, Abel's
identity and the leading coefficients give
\[
  W_z[Y_{01},Y_{02}]
  =-\bar\gamma\,z^{-\bar\delta}
    \ee^{\bar\gamma/z-\bar\epsilon z}.
\]
The two solutions are therefore independent for $\bar\gamma\ne0$.

To obtain the solutions at infinity we set $t=1/z$.  Once the leading
algebraic or exponential behavior has been factored out, the equation
is again of the standard form.  With
$\rho_{\rm H}=\bar\alpha/\bar\epsilon$ we find
\begin{align}
  \widetilde Y_{\infty1}(z)
   &=\ee^{\bar\gamma/z}z^{-\rho_{\rm H}}\,
      \mathrm{HeunD}_0\bigl[
      \bar q+\bar\gamma\bar\epsilon
                  -\rho_{\rm H}(\rho_{\rm H}+1-\bar\delta),
      \bar\gamma(\rho_{\rm H}+2-\bar\delta),-\bar\epsilon,
      2-\bar\delta+2\rho_{\rm H},\bar\gamma,1/z\bigr],
  \label{eq:heund0_infinity_first}\\
  \widetilde Y_{\infty2}(z)
   &=\ee^{-\bar\epsilon z}z^{\rho_{\rm H}-\bar\delta}\,
      \mathrm{HeunD}_0\bigl[
      \bar q+\bar\gamma\bar\epsilon
                  -(\rho_{\rm H}-\bar\delta)(\rho_{\rm H}-1),
      \bar\gamma(\rho_{\rm H}-\bar\delta),\bar\epsilon,
      2+\bar\delta-2\rho_{\rm H},-\bar\gamma,1/z\bigr].
  \label{eq:heund0_infinity_second}
\end{align}
The leading behaviors are $z^{-\bar\alpha/\bar\epsilon}$ and
$\ee^{-\bar\epsilon z}z^{\bar\alpha/\bar\epsilon-\bar\delta}$, and the
factor $\ee^{\bar\gamma/z}=1+\OO(z^{-1})$ in
\eqref{eq:heund0_infinity_first} does not affect the normalization.
All four series are generated by the recurrence
\eqref{eq:heund0_recurrence} with the appropriate parameters.

\paragraph{Normal-form solutions in terms of $\mathrm{HeunD}_0$.}
Multiplication by the factor $G(z)$ of \eqref{eq:dche_standard_gauge}
gives solutions of the normal form.  They take a symmetric form when
exponential factors that are analytic at the respective singularity are
absorbed into the functions
$\mathrm{HeunD}_0$.\footnote{\label{fn:heund0_identities}%
The functions $\mathrm{HeunD}_0$ obey the identities
\[
  \begin{aligned}
  \mathrm{HeunD}_0[\bar q,\bar\alpha,\bar\gamma,\bar\delta,\bar\epsilon,z]
    &=\ee^{-\bar\epsilon z}\,
      \mathrm{HeunD}_0[\bar q+\bar\gamma\bar\epsilon,
      \bar\alpha-\bar\delta\bar\epsilon,
      \bar\gamma,\bar\delta,-\bar\epsilon,z],\\
  \mathrm{HeunD}_0[\bar q,\bar\alpha,\bar\gamma,\bar\delta,\bar\epsilon,z]
    &=\mathrm{HeunD}_0[\bar q,\bar\alpha/\bar\epsilon,
      \bar\gamma\bar\epsilon,\bar\delta,1,\bar\epsilon z],
      \qquad\bar\epsilon\ne0.
  \end{aligned}
\]
Both follow by substitution into \eqref{eq:dche_standard_form} and hold
for the formal series and for their Borel sums.  In the second
identity the direction of summation changes from $\theta$ to
$\theta+\arg\bar\epsilon$.  Applied to the variable $t=1/z$, the first
identity gives the prefactor and the parameter shifts in
\eqref{eq:heund0_infinity_first}.}
We define
\begin{equation}
  q_{j,\pm}=\frac{\lambda}{2}-u-m_j(m_j\pm 1),
  \qquad j=1,2.
  \label{eq:normal_heund0_q}
\end{equation}
With $s_\pm=m_1+m_2\pm 1$, the four solutions of the normal form are
\begin{align}
  \Phi_{0,\pm}(z)
   &=\ee^{\pm(L_1z+L_2/z)/2}z^{1\pm m_2}\,\mathrm{HeunD}_0[
       q_{2,\pm},L_1s_\pm,\mp L_2,
       2\pm 2m_2,\pm L_1,z],
  \label{eq:normal_zero_heund0}\\
  \Phi_{\infty,\pm}(z)
   &=\ee^{\pm(L_1z+L_2/z)/2}z^{\mp m_1}\,\mathrm{HeunD}_0[
       q_{1,\pm},L_2s_\pm,\mp L_1,
       2\pm 2m_1,\pm L_2,1/z].
  \label{eq:normal_infinity_heund0}
\end{align}
These expressions follow from the solutions of the standard form or by
direct substitution.  The additional factor $\ee^{\tau L_1z/2}$ at
$z=0$, or $\ee^{\tau L_2/(2z)}$ at $z=\infty$, is analytic there and
equal to one at the singularity, so that the solutions are normalized
as
\begin{align}
  \Phi_{0,\tau}(z)
   &\sim\ee^{\tau L_2/(2z)}z^{1+\tau m_2}
       [1+\OO(z)],&&z\to0,
       \label{eq:dche_zero_solutions}\\
  \Phi_{\infty,\tau}(z)
   &\sim\ee^{\tau L_1z/2}z^{-\tau m_1}
       [1+\OO(z^{-1})],&&z\to\infty.
       \label{eq:dche_infinity_solutions}
\end{align}
These asymptotic expansions hold on closed subsectors of the sectors of
summation.

For the Razavy parameters \eqref{eq:razavy_dche_dictionary} we define
\begin{equation}
  q_\tau=\frac{\kappa^2}{2}-E-\frac{(\beta+\tau)^2}{4}.
  \label{eq:razavy_heund0_q}
\end{equation}
The solutions then take the form
\begin{align}
  \Phi_{0,\pm}(z)
   &=\ee^{\pm(z+\kappa^2/z)/2}z^{1\pm\beta/2}\,
     \mathrm{HeunD}_0
     [q_\pm,\beta\pm 1,\mp\kappa^2,2\pm\beta,\pm 1,z],
     \label{eq:razavy_zero_heund0}\\
  \Phi_{\infty,\pm}(z)
   &=\ee^{\pm(z+\kappa^2/z)/2}z^{\mp\beta/2}\,
     \mathrm{HeunD}_0\!\left[
     q_\pm,\kappa^2(\beta\pm 1),\mp 1,
     2\pm\beta,\pm\kappa^2,1/z\right].
     \label{eq:razavy_infinity_heund0}
\end{align}
The bound-state condition involves only the two solutions with
$\tau=-1$, but both signs are needed to form a basis at each
singularity.

\paragraph{Borel summation and sectorial bases.}
The series constructed above are expansions at irregular singularities
of rank one.  Once the exponential and algebraic prefactors have been
removed, their coefficients grow at most factorially,
$|c_n|\le KA^nn!$ with constants $K,A>0$.  They are therefore of Gevrey
order one, and their Borel transforms converge near the origin.  The
theory of summability then provides analytic solutions in sectors
\cite{Balser:1994,LodayRichaudRemy:2011}.  For a formal series
$\widetilde F(t)=\sum_{n\ge0}c_nt^n$ we define the Borel transform and
its Laplace transform in the direction $\theta$ by \cite{Sauzin:2014}
\begin{align}
  \mathcal B\widetilde F(\xi)
   &=\sum_{n=0}^{\infty}\frac{c_n}{n!}\xi^n,
     \label{eq:local_borel_transform}\\
  (\mathcal L_\theta\widetilde F)(t)
   &=\frac1t\int_0^{\ee^{\ii\theta}\infty}
       \ee^{-\xi/t}\mathcal B\widetilde F(\xi)\,\dd\xi
    =\frac{\ee^{\ii\theta}}t\int_0^\infty
       \ee^{-s\ee^{\ii\theta}/t}
       \mathcal B\widetilde F(s\ee^{\ii\theta})\,\dd s.
     \label{eq:local_borel_laplace}
\end{align}
The integral involves the analytic continuation of the Borel transform
along the ray $\arg\xi=\theta$.  It is defined for
$|\arg t-\theta|<\pi/2$ and sufficiently small $|t|$, provided this
continuation grows at most exponentially.  If the ray contains
singularities, the contour is deformed above or below them, which
defines the lateral sums $\mathcal L_{\theta^+}$ and
$\mathcal L_{\theta^-}$.  Accordingly, we define
\begin{equation}
  \mathrm{HeunD}_0^{(\theta)}
   [\bar q,\bar\alpha,\bar\gamma,\bar\delta,\bar\epsilon,t]
  :=\mathcal L_\theta\!\bigl(
      \mathrm{HeunD}_0
       [\bar q,\bar\alpha,\bar\gamma,\bar\delta,\bar\epsilon,\cdot]
     \bigr)(t).
  \label{eq:heund0_sectorial_definition}
\end{equation}
We omit the direction $\theta$ once it has been fixed.  For the
solutions at infinity the variable is $t=1/z$.  The functions defined
in this way near the two singularities are continued analytically to a
common regular point, where the Laplace integrals themselves need not
converge.

For the Razavy parameters the recessive solutions decay in the sectors
\begin{equation}
  \mathcal S_0=\{\,|\arg(\kappa^2/z)|<\pi/2\,\},
  \qquad
  \mathcal S_\infty=\{\,|\arg z|<\pi/2\,\}.
  \label{eq:physical_stokes_sectors}
\end{equation}
By \eqref{eq:heund0_recurrence}, the Borel transform of the series
\eqref{eq:heund0_formal_series} can be singular at finite $\xi\neq0$
only at $\xi=-\bar\gamma$.  The series in \eqref{eq:normal_zero_heund0}
and \eqref{eq:normal_infinity_heund0} can therefore be singular at
$\xi=\tau L_2$ and $\xi=\tau L_1$, respectively, in accordance with the
exponential factors that separate the two solutions at each
singularity \cite{LodayRichaudRemy:2011}.  The recessive solutions,
$\tau=-1$, are thus Borel summable along the positive real axis,
whereas the dominant ones, $\tau=+1$, in general require lateral sums.
The two lateral sums of a dominant solution differ by a multiple of the
recessive solution, which is itself unambiguous
\cite[sec.~1]{Schaefke:1984}.

A superscript $\pm$ on a solution or on a connection coefficient refers
to the continuation $z\to z\ee^{\pm\ii0}$.  Since the inversion
$z\mapsto\kappa^2/z$ reverses this sign, we take the basis at infinity
in the sector $0<\arg z<\theta_0$ and the basis at zero in
$-\theta_0<\arg z<0$, for small $\theta_0>0$, or the opposite pair.  The
inversion exchanges these two sectors, so that reflection relates the
solutions without crossing a Stokes ray.  The directions of summation
are the arguments of $t=z$ and $t=1/z$, respectively.  For real
parameters the two choices are complex conjugates of each other, and
their recessive solutions coincide on $z>0$.

For numerical evaluation the Borel transform can be continued by
diagonal Pad\'e approximants $P_N=U_N/V_N$, whose numerator and
denominator have degree at most $N$, with $V_N(0)=1$, and are fixed by
the coefficients $c_0,\ldots,c_{2N}$,
\begin{equation}
  \begin{aligned}
    P_N(\xi)&=\frac{U_N(\xi)}{V_N(\xi)},\qquad
      V_N(\xi)\mathcal B\widetilde F(\xi)-U_N(\xi)
        =\OO(\xi^{2N+1}),\\
    F_\theta^{[N]}(t)&=\frac1t\int_0^{\ee^{\ii\theta}\infty}
       \ee^{-\xi/t}P_N(\xi)\,\dd\xi.
  \end{aligned}
  \label{eq:local_borel_pade}
\end{equation}
The function $F_\theta^{[N]}$ approximates
$\mathrm{HeunD}_0^{(\theta)}$, with the contour chosen according to the
lateral prescription.  Its accuracy can be monitored by varying $N$ and
the matching point and by testing the constancy of the Wronskians.

\paragraph{Connection matrix and spectral determinant.}
With the sectors fixed as above, we define the connection matrix $C$ by
\begin{equation}
  \begin{pmatrix}\Phi_{0,+}\\\Phi_{0,-}\end{pmatrix}
  =C\begin{pmatrix}\Phi_{\infty,+}\\\Phi_{\infty,-}\end{pmatrix},
  \qquad
  \Phi_{0,\tau}=C_{\tau,+}\Phi_{\infty,+}
                  +C_{\tau,-}\Phi_{\infty,-}.
  \label{eq:dche_connection_matrix}
\end{equation}
In terms of the Wronskian $W_z[f,g]=fg'-f'g$, where primes denote
derivatives with respect to $z$,
\begin{equation}
  C=\frac{1}{W_z[\Phi_{\infty,+},\Phi_{\infty,-}]}
  \begin{pmatrix}
    W_z[\Phi_{0,+},\Phi_{\infty,-}]&
    W_z[\Phi_{\infty,+},\Phi_{0,+}]\\[2pt]
    W_z[\Phi_{0,-},\Phi_{\infty,-}]&
    W_z[\Phi_{\infty,+},\Phi_{0,-}]
  \end{pmatrix}.
  \label{eq:local_wronskian_connection_matrix}
\end{equation}
The Wronskians are evaluated with the analytically continued solutions
at a common regular point.  They are constant because
\eqref{eq:canonical_dche} has no first-derivative term.  The
normalizations \eqref{eq:dche_zero_solutions} and
\eqref{eq:dche_infinity_solutions} give
\begin{equation}
  W_z[\Phi_{0,+},\Phi_{0,-}]=L_2,
  \qquad W_z[\Phi_{\infty,+},\Phi_{\infty,-}]=-L_1,
  \qquad \det C=-\frac{L_2}{L_1}.
  \label{eq:local_wronskian_normalisations}
\end{equation}
The Wronskians that involve both singularities contain the connection
data and are not determined by the local expansions.

In terms of the wavefunction we define
\begin{align}
  \psi_L(x)=z^{-1/2}\Phi_{0,-}(z)
   &\sim\ee^{-\kappa^2/(2z)}z^{(1-\beta)/2},&&z\to0,
   \label{eq:left_jost_asymptotic}\\
  \psi_R(x)=z^{-1/2}\Phi_{\infty,-}(z)
   &\sim\ee^{-z/2}z^{(\beta-1)/2},&&z\to\infty.
   \label{eq:right_jost_asymptotic}
\end{align}
For real $E$ these are the real solutions that are square integrable at
$x\to-\infty$ and $x\to+\infty$, respectively, normalized as indicated.
The transformation \eqref{eq:liouville_transformation} preserves the
Wronskian, $W_x[\psi_L,\psi_R]=W_z[\Phi_{0,-},\Phi_{\infty,-}]$.  We
define the spectral determinant
\begin{equation}
  \mathcal D(E;\kappa,\beta)
  =W_z[\Phi_{0,-},\Phi_{\infty,-}]
  =-C_{-,+},
  \label{eq:heun_spectral_determinant}
\end{equation}
where we have used $L_1=1$.  It vanishes exactly on the spectrum,
\begin{equation}
  \mathcal D(E;\kappa,\beta)=0
  \quad\Longleftrightarrow\quad C_{-,+}=0
  \quad\Longleftrightarrow\quad E\in\operatorname{spec}\Ham.
  \label{eq:wronskian_quantisation}
\end{equation}
At a zero the two recessive solutions are proportional and define a
wavefunction in $L^2(\RR)$.  The determinant does not depend on the
lateral prescription for the dominant solutions.

\paragraph{Reflection and parity.}
On solutions of the normal form the reflection $x\mapsto-x$ acts as
\begin{equation}
  (\mathsf P\Phi)(z)=\frac z\kappa
       \Phi\!\left(\frac{\kappa^2}{z}\right),
  \qquad \mathsf P^2=1.
  \label{eq:dche_reflection_operator}
\end{equation}
The prefactor follows from $\psi=z^{-1/2}\Phi$.  Comparing the leading
behavior of the recessive solutions gives
\begin{equation}
  \mathsf P\Phi_{0,-}=\kappa^{1-\beta}\Phi_{\infty,-},
  \qquad
  \mathsf P\Phi_{\infty,-}=\kappa^{\beta-1}\Phi_{0,-}.
  \label{eq:reflection_on_recessive_bases}
\end{equation}
The rescaled solutions
\begin{equation}
  \hat\Phi_0=\kappa^{(\beta-1)/2}\Phi_{0,-},
  \qquad
  \hat\Phi_\infty=\kappa^{(1-\beta)/2}\Phi_{\infty,-}
  \label{eq:reflection_symmetric_normalisation}
\end{equation}
are therefore exchanged by $\mathsf P$.  At an eigenvalue,
\begin{equation}
  \hat\Phi_0=\epsilon\hat\Phi_\infty,
  \qquad \epsilon=\kappa^{\beta-1}C_{-,-}=\pm1,
  \label{eq:local_connection_parity}
\end{equation}
and $\epsilon$ is the parity of the bound state.  Equivalently, at the
fixed point $z=\kappa$ of the reflection,
\begin{equation}
  \psi_R'(0)=0\quad(\epsilon=+1,\ \mathrm{even}),
  \qquad
  \psi_R(0)=0\quad(\epsilon=-1,\ \mathrm{odd}),
  \label{eq:parity_boundary_conditions}
\end{equation}
where the prime denotes a derivative with respect to $x$.

Reflection also factorizes the determinant for arbitrary $E$.  Let
$f(z)=\Phi_{\infty,-}(z)$, so that $\psi_R(x)=z^{-1/2}f(z)$ and
\[
  \frac{\dd\psi_R}{\dd x}
  =z^{-1/2}\left[zf'(z)-\frac12f(z)\right],
\]
where $f'$ denotes a derivative with respect to $z$.  We define
\begin{equation}
  \mathcal D_{\rm o}(E;\kappa,\beta)=f(\kappa),
  \qquad
  \mathcal D_{\rm e}(E;\kappa,\beta)
    =\kappa f'(\kappa)-\frac12f(\kappa).
  \label{eq:local_parity_determinants}
\end{equation}
These functions satisfy $\mathcal D_{\rm o}=\sqrt\kappa\,\psi_R(0)$ and
$\mathcal D_{\rm e}=\sqrt\kappa\,\psi_R'(0)$.  Their zeros are the
eigenvalues of the half-line problems with Dirichlet and Neumann
conditions at $x=0$, that is, the odd and the even levels.

For the full determinant we use
$\psi_L(x)=\kappa^{1-\beta}\psi_R(-x)$, which follows from
\eqref{eq:reflection_on_recessive_bases} and implies
$\psi_L(0)=\kappa^{1-\beta}\psi_R(0)$ and
$\psi_L'(0)=-\kappa^{1-\beta}\psi_R'(0)$.  The two terms of
$W_x[\psi_L,\psi_R]=\psi_L\psi_R'-\psi_L'\psi_R$ then add, and we
obtain
\begin{equation}
  \mathcal D(E;\kappa,\beta)
    =2\kappa^{1-\beta}\psi_R(0)\psi_R'(0)
    =2\kappa^{-\beta}\mathcal D_{\rm o}(E;\kappa,\beta)
                          \mathcal D_{\rm e}(E;\kappa,\beta).
  \label{eq:local_parity_factorisation}
\end{equation}
This identity holds for all $E$.  The two factors vanish on the odd and
on the even levels, respectively, and they have no common zero, since a
solution with $\psi_R(0)=\psi_R'(0)=0$ vanishes identically.

In section~\ref{sec:irregular_block} we evaluate $C_{-,+}$ in terms of
the irregular block.  Appendix~\ref{app:recurrence_connection} gives an
independent representation of the same determinant as the limit of a
recurrence, following \cite{Schaefke:1984,Wolf:1998}.

\subsection{The irregular-block connection formula}
\label{sec:irregular_block}

We first define the Floquet exponent.  Since $V(x+2\pi\ii)=V(x)$, the
continuation of the solutions of the Schr\"odinger equation from $x$ to
$x+2\pi\ii$ defines a monodromy matrix $M_\psi$, and the constancy of
the Wronskian gives $\det M_\psi=1$.  Its eigenvalues define the
characteristic exponent $\nu$ \cite{AbadGomezSesma:2008},
\begin{equation}
  \operatorname{spec}M_\psi
  =\{\ee^{2\pi\ii\nu},\ee^{-2\pi\ii\nu}\},
  \qquad
  \psi_\nu(x+2\pi\ii)=\ee^{2\pi\ii\nu}\psi_\nu(x).
\end{equation}
This determines $\nu$ only up to a sign and integer shifts, and we
choose the branch on which $E\to-\nu^2$ as $\kappa\to0$.  The exponent
is then a function of $E$, $\kappa$, and $\beta$.  The factor in
$\psi=\ee^{-\kappa\cosh x}\phi$ is periodic and does not change the
multipliers, and the Floquet series
$\phi_\nu=\sum_{m\in\mathbb Z}c_m\ee^{(\nu+m)x}$ leads to the
recurrence of appendix~\ref{app:block_recursion}.  For
$\Phi=z^{1/2}\psi$ a circuit around $z=0$ gives an additional sign, so
that the multipliers are $-\ee^{\pm2\pi\ii\nu}$.

The connection matrix of \eqref{eq:canonical_dche} can be written in
terms of the classical limit of an irregular Virasoro block of rank one
\cite{Bonelli:2023}, in which $\nu$ plays the role of the intermediate
momentum.  The classical block is defined by
\begin{equation}
  \mathcal F(m_1,\nu,m_2;\lambda)
  =\lim_{\mathfrak b\to0}\mathfrak b^2
   \log\!\left[(\Lambda_1\Lambda_2)^{-\Delta_\alpha}
   {}_1\mathfrak F_1
   (\mu_1,\alpha,\mu_2;\Lambda_1\Lambda_2)\right],
  \label{eq:classical_irregular_block_definition}
\end{equation}
with
\begin{equation}
  \mathfrak b\alpha\to\nu,\qquad
  \mathfrak b\mu_i\to m_i,\qquad
  \mathfrak b\Lambda_i\to L_i,\qquad
  \lambda=L_1L_2,
  \label{eq:classical_block_scaling}
\end{equation}
and the Liouville conventions
\begin{equation}
  Q=\mathfrak b+\mathfrak b^{-1},
  \qquad
  \Delta_\alpha=\frac{Q^2}{4}-\alpha^2.
  \label{eq:liouville_dimension_convention}
\end{equation}
The block ${}_1\mathfrak F_1$ has an explicit expansion as a sum over
Young diagrams \cite{Bonelli:2023}.  The energy is related to $\nu$ by
the Matone-type relation \cite{Matone:1995,Flume:2004}
\begin{equation}
  E=-\nu^2+\lambda\partial_\lambda
  \mathcal F(\nu,\beta;\lambda),
  \qquad \lambda=\kappa^2,
  \label{eq:matone_energy_relation}
\end{equation}
where we suppress the masses $m_1=m_2=\beta/2$.  The first term of the
block is
\begin{equation}
  \mathcal F(\nu,\beta;\lambda)
  =-\frac{\beta^2}{2(4\nu^2-1)}\lambda+\OO(\lambda^2).
  \label{eq:irregular_block_first_term}
\end{equation}
The next coefficient and a recursion that determines all of them are
given in appendix~\ref{app:block_recursion}.  The coefficients are
rational functions of $\nu$ with poles at $\nu=\pm r/2$,
$r\in\mathbb Z^+$, which we call Kac poles; those of
\eqref{eq:irregular_block_first_term} are at $\nu=\pm\tfrac12$.  They
are the classical counterparts of the poles of Zamolodchikov's
recursion at degenerate intermediate weights
\cite[sec.~2.2]{ChoCollierYin:2017}, where the Gram matrix of the
intermediate Verma module is singular.  Since the classical block is
the logarithm of a quantum block, its coefficients can have poles of
higher order.  The Kac poles should be distinguished from the poles of
the gamma functions in the connection coefficients.

Specializing eq.~(4.4.10) of \cite{Bonelli:2023} to the Razavy
parameters \eqref{eq:razavy_dche_dictionary}, we find for the
coefficient of the growing solution at infinity
\begin{equation}
  C_{-,+}=\mathcal N_0
  \sum_{\sigma=\pm1}
  \frac{\Gamma(-2\sigma\nu)\Gamma(1-2\sigma\nu)}
       {\Gamma\!\left(\frac{1-\beta}{2}-\sigma\nu\right)^2}
  \kappa^{2\sigma\nu}
  \exp\!\left[-\frac{\sigma}{2}\partial_\nu\mathcal F\right].
  \label{eq:razavy_growing_coefficient}
\end{equation}
The normalization factor is
\begin{equation}
  \mathcal N_0=\kappa^{1-\beta}
  \exp\!\left[-\frac12(\partial_{m_1}+\partial_{m_2})
  \mathcal F(m_1,\nu,m_2;\kappa^2)\right]_{m_1=m_2=\beta/2}.
  \label{eq:dche_common_normalisation}
\end{equation}
The derivatives with respect to the masses are taken at fixed $\nu$ and
$\lambda$.  The factor $\mathcal N_0$ accounts for the difference
between the normalization of the conformal-block solutions and that of
\eqref{eq:dche_zero_solutions} and \eqref{eq:dche_infinity_solutions}.
Indeed, for $L_1=1$ the solutions $\mathcal U_{0,-}$ and
$\mathcal U_{\infty,+}$ of
\cite[eqs.~(3.4.17), (3.4.20)]{Bonelli:2023} are related to ours by
\begin{equation}
  \Phi_{0,-}=L_2^{1/2-m_2}
      \ee^{-\frac12\partial_{m_2}\mathcal F}\mathcal U_{0,-},
  \qquad
  \Phi_{\infty,+}=
      \ee^{\frac12\partial_{m_1}\mathcal F}\mathcal U_{\infty,+}.
\end{equation}
If $\mathcal K_{-,+}$ denotes the corresponding connection coefficient
in that basis, then
\begin{equation}
  C_{-,+}=L_2^{1/2-m_2}
     \ee^{-\frac12(\partial_{m_1}+\partial_{m_2})\mathcal F}
     \mathcal K_{-,+}.
\end{equation}
The sum over $\sigma$ in \eqref{eq:razavy_growing_coefficient} is
$\mathcal K_{-,+}$, and $L_2^{1/2-m_2}=\kappa^{1-\beta}$ gives
\eqref{eq:dche_common_normalisation}.  Since
$\partial_\beta\mathcal F=(\partial_{m_1}\mathcal F+
\partial_{m_2}\mathcal F)/2$ at equal masses, we can also write
\begin{equation}
  \mathcal N_0=\kappa^{1-\beta}
      \exp[-\partial_\beta\mathcal F(\nu,\beta;\kappa^2)],
\end{equation}
with the derivative taken at fixed $\nu$ and $\kappa$.  Away from
resonance $\mathcal N_0$ is finite and nonzero, so that it does not
affect the zeros of $C_{-,+}$.

For sufficiently small $\kappa$, and away from the poles of the gamma
functions and the Kac poles of the block, the condition $C_{-,+}=0$ can
be written as
\begin{equation}
  \begin{aligned}
  X(\nu;\kappa,\beta)&=
  \kappa^{2\nu}\exp\!\left[-\frac12\partial_\nu\mathcal F\right]
  \frac{\Gamma(-2\nu)
  \Gamma\!\left(\frac{1-\beta}{2}+\nu\right)}
  {\Gamma(2\nu)
  \Gamma\!\left(\frac{1-\beta}{2}-\nu\right)},\\
  \mathcal D(E;\kappa,\beta)=0
  &\quad\Longleftrightarrow\quad X(\nu;\kappa,\beta)^2=1.
  \end{aligned}
  \label{eq:heun_exact_quantisation}
\end{equation}
Indeed, by $\Gamma(1+z)=z\Gamma(z)$ the two terms of
\eqref{eq:razavy_growing_coefficient} have the ratio $-X^2$.  The
condition $X^2=1$ does not distinguish the two parities, which we
identify in the weak-coupling limit.  Near a Morse level with
noninteger $r_n=\beta-1-2n>0$, the pole of
$\Gamma\bigl(\frac{1-\beta}{2}+\nu\bigr)$ gives
\begin{equation}
  X\sim\frac{\Gamma(\beta-n)}{r_n n!\,\Gamma(r_n)^2}
       \frac{\kappa^{r_n}}{\delta\nu},\qquad
  \delta\nu=\nu-\frac{r_n}{2}.
\end{equation}
The residue is positive, while $\partial_\nu E=-r_n+o(1)<0$.
The solution with $X=+1$ is therefore the lower member of the doublet,
which is the even state by the ordering of
section~\ref{sec:double_well_chamber}, and the solution with $X=-1$ is
the odd state.  By continuity this assignment holds along each branch
of eigenvalues.  With $\epsilon$ the parity defined in
\eqref{eq:local_connection_parity}, the quantization condition is
therefore
\begin{equation}
  \kappa^{2\nu}\exp\!\left[-\frac12\partial_\nu\mathcal F\right]
  \frac{\Gamma(-2\nu)
  \Gamma\!\left(\frac{1-\beta}{2}+\nu\right)}
  {\Gamma(2\nu)
  \Gamma\!\left(\frac{1-\beta}{2}-\nu\right)}
  =\epsilon,
  \qquad
  \epsilon=+1\ (\mathrm{even}),\quad
  \epsilon=-1\ (\mathrm{odd}).
  \label{eq:parity_resolved_heun_quantisation}
\end{equation}
Together with \eqref{eq:matone_energy_relation}, it determines the
energies in each parity sector.  At integer $r_n$ the expansion of the
block in powers of $\kappa^2$ must be reorganized, as discussed in
section~\ref{sec:resonant_block}.  Outside the domain of convergence of
the block expansion we understand these equations as continued along
the branch selected by the Morse limit.

\subsection{Level splitting at noninteger exponent}
\label{sec:heund_splitting}

For the $n$-th Morse level we set
\begin{equation}
  r_n=\beta-1-2n>0,
  \qquad \mu_n=\frac{r_n}{2}.
  \label{eq:morse_monodromy}
\end{equation}
For noninteger $r_n$ we write $\nu=\mu_n+\delta\nu$ and expand
\eqref{eq:parity_resolved_heun_quantisation} about the pole at
$\nu=\mu_n$, which gives
\begin{equation}
  \frac{(r_n+1)_n}{n!\,\Gamma(r_n)}
  \frac{\kappa^{r_n}}{\delta\nu}[1+o(1)]=\epsilon,
  \label{eq:gamma_residue_quantisation}
\end{equation}
and hence
\begin{equation}
  \displaystyle
  \Delta E_n\equiv E_{n,\mathrm o}-E_{n,\mathrm e}
  =2\frac{\Gamma(\beta-n)}
  {n!\,\Gamma(\beta-1-2n)^2}
  \kappa^{\beta-1-2n}[1+o(1)].
  \label{eq:generic_splitting}
\end{equation}

The connection formula also determines the corrections to
\eqref{eq:generic_splitting} in powers of $\kappa^2$.  We write
\begin{equation}
  \mathcal F(\nu,\beta;\lambda)
  =\sum_{k\geq1}f_k(\nu,\beta)\lambda^k,
  \qquad
  \mathcal A_{r,n}:=\frac{\Gamma(n+r+1)}{n!\,\Gamma(r)^2},
  \qquad \mathcal A_n:=\mathcal A_{r_n,n}.
  \label{eq:block_series_and_morse_residue}
\end{equation}
Here $r>0$ and $n\in\mathbb Z_{\geq0}$.  Since $n+r_n+1=\beta-n$, the
coefficient of the leading gap \eqref{eq:generic_splitting} is
$2\mathcal A_n$ and the residue in
\eqref{eq:gamma_residue_quantisation} is $\mathcal A_n/r_n$.  The two
are related by $\partial_\nu E\to-r_n$.  At integer $r$ the coefficient
$\mathcal A_{r,n}$ reduces to the factorial expression of
section~\ref{sec:qes_splitting}.

The series $\mathcal F$ is related to the expansion of the
characteristic values of the Whittaker--Hill equation.  With
$h=2\ii\kappa$, $\mu=-\ii\beta/2$, $\nu_{\rm PP}=2\nu$, and
$\lambda_{\rm PP}=-4E$, the coefficients $f_1$ and $f_2$ reproduce the
terms of order $h^2$ and $h^4$ of the characteristic value at
noninteger exponent given in \cite{PiatekPietrykowski:2016}.  To first
order in $\kappa^{r_n}$ and to all orders in $\lambda=\kappa^2$, the
pole of the gamma function in
\eqref{eq:parity_resolved_heun_quantisation} gives
\begin{equation}
  \delta\nu_\epsilon
  =\epsilon\,\frac{\mathcal A_n}{r_n}\kappa^{r_n}
  \exp\!\left[-\frac12\partial_\nu\mathcal F
  (\mu_n,\beta;\kappa^2)\right]
  +\OO\!\left(\kappa^{2r_n}|\log \kappa|\right).
  \label{eq:subleading_floquet_shift}
\end{equation}
The remaining gamma functions and the factor $\kappa^{2\delta\nu}$
contribute to the splitting only from three traversals on.  Inserting
\eqref{eq:subleading_floquet_shift} into the Matone relation
\eqref{eq:matone_energy_relation}, we obtain the one-traversal
contribution in closed form,
\begin{equation}
  \Delta E_n^{(1)}
  =2\frac{\mathcal A_n}{r_n}\kappa^{r_n}
  \left[r_n-
  \partial_\nu\!\left(\lambda\partial_\lambda\mathcal F\right)
  \right]
  \exp\!\left(-\frac12\partial_\nu\mathcal F\right)
  \bigg|_{\nu=\mu_n,\,\lambda=\kappa^2}.
  \label{eq:one_traversal_gap_from_block}
\end{equation}
The exponential is the fluctuation factor of the shift
$\delta\nu_\epsilon$, and the bracket converts this shift into an
energy difference.

Let a prime denote $\partial_\nu$ at $\nu=\mu_n$.  With
\eqref{eq:irregular_block_first_term} and
\eqref{eq:irregular_block_second_term}, the expansion of
\eqref{eq:one_traversal_gap_from_block} gives
\begin{align}
  \Delta E_n
  ={}&2\frac{\Gamma(\beta-n)}{n!\,\Gamma(r_n)^2}\kappa^{r_n}
  \left[1+c_{1,n}\kappa^2+c_{2,n}\kappa^4+\OO(\kappa^6)\right]
  +\OO\!\left(\kappa^{3r_n}|\log \kappa|^2\right),
  \label{eq:generic_splitting_subleading}\\
  c_{1,n}
  ={}&-\left(\frac12+\frac1{r_n}\right)f_1'
  =-\frac{\beta^2(r_n+2)}{(r_n^2-1)^2},
  \label{eq:generic_splitting_c1}\\
  c_{2,n}
  ={}&\left(\frac18+\frac1{2r_n}\right)(f_1')^2
  -\left(\frac12+\frac2{r_n}\right)f_2'.
  \label{eq:generic_splitting_c2}
\end{align}
The explicit form of $c_{2,n}$ is given in
\eqref{eq:generic_splitting_c2_explicit}.  The leading contribution to
the last term of \eqref{eq:generic_splitting_subleading} is
\begin{equation}
  \Delta E_n^{(3)}
  =2\frac{\Gamma(\beta-n)}{n!\,\Gamma(r_n)^2}
  d_{{\rm mt};r_n,n}(\kappa)\,\kappa^{3r_n}+\cdots,
  \label{eq:generic_three_traversal_gap}
\end{equation}
where $d_{{\rm mt};r,n}(\kappa)$, given in
\eqref{eq:generic_multitraversal_coefficient}, is a quadratic polynomial
in $\log\kappa$.  Equation \eqref{eq:generic_splitting_subleading} thus
contains two sectors of the transseries: the bracket is the fluctuation
series of one traversal, and the last term is the first multi-traversal
contribution.  Their relative order depends on $r_n$.  For $r_n>2$ the
terms $c_{1,n}\kappa^2$ and $c_{2,n}\kappa^4$ are the first two
corrections to the gap.  For $1<r_n<2$ the three-traversal term, of
relative order $\kappa^{2r_n}$, lies between them, and for $0<r_n<1$ it
is larger than the $\kappa^2$ term.  In general the sector with $2k+1$
traversals, of relative order $\kappa^{2kr_n}$, dominates the
$\kappa^2$ correction when $kr_n<1$.

These results hold for fixed noninteger $r_n>0$, uniformly on compact
sets of parameters that stay away from $r_n=0$ and from the positive
integers.  At each positive integer the block coefficients of
sufficiently high order have Kac poles.  For integer $r_n\geq2$ the
coefficient \eqref{eq:generic_splitting_c1} is finite and agrees with
the spin-tunneling correction of \cite{Ulyanov:1999}, but the complete
expansion requires the treatment of section~\ref{sec:resonance_qes}.
At $r_n=1$ already $f_1$ is singular, and the splitting is given by
\eqref{eq:qes_r1_splitting}.  Similarly, $c_{2,n}$ has poles at
$r_n=1$ and $r_n=2$.

The numerical spectra of figure~\ref{fig:razavy_level_examples} provide
a test of \eqref{eq:generic_splitting_subleading} at noninteger
exponent.  For $\beta=6.8$ and $\kappa=0.05$, the one-traversal series
with the coefficients \eqref{eq:generic_splitting_c1} and
\eqref{eq:generic_splitting_c2_explicit} gives
$\Delta E_0=3.85063\times10^{-9}$ and
$\Delta E_1=8.80979\times10^{-5}$, to be compared with the shooting
values $3.85064\times10^{-9}$ and $8.80979\times10^{-5}$.  For the
third doublet, with $r_2=1.8$, the same series gives $0.0852$ and the
three-traversal term \eqref{eq:generic_three_traversal_gap} raises it
to $0.0880$, while the numerical gap is $0.0875$.  The slower
convergence at small $r_n$ reflects the ordering of sectors described
above.

In the next section we describe the same splitting in exact WKB, where
the exponent $r_n$ arises from the logarithmic divergence of the
barrier period.

\section{Exact WKB, quantum periods, and transseries}
\label{sec:exact_wkb}

In exact WKB the spectrum is determined by the quantum periods of the
curve of section~\ref{sec:spectral_curve}.  The period of the well
determines the perturbative levels, and the period of the barrier
determines the splitting.  We apply the exact quantization condition of
a symmetric double well below the barrier, evaluate the Razavy periods,
and study their limit $\kappa\to0^+$.  In this limit the barrier period
diverges logarithmically, and the instanton factor turns into the power
of $\kappa$ found in section~\ref{sec:heund}.

\subsection{Quantum periods and Stokes geometry}
\label{sec:wkb_conventions}

We restore $\hbar$ in \eqref{eq:razavy_hamiltonian}.  Let
\begin{equation}
  p(x;E)=\sqrt{E-V(x)}
  \label{eq:classical_wkb_momentum}
\end{equation}
be positive on the classically allowed interval of the right well.  We
use the WKB ansatz
\begin{equation}
  \psi(x)=\exp\!\left[
  \frac{\ii}{\hbar}\int^x P(x';\hbar)\,\dd x'\right].
  \label{eq:wkb_riccati_ansatz}
\end{equation}
The Schr\"odinger equation then becomes the Riccati equation
\begin{equation}
  P^2-\ii\hbar P'=E-V.
  \label{eq:wkb_riccati}
\end{equation}
Its two formal solutions can be written as
\begin{equation}
  P^{(\pm)}=\pm P_{\rm even}+P_{\rm odd},
  \qquad
  P_{\rm odd}=\frac{\ii\hbar}{2}\,\partial_x\log P_{\rm even},
  \label{eq:wkb_even_odd_decomposition}
\end{equation}
which gives the WKB wavefunctions
\begin{equation}
  \psi_\pm(x)=\frac{1}{\sqrt{P_{\rm even}(x)}}
  \exp\!\left[\pm\frac{\ii}{\hbar}
  \int^x P_{\rm even}(x')\dd x'\right].
  \label{eq:exact_wkb_wavefunctions}
\end{equation}
The subscripts refer to even and odd powers of $\hbar$.  The expansion
of $P_{\rm even}$ starts with
\begin{equation}
  P_{\rm even}(x;\hbar)
  =\sum_{k\geq0}\hbar^{2k}P_{2k}(x),
  \qquad
  P_0=p,
  \qquad
  P_2=-\frac{Q''}{8Q^{3/2}}
      +\frac{5(Q')^2}{32Q^{5/2}},
  \quad Q=E-V.
  \label{eq:even_wkb_momentum}
\end{equation}
For a cycle $\gamma$ on the curve of section~\ref{sec:spectral_curve}
we define the quantum period
\begin{equation}
  \Pi_\gamma(E,\hbar)
  =\oint_\gamma P_{\rm even}\dd x
  =\sum_{k\geq0}\hbar^{2k}\Pi_{\gamma,2k}(E),
  \qquad
  \Pi_{\gamma,2k}=\oint_\gamma P_{2k}\dd x.
  \label{eq:quantum_period_definition}
\end{equation}
These series in $\hbar$ are in general divergent, and they enter the
exact quantization condition through their lateral Borel sums.  After
an integration by parts on the closed cycle, the first correction
becomes
\begin{equation}
  \Pi_{\gamma,2}
  =\frac1{48}\oint_\gamma
  \frac{V''(x)}{[E-V(x)]^{3/2}}\,\dd x.
  \label{eq:first_quantum_period}
\end{equation}
The contour encircles the turning points, where the integrand has
nonintegrable singularities.

With the cycles of section~\ref{sec:spectral_curve}, we define
\begin{equation}
  t(E,\hbar)=\frac{\Pi_A(E,\hbar)}{2\pi},
  \qquad
  t_D(E,\hbar)=-\ii\Pi_B(E,\hbar),
  \qquad
  \langle A,B\rangle=1,
  \label{eq:action_conventions}
\end{equation}
where $\langle A,B\rangle$ is the intersection number of the two cycles
on the curve in the variables $(w,v)$, and the orientation is chosen
such that $t_D>0$ below the barrier for real $\hbar$.\footnote{%
The curve $\Sigma_E$ of \eqref{eq:razavy_spectral_curve} is the
quotient of this curve by the reflection $w\mapsto1/w$.  The quotient
map is $u=(w+w^{-1})/2$, $y=(w-w^{-1})p/(2\kappa)$, under which
$p\,\dd x=\kappa y\,\dd u/(u^2-1)$.  The $A$-cycle is mapped once onto
the cycle around $[u_-,u_+]$, whereas the $B$-cycle is mapped twice
onto the cycle around $[1,u_-]$.  These multiplicities are included in
the period integrals below.}
We work in the region
\begin{equation}
  \mathcal C_{\rm sub}:
  \qquad \kappa>0,\quad \beta>2\kappa,\quad
  \Vmin<E<\Vtop,\quad \arg\hbar=0^\pm.
  \label{eq:physical_wkb_chamber}
\end{equation}
Above the barrier top the Stokes geometry is different and requires a
separate analysis
\cite{MisumiPazarbasi:2025,MisumiPazarbasi:2026}.

The action for one traversal of the barrier is
\begin{equation}
  S_B(E)=\int_{-x_-}^{x_-}\sqrt{V(x)-E}\,\dd x.
  \label{eq:barrier_action}
\end{equation}
For $\hbar=|\hbar|\ee^{\ii\vartheta}$, the Stokes curves emanating from
a turning point $x_t$ are defined by
\begin{equation}
  \operatorname{Im}\!\left[
    \ee^{-\ii\vartheta}
    \int_{x_t}^{x}\sqrt{V(z)-E}\,\dd z
  \right]=0.
  \label{eq:stokes_curve_definition}
\end{equation}
We follow the conventions of the exact WKB literature for Stokes and
anti-Stokes curves, which are interchanged in part of the physics
literature \cite{Sueishi:2020}.

\begin{figure}[t]
  \centering
  \includegraphics[width=\textwidth]{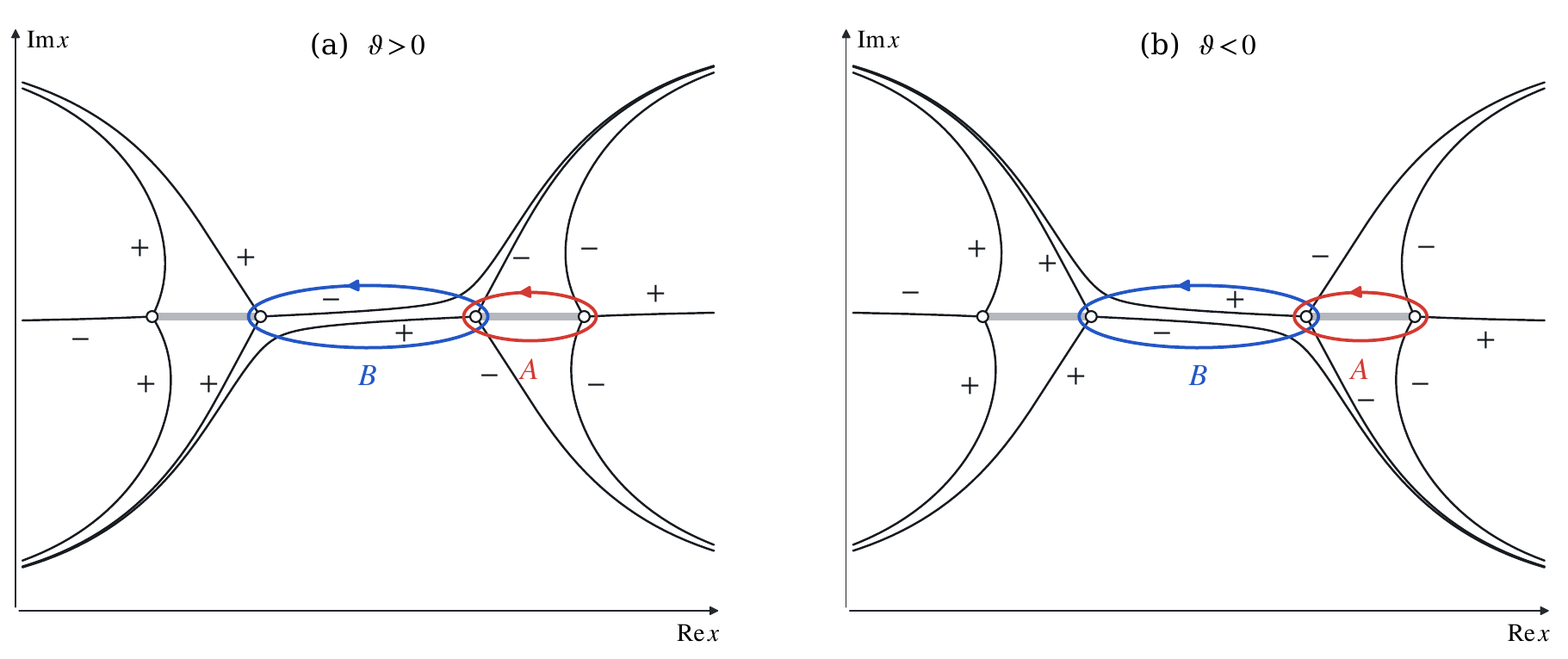}
  \caption{Stokes graphs of the Razavy potential in the strip
  $-\pi<\operatorname{Im}x\leq\pi$ for $\kappa=0.5$, $\beta=4$, and
  $E=-3$.  The two panels correspond to $\vartheta=\arg\hbar>0$ and
  $\vartheta<0$, respectively.  The thin dark curves are the Stokes
  curves \eqref{eq:stokes_curve_definition} emanating from the four
  turning points.  Thick gray segments mark the classically allowed
  intervals, which are taken as branch cuts.  The $A$-cycle (red) and
  the $B$-cycle (blue) are shown in projection, and the signs $\pm$
  label the dominant WKB solution.}
  \label{fig:razavy-stokes-graph}
\end{figure}

For $\Vmin<E<\Vtop$ the strip $-\pi<\operatorname{Im}x\leq\pi$ contains
the four simple turning points $-x_+<-x_-<x_-<x_+$.  At $\vartheta=0$
the forbidden interval $[-x_-,x_-]$ is a saddle connection, and the two
allowed intervals are anti-Stokes lines.  The saddle connection is
resolved for $\vartheta\to0^\pm$.  The two resolutions correspond to
the two lateral Borel sums and are shown in
figure~\ref{fig:razavy-stokes-graph}.

Since the $A$-cycle encircles the allowed interval of the right well
and the $B$-cycle the forbidden interval, the classical periods are
\begin{equation}
  \Pi_{A,0}
  =2\int_{x_-}^{x_+}\sqrt{E-V(x)}\,\dd x,
  \qquad
  \Pi_{B,0}=2\ii S_B(E),
  \qquad
  t_{D,0}=2S_B(E).
  \label{eq:stokes_graph_cycle_normalisation}
\end{equation}
One traversal of the barrier thus contributes the factor
$\exp[-t_{D,0}/(2\hbar)]=\exp[-S_B/\hbar]$, the square root of the
classical Voros multiplier of the $B$-cycle.

The boundary conditions are imposed along the Stokes curves that run to
$x\to\pm\infty$.  In the coordinate $z=\kappa\ee^x$ the WKB exponents
behave there as
\begin{equation}
  \int^z\sqrt{V(\zeta)-E}\,\frac{\dd\zeta}{\zeta}
  \sim \pm\frac{z}{2}\quad(z\to\infty),
  \qquad
  \int^z\sqrt{V(\zeta)-E}\,\frac{\dd\zeta}{\zeta}
  \sim \pm\frac{\kappa^2}{2z}\quad(z\to0),
  \label{eq:irregular_end_actions}
\end{equation}
where the signs depend on the sheet and $V(\zeta)$ denotes
\eqref{eq:razavy_morse_expansion}.  They select the recessive solutions
of section~\ref{sec:heund}, and the quantization condition in
$\mathcal C_{\rm sub}$ then involves only the periods of the cycles $A$
and $B$.

\subsection{Exact quantization condition and transseries}
\label{sec:exact_quantization}

We now impose the boundary conditions in the region
\eqref{eq:physical_wkb_chamber}.  The connection formulas across the
Stokes curves give a quantization condition, which factorizes under
reflection into an even and an odd part.  Expanding about the
perturbative levels then generates the transseries for the energies.

\paragraph{Stokes phenomenon and quantization condition.}
With $Q=E-V$, the leading WKB exponent normalized at a simple turning
point $x_t$ is
\begin{equation}
  \chi_t(x)=\frac{\ii}{\hbar}\int_{x_t}^{x}\sqrt{Q(x')}\,\dd x'.
  \label{eq:wkb_stokes_exponent}
\end{equation}
The Stokes curves \eqref{eq:stokes_curve_definition} are the curves
$\operatorname{Im}\chi_t=0$.  On a curve labeled $+$ in
figure~\ref{fig:razavy-stokes-graph} one has
$\operatorname{Re}\chi_t>0$ and $\psi_+$ is dominant, whereas on a
curve labeled $-$ the solution $\psi_-$ is dominant.  The coefficients
of a solution in the WKB basis are constant within each Stokes region
and change across a Stokes curve according to the standard connection
formulas \cite[secs.~III.B--III.C]{Sueishi:2020}.

Let $\mathcal S_s$ denote the lateral Borel summation of a formal
series in $\hbar$, with the contour above the positive real axis for
$s=+1$ and below it for $s=-1$.  These are the boundary values from
$\arg\hbar=0^+$ and $0^-$, respectively.  In contrast to
$\mathcal L_\theta$ of section~\ref{sec:local-solutions}, which acts on
series in $z$ or $1/z$, the operation $\mathcal S_s$ acts on the
expansions of the quantum periods.  We set
\begin{equation}
  t^s=\mathcal S_s t,\qquad t_D^s=\mathcal S_s t_D,\qquad
  \mathcal V_A^s=\exp\!\left(\frac{2\pi\ii}{\hbar}t^s\right),\qquad
  \mathcal V_B^s=\exp\!\left(-\frac{t_D^s}{\hbar}\right).
  \label{eq:wkb_voros_translation}
\end{equation}
In the region \eqref{eq:physical_wkb_chamber} the configuration of the
four simple turning points and of the Stokes curves is the same as for
the symmetric quartic double well, and the connection formulas of
\cite[eqs.~(82)--(84)]{Sueishi:2020} apply with the Razavy quantum
periods.

With our orientation of the $A$-cycle, the Voros multipliers of the
left well, the barrier, and the right well are
$(\mathcal V_A^s)^{-1}$, $\mathcal V_B^s$, and $\mathcal V_A^s$.  Up to
nonvanishing factors, the coefficient of the growing solution at
$x\to+\infty$ vanishes if and only if
\begin{equation}
  \mathfrak F^s(E)
  :=\bigl[1+(\mathcal V_A^s)^s\bigr]^2+\mathcal V_B^s=0,
  \qquad s=\pm1.
  \label{eq:wkb_scalar_secular_condition}
\end{equation}
With $\tau_s=\exp[-t_D^s/(2\hbar)]$, the factor for one traversal of
the barrier, this condition factorizes into
\begin{equation}
  \mathfrak D_\epsilon^s(E,\hbar)
  :=1+(\mathcal V_A^s)^s-s\ii\epsilon\tau_s=0,
  \qquad \epsilon=\pm1.
\end{equation}
The parity is assigned as in section~\ref{sec:local-solutions},
\begin{equation}
  \epsilon=+1\quad\hbox{(even, $\psi'(0)=0$)},\qquad
  \epsilon=-1\quad\hbox{(odd, $\psi(0)=0$)}.
  \label{eq:wkb_parity_convention}
\end{equation}
In terms of the lateral quantum periods the condition reads
\begin{equation}
  \mathfrak D_\epsilon^\pm(E,\hbar)
  =1+\exp\!\left(\pm\frac{2\pi\ii}{\hbar}t^\pm\right)
   \mp\ii\epsilon\exp\!\left(-\frac{t_D^\pm}{2\hbar}\right)=0.
  \label{eq:exact_wkb_quantisation}
\end{equation}
This is the quantization condition of a symmetric double well given in
\cite[eq.~(4.2)]{vanSpaendonck:2024}, with the Razavy periods.  That
reference uses the kinetic term $-\hbar^2\partial_x^2/2$, which amounts
to replacing $V$ and $E$ by $V/2$ and $E/2$.

\paragraph{Period shift and its recursive expansion.}
It is convenient to use $t$ instead of $E$ as the independent variable.
Let $E(t,\hbar)$ be the inverse function of the quantum $A$-period, and
set
\begin{equation}
  t_D(t,\hbar):=t_D(E(t,\hbar),\hbar),\qquad
  q(t,\hbar):=\exp\!\left[-\frac{t_D(t,\hbar)}{\hbar}\right].
  \label{eq:wkb_transseries_q}
\end{equation}
These functions are constructed as formal series and then summed with a
fixed lateral prescription, which we do not indicate on $E$, $t_D$, and
$q$.  In this subsection primes denote derivatives with respect to $t$
at fixed $\hbar$, $\kappa$, and $\beta$, so that
\begin{equation}
  E'=\frac1{\partial_E t},\qquad
  t_D'=\frac{\partial_E t_D}{\partial_E t},\qquad
  \partial_t q=-\frac{t_D'}{\hbar}q.
  \label{eq:wkb_inverse_period_derivatives}
\end{equation}

Without the barrier term both parities obey the perturbative
quantization condition
\begin{equation}
  t_N=\hbar\left(N+\frac12\right),\qquad
  t(E_N^{\rm pert},\hbar)=t_N,\qquad N\in\mathbb Z_{\geq0}.
  \label{eq:perturbative_bohr_sommerfeld}
\end{equation}
Here $N$ is the level number introduced in
section~\ref{sec:two_limits}.  The perturbative energy
$E_N^{\rm pert}$ depends on the lateral prescription and is in general
ambiguous by itself.  We define
\begin{equation}
  q_N=q(t_N,\hbar),\qquad
  \tau_N=q_N^{1/2}
  :=\exp\!\left[-\frac{t_D(E_N^{\rm pert},\hbar)}{2\hbar}\right].
  \label{eq:instanton_counting_parameter}
\end{equation}
Since \eqref{eq:exact_wkb_quantisation} contains the factor for a
single traversal, we make the ansatz
\begin{equation}
  \hat t=t+\Delta t_\epsilon^s(t,\hbar),\qquad
  \Delta t_\epsilon^s
    =\sum_{n\geq1}\lambda_\epsilon^{(n),s}(t,\hbar)q^{n/2}.
  \label{eq:wkb_period_shift_ansatz}
\end{equation}
At $t=t_N$ the perturbative exponential equals $-1$, and
\eqref{eq:exact_wkb_quantisation} becomes
$\exp(2\pi\ii s\Delta t/\hbar)
 =1-s\ii\epsilon\exp[-t_D(t+\Delta t,\hbar)/(2\hbar)]$.
Taking the logarithm on the branch that vanishes at $q^{1/2}=0$, we
obtain
\begin{equation}
  \Delta t_\epsilon^s
   =\frac{s\hbar}{2\pi\ii}\log\!\left[
       1-s\ii\epsilon
       \exp\!\left(-\frac{t_D(t+\Delta t_\epsilon^s,\hbar)}{2\hbar}\right)
                                  \right].
  \label{eq:wkb_period_shift_implicit}
\end{equation}
As in \cite[eqs.~(4.3)--(4.5)]{vanSpaendonck:2024}, the coefficient
functions are computed for continuous $t$ and evaluated at $t_N$ at the
end.

The barrier period on the right-hand side of
\eqref{eq:wkb_period_shift_implicit} is evaluated at the shifted
argument.  With $\tau=q^{1/2}$ its expansion starts with
\begin{equation}
  \ee^{-t_D(t+\Delta t)/(2\hbar)}
  =\tau\left[1-\frac{t_D'}{2\hbar}\Delta t
       +\left(\frac{(t_D')^2}{8\hbar^2}
                    -\frac{t_D''}{4\hbar}\right)(\Delta t)^2
       +\cdots\right].
  \label{eq:wkb_shifted_barrier_expansion}
\end{equation}
Expanding the logarithm to second order in $\tau$ gives
\begin{align}
  \Delta t_\epsilon^s
  &=-\frac{\epsilon\hbar}{2\pi}\tau
    +\frac{\epsilon t_D'}{4\pi}\Delta t_\epsilon^s\tau
    -\frac{s\ii\hbar}{4\pi}\tau^2
    +\OO_{\rm tr}(\tau^3),\nonumber\\
  \lambda_\epsilon^{(1),s}
  &=-\frac{\epsilon\hbar}{2\pi},\qquad
  \lambda_\epsilon^{(2),s}
    =\frac{\hbar[-t_D'(t,\hbar)-2\pi s\ii]}{8\pi^2}.
  \label{eq:wkb_first_shift_coefficients}
\end{align}
Here $\OO_{\rm tr}(\tau^n)$ denotes terms with at least $n$ powers of
$\tau$, whose coefficients are not expanded in $\hbar$ or $\kappa$ and
can contain logarithms when the curve degenerates.  The first
coefficient is odd in $\epsilon$.  The second is independent of the
parity and contains the lateral ambiguity through the term proportional
to $s$.

The higher coefficients follow recursively.  With a formal variable
$\rho$ and $\Delta_{<n}(\rho)=\sum_{j=1}^{n-1}
\lambda_\epsilon^{(j),s}\rho^j$,
\begin{equation}
  \lambda_\epsilon^{(n),s}
  =\frac{s\hbar}{2\pi\ii\,n!}
    \left.\partial_\rho^{\,n}
    \log\!\left[1-s\ii\epsilon\rho
       \exp\!\left(-\frac{
       t_D(t+\Delta_{<n}(\rho),\hbar)-t_D(t,\hbar)}{2\hbar}
           \right)\right]\right|_{\rho=0}.
  \label{eq:wkb_shift_coefficient_recursion}
\end{equation}
The right-hand side, with $\Delta_{<1}=0$, involves only the
coefficients already determined and finitely many derivatives of $t_D$
with respect to $t$.

\paragraph{Energy transseries, splitting and lateral ambiguity.}
The energy is obtained by evaluating $E(t,\hbar)$ at the shifted
period,
\begin{equation}
  \hat E_\epsilon^s(t,\hbar)
    =E(t+\Delta t_\epsilon^s(t,\hbar),\hbar).
  \label{eq:wkb_energy_composition}
\end{equation}
A Taylor expansion with \eqref{eq:wkb_first_shift_coefficients} then
gives
\begin{equation}
  \hat E_\epsilon^s =E-\epsilon\frac{\hbar}{2\pi}E' q^{1/2}
  +\left[\frac{\hbar^2E''-\hbar t_D'E'}{8\pi^2} -\frac{s\ii\hbar}{4\pi}E'\right]q
  +\OO_{\rm tr}(q^{3/2}).
  \label{eq:wkb_energy_two_traversal}
\end{equation}
All functions on the right-hand side are evaluated at $t$.  The term of
order $q^{1/2}$ splits the doublet, and the term of order $q$ shifts
both levels by the same amount.  The imaginary part of the latter is
compensated by the lateral ambiguity of the perturbative energy and of
the higher sectors.

The separation into even and odd powers of the traversal factor can be
made explicit.  With
$w=\exp[-t_D(t+\Delta t_\epsilon^s,\hbar)/(2\hbar)]$ and the branches
of the logarithm and of the arctangent that vanish at $w=0$,
\begin{equation}
  \log(1-s\ii\epsilon w)
  =\frac12\log(1+w^2)-s\ii\epsilon\arctan w,
  \qquad s,\epsilon=\pm1.
\end{equation}
Equation \eqref{eq:wkb_period_shift_implicit} then becomes
\begin{equation}
  \Delta t_\epsilon^s
   =\frac{s\hbar}{4\pi\ii}\log(1+w^2)
      -\frac{\epsilon\hbar}{2\pi}\arctan w.
  \label{eq:wkb_shift_ddp_parity_split}
\end{equation}
The first term is even in $w$ and depends on the lateral sign $s$ but
not explicitly on the parity, whereas the second is odd in $w$ and
depends on $\epsilon$ but not explicitly on $s$
\cite[eq.~(4.23)]{vanSpaendonck:2024}.  The two terms are coupled,
because $w$ itself depends on $\Delta t_\epsilon^s$.  For example, the
term proportional to $t_D'$ in \eqref{eq:wkb_first_shift_coefficients}
arises from inserting the leading shift into the arctangent.  The two
lateral transseries are related by the Delabaere--Dillinger--Pham
transformation, and the real spectrum is obtained by median summation
\cite{Delabaere:1997,Sueishi:2020,vanSpaendonck:2024}.  This
prescription is understood in all formulas below that carry no lateral
superscript.

At $t=t_N$ the two levels are
\begin{equation}
  E_{N,\epsilon}
  =E_N^{\rm pert}
   -\epsilon\frac{\hbar}{2\pi}
      \left(\frac{\partial E}{\partial t}\right)_{\!N}
      \exp\!\left[-\frac{t_D(E_N^{\rm pert},\hbar)}{2\hbar}\right]
   +\OO_{\rm tr}(q_N).
  \label{eq:parity_resolved_energy_shift}
\end{equation}
Since $\partial E/\partial t>0$, the even level is the lower one.  The
recursion \eqref{eq:wkb_shift_coefficient_recursion} depends on the
parity only through the combination $\epsilon\rho$, so that
$\lambda_{-\epsilon}^{(n),s}=(-1)^n\lambda_\epsilon^{(n),s}$, and the
sectors with an even number of traversals cancel in the gap,
\begin{equation}
  \Delta E_N\equiv E_{N,\mathrm o}-E_{N,\mathrm e}
  =\frac{\hbar}{\pi}
      \left(\frac{\partial E}{\partial t}\right)_{\!N}
      \exp\!\left[-\frac{t_D(E_N^{\rm pert},\hbar)}{2\hbar}\right]
      +\OO_{\rm tr}(q_N^{3/2}).
  \label{eq:exact_wkb_leading_splitting}
\end{equation}
The one-traversal term involves the Borel-resummed quantum periods to
all orders in $\hbar$, and its first correction is of relative order
$q_N$.

In general the energies have the transseries structure
\begin{equation}
  E_{N,\epsilon}
    =E_N^{\rm pert}
     +\sum_{m\geq1}\epsilon^m q_N^{m/2}
       \mathcal E_{N,m}(\hbar;\kappa,\beta).
  \label{eq:parity_transseries_structure}
\end{equation}
The coefficients $\mathcal E_{N,m}$ are fluctuation series in $\hbar$,
summed with the prescription inherited from the periods.
Terms with odd $m$ distinguish the two parities, and terms with even
$m$ shift both levels.  The Borel ambiguity of the perturbative series
is related to the sector $m=2$, whose classical action is
$t_{D,0}=2S_B$, and not to the single traversal that governs the gap
\cite{Delabaere:1997,Sueishi:2020,vanSpaendonck:2024}.  The expansion
\eqref{eq:parity_transseries_structure} is not uniform as the energy
approaches the barrier top, and above it the cycles and the
quantization condition change \cite{MisumiPazarbasi:2025}.  We now
compute the periods of the Razavy potential and study their limit
$\kappa\to0^+$.

\subsection{Classical periods and the first quantum correction}
\label{sec:wkb_periods}

To evaluate \eqref{eq:exact_wkb_quantisation} we need the periods of
the Razavy potential.  At leading order
\eqref{eq:classical_momentum_u} gives
\begin{align}
  t_0(E)
  &=\frac{\kappa}{\pi}\int_{u_-}^{u_+}
  \frac{\sqrt{(u_+-u)(u-u_-)}}{\sqrt{(u-1)(u+1)}}\,\dd u,
  \label{eq:classical_A_period_integral}\\
  t_{D,0}(E)
  &=4\kappa\int_1^{u_-}
  \frac{\sqrt{(u_--u)(u_+-u)}}{\sqrt{(u-1)(u+1)}}\,\dd u.
  \label{eq:classical_B_period_integral}
\end{align}
We introduce the parameters
\begin{equation}
  n_A=\frac{u_+-u_-}{u_+-1},
  \qquad
  n_B=\frac{u_--1}{u_-+1},
  \qquad
  m_A=\frac{2(u_+-u_-)}{(u_+-1)(u_-+1)},
  \qquad
  m_B=1-m_A.
  \label{eq:elliptic_moduli}
\end{equation}
The two periods then reduce to complete elliptic integrals,
\begin{align}
  t_0(E)
  ={}&\frac{\kappa}{\pi\sqrt{(u_+-1)(u_-+1)}}
  \Big\{
  (u_--1)\big[(u_++u_-)\Pell(n_A|m_A)
             -(u_+-1)\K(m_A)\big]
  \nonumber\\[-1mm]
  &\hspace{39mm}
  -(u_+-1)(u_-+1)\Eell(m_A)
  \Big\},
  \label{eq:A_period_complete_elliptic_main}\\[1mm]
  t_{D,0}(E)
  ={}&\frac{4\kappa}{\sqrt{(u_+-1)(u_-+1)}}
  \Big\{
  2u_+(u_-+1)\K(m_B)
  -2(u_++u_-)\Pell(n_B|m_B)
  \nonumber\\[-1mm]
  &\hspace{39mm}
  -(u_+-1)(u_-+1)\Eell(m_B)
  \Big\}.
  \label{eq:B_period_complete_elliptic_main}
\end{align}
Here $\K$, $\Eell$, and $\Pell$ are the complete elliptic integrals of
the first, second, and third kind with parameter $m=k^2$.  In the
region \eqref{eq:physical_wkb_chamber} one has
\begin{equation}
  0<m_A<n_A<1,
  \qquad
  0<n_B<m_B<1,
  \label{eq:elliptic_principal_branch_ordering}
\end{equation}
so that all functions in \eqref{eq:A_period_complete_elliptic_main} and
\eqref{eq:B_period_complete_elliptic_main} are real.  The derivation is
given in appendix~\ref{app:elliptic_reduction}.

The prefactor of the splitting involves the derivatives of the periods
with respect to the energy, which follow directly from
\eqref{eq:classical_A_period_integral} and
\eqref{eq:classical_B_period_integral},
\begin{align}
  \frac{\partial t_0}{\partial E}
  &=\frac{\K(m_A)}
  {\pi \kappa\sqrt{(u_+-1)(u_-+1)}},
  \label{eq:A_period_derivative}\\
  \frac{\partial t_{D,0}}{\partial E}
  &=-\frac{4\K(m_B)}
  {\kappa\sqrt{(u_-+1)(u_+-1)}}.
  \label{eq:B_period_derivative}
\end{align}
At the bottom of the well \eqref{eq:A_period_derivative} tends to
$1/\Omega_0$, as it must for a harmonic oscillator.

The quantum corrections to the periods are integrals of meromorphic
differentials on the same curve, and they can be obtained from the
classical periods by differential operators in the parameters.  In
terms of the variables $\xi$ and $\eta$ of
\eqref{eq:turning_point_parameters}, the first correction is
\begin{equation}
  \Pi_{\gamma,2}=\mathcal L_2 \Pi_{\gamma,0},
  \qquad
  \mathcal L_2=
  \frac{1}{24\kappa^2}\left[
  -2\partial_\eta^2
  -\frac{3\xi}{\eta}\partial_\eta\partial_\xi
  -\frac{\xi^2-1}{\eta^2}\partial_\xi^2
  \right],
  \qquad \gamma=A,B.
  \label{eq:first_quantum_period_operator_main}
\end{equation}
In particular, $t_2=\mathcal L_2t_0$ and
$t_{D,2}=\mathcal L_2t_{D,0}$.  The derivatives are taken at fixed
$\kappa$, and the derivation is given in
appendix~\ref{app:quantum_periods}.  The higher corrections can be
reduced to the classical periods in the same way.

Inverting the quantum $A$-period gives the perturbative energy,
\begin{equation}
  E_N^{\rm pert} =\Vmin+\hbar\Omega_0\left(N+\frac12\right) -\frac{\hbar^2}{\Omega_0^2}
  \left[ \frac{\beta^2}{4}(2N+1)^2 +\kappa^2(2N^2+2N+1) \right]+\OO(\hbar^3).
  \label{eq:local_energy_two_loops}
\end{equation}
This agrees with second-order Rayleigh--Schr\"odinger perturbation
theory about either minimum.
As $\kappa\to0$ the series reduces to the Morse value
$-[\beta-\hbar(2N+1)]^2/4$.

If the periods in \eqref{eq:exact_wkb_leading_splitting} are replaced by
their classical parts, the splitting becomes
\begin{equation}
  \Delta E_N^{\rm sc}
  =\frac{\hbar \kappa
  \sqrt{(u_+-1)(u_-+1)}}{\K(m_A)}
  \exp\!\left[-\frac{t_{D,0}}{2\hbar}\right],
  \label{eq:classical_one_instanton_splitting}
\end{equation}
with all quantities evaluated at $E_N^{\rm pert}$.  Near the bottom of
the well the prefactor tends to $\hbar\Omega_0/\pi$.  The corrections
to this leading semiclassical approximation come from the higher terms
of the quantum periods.

\subsection{Logarithmic growth of the barrier action}
\label{sec:wkb_small_kappa}

As $\kappa\to0^+$ at fixed $\beta$, $\hbar$, and level index $n$, the
wells move apart while the Morse levels stay finite.  The logarithmic
separation and the resulting power-law splitting were noticed in the
studies of the double Morse potential and of spin tunneling
\cite{UlyanovZaslavskii:1992,Konwent:1998,Becker:2017}.  To see how the
power law arises from the periods, we determine the divergent and the
finite part of the barrier action.  We parametrize the energy by
$r=2\sqrt{-E}/\hbar$, which equals the exponent $r_n$ at a Morse level, and write
\begin{equation}
  E=-\frac{\hbar^2r^2}{4},
  \qquad 0<\hbar r<\beta,
  \qquad d=\sqrt{\beta^2-\hbar^2r^2}.
  \label{eq:singular_wkb_energy}
\end{equation}
The expansion of the integral \eqref{eq:barrier_action} for
$\kappa\to0$ gives
\begin{align}
  S_B(E;\kappa)
  &=\hbar r\log\frac1\kappa+\mathcal C_0(\beta,r;\hbar)
  +\OO\!\left(\kappa^2\log\frac1\kappa\right),
  \label{eq:barrier_action_degeneration}\\
  \mathcal C_0(\beta,r;\hbar)
  &=\hbar r\left[
    \log\frac{2\hbar^2r^2}{d}-1
  \right]
  +\frac{\beta}{2}\log\frac{\beta-\hbar r}{\beta+\hbar r}.
  \label{eq:barrier_finite_part}
\end{align}
The derivation is given in appendix~\ref{app:elliptic_reduction}.
Equivalently,
\begin{equation}
  t_{D,0}
  =2\hbar r\log\frac1\kappa+2\mathcal C_0(\beta,r;\hbar)
  +\OO\!\left(\kappa^2\log\frac1\kappa\right),
  \qquad
  \frac{\partial E}{\partial t_0}\rightarrow\hbar r.
  \label{eq:dual_period_degeneration}
\end{equation}
The leading semiclassical splitting
\eqref{eq:classical_one_instanton_splitting} therefore becomes
\begin{equation}
  \Delta E^{\rm sc}(r)
  \sim\frac{\hbar^2r}{\pi}
  \exp\!\left[-\frac{\mathcal C_0(\beta,r;\hbar)}{\hbar}\right]
  \kappa^r.
  \label{eq:classical_power_law}
\end{equation}
The power of $\kappa$ is the instanton factor of a barrier whose action
grows as $\log(1/\kappa)$.  The classical periods thus give the
exponent correctly, whereas the coefficient at fixed $\hbar$ receives
contributions from all orders of the quantum periods.

At the $n$-th Morse level,
\begin{equation}
  r_n=\frac{\beta}{\hbar}-1-2n>0,
  \qquad
  E_n^{(0)}=-\frac{\hbar^2r_n^2}{4},
  \qquad r=r_n.
  \label{eq:morse_exponent_general_hbar}
\end{equation}
For $\hbar=1$ this gives the power $\kappa^{r_n}$ found in
section~\ref{sec:heund_splitting}.  At the QES values $\beta/\hbar=M$
the exponent is the integer $M-1-2n$, which is also the number of steps
that connect the two spin states in section~\ref{sec:qes_splitting}.

The first quantum correction does not modify the coefficient of the
logarithm.  Applying \eqref{eq:first_quantum_period_operator_main} to
the barrier period, we find the finite limit
\begin{equation}
  \lim_{\kappa\to0^+}t_{D,2}
  =\frac{\beta^2-2\hbar^2r^2}
  {3\hbar r(\beta^2-\hbar^2r^2)}.
  \label{eq:first_B_period_degeneration}
\end{equation}
We have not determined the limits of the higher corrections
$t_{D,2k}$, and a finite number of them does not give the exact
coefficient at fixed $\hbar$.  The exact coefficient follows from the
connection formula of section~\ref{sec:heund} at noninteger exponent
and from the two-mode reduction of section~\ref{sec:resonance_qes} at
integer exponent.  With the rescaling
\eqref{eq:hbar_rescaling_dictionary}, both give
\begin{equation}
  \Delta E_n
  \sim
  2\hbar^2
  \frac{\Gamma(\beta/\hbar-n)}
  {n!\,\Gamma(r_n)^2}
  \left(\frac{\kappa}{\hbar}\right)^{r_n},
  \qquad
  r_n>0.
  \label{eq:wkb_to_morse_matching}
\end{equation}
For $\hbar=1$ this reduces to
\begin{equation}
  \Delta E_n
  \sim
  2\frac{\Gamma(\beta-n)}
  {n!\,\Gamma(r_n)^2}\kappa^{r_n}.
  \label{eq:wkb_to_morse_matching_hbar_one}
\end{equation}
The two descriptions are thus complementary: the barrier period
explains the power of $\kappa$, and the connection formula determines
its coefficient.  In the next subsection we examine how well the first
terms of the WKB expansion approximate this coefficient.

\subsection{Finite-order WKB approximations and the QES doublets}
\label{sec:wkb_amplitude_comparison}

A WKB approximation of finite order gives the correct power of
$\kappa$ but only an approximation to its coefficient.  To quantify the
difference we set $\hbar=1$, fix a Morse level, and write
\begin{equation}
  r=r_n=\beta-1-2n>0,\qquad
  C_{\beta,n}:=\lim_{\kappa\to0^+}
  \frac{\Delta E_n}{\kappa^r}
  =\frac{2\Gamma(\beta-n)}{n!\,\Gamma(r)^2}.
  \label{eq:exact_power_coefficient}
\end{equation}
In the notation of \eqref{eq:block_series_and_morse_residue},
$C_{\beta,n}=2\mathcal A_n$.  At $\beta=M\in\mathbb Z^+$ this is the
spin-tunneling coefficient
$C_{M,n}^{\rm QES}=2\mathcal A_{r,n}$ derived in
section~\ref{sec:qes_splitting}.

\paragraph{Finite-order coefficients.}
We write $\mathcal C_0(\beta,r)$ for $\mathcal C_0(\beta,r;1)$.  In the
Morse limit the classical action of the well tends to
$t_0(E)=(\beta-2\sqrt{-E})/2$, so that $\partial_Et_0\to1/r$ at
$E=-r^2/4$, and
\eqref{eq:classical_one_instanton_splitting} gives
\begin{equation}
  C_{\beta,n}^{(1,0)}
  =\frac r\pi\ee^{-\mathcal C_0(\beta,r)}
  =\frac r\pi\ee^r
    \left(\frac{\sqrt{\beta^2-r^2}}{2r^2}\right)^r
    \left(\frac{\beta+r}{\beta-r}\right)^{\beta/2}.
  \label{eq:classical_power_coefficient}
\end{equation}
The first superscript is the number of traversals, and the second is
the highest order in $\hbar$ retained in the barrier period.  Thus
$C^{(1,0)}$ involves only the classical periods.

Keeping $t_{D,0}+\hbar^2t_{D,2}$ in the exponent and setting $\hbar=1$
gives the next approximation,
\begin{equation}
  C_{\beta,n}^{(1,2)}
  :=C_{\beta,n}^{(1,0)}
  \exp\!\left[-\frac{\beta^2-2r^2}{6r(\beta^2-r^2)}\right].
  \label{eq:first_corrected_power_coefficient}
\end{equation}
At the same order one should also include the corrections to the
perturbative energy and to the factor $1/\partial_Et$.  In
appendix~\ref{app:wkb_amplitudes} we show that they do not contribute
in the limit, since $t_2=\OO(\kappa^2)$ and the induced shift of the
classical barrier action is $\OO(\kappa^2\log(1/\kappa))$.  The
correction in \eqref{eq:first_corrected_power_coefficient} thus comes
entirely from \eqref{eq:first_B_period_degeneration}.  We keep it in
the exponent; expanding the exponential would change the result only at
higher order.

\paragraph{Comparison at the QES values.}
Table~\ref{tab:wkb_qes_amplitudes} compares both approximations with the
exact coefficients of the first six QES doublets.  The correction
$t_{D,2}$ improves the agreement in five of the six cases.  The
exception is $(M,n)=(4,0)$, where the classical error of $-0.5\%$
happens to be small and becomes $+1.1\%$.  Over this set the largest
error decreases from $16.4\%$ to $1.1\%$.  These are errors of the
coefficient in the limit $\kappa\to0^+$, and they do not decrease with
$\kappa$.

\begin{table}[!htbp]
  \centering
  \caption{Coefficients of the leading splitting
  $\Delta E_n\sim C_{M,n}\kappa^r$ at $\hbar=1$ and $\beta=M$, with
  $r=M-1-2n$.  The exact values $C^{\rm QES}$ are given by
  \eqref{eq:exact_power_coefficient}, and the two WKB approximations by
  \eqref{eq:classical_power_coefficient} and
  \eqref{eq:first_corrected_power_coefficient}.  The errors are
  $100\,(C/C^{\rm QES}-1)$.}
  \label{tab:wkb_qes_amplitudes}
  \begingroup
  \small
  \setlength{\tabcolsep}{5pt}
  \renewcommand{\arraystretch}{1.15}
  \begin{tabular*}{\textwidth}{@{\extracolsep{\fill}}ccc r rr rr@{}}
    \toprule
    $M$ & $n$ & $r$ & $C^{\rm QES}$
      & \multicolumn{2}{c}{Classical WKB}
      & \multicolumn{2}{c}{First corrected WKB} \\
    \cmidrule(lr){5-6}\cmidrule(l){7-8}
    & & & & $C^{(1,0)}$ & Error (\%)
          & $C^{(1,2)}$ & Error (\%) \\
    \midrule
    2 & 0 & 1 & $2$       & 2.248001  & $+12.4000$ & 2.011600  & $+0.5800$ \\
    3 & 0 & 2 & $4$       & 4.108792  & $+2.7198$  & 4.040879  & $+1.0220$ \\
    4 & 0 & 3 & $3$       & 2.984561  & $-0.5146$  & 3.032313  & $+1.0771$ \\
    4 & 1 & 1 & $4$       & 4.654336  & $+16.3584$ & 3.983830  & $-0.4042$ \\
    5 & 0 & 4 & $4/3$     & 1.304907  & $-2.1320$  & 1.347888  & $+1.0916$ \\
    5 & 1 & 2 & $12$      & 12.836611 & $+6.9718$  & 11.999212 & $-0.0066$ \\
    \bottomrule
  \end{tabular*}
  \endgroup
\end{table}

The quantum correction can be checked analytically.  With
$a_\pm=(\beta\pm r)/2$ the exact coefficient is
$2\Gamma(a_++1/2)/[\Gamma(a_-+1/2)\Gamma(r)^2]$.  The leading terms of
the Stirling expansions of the gamma functions reproduce
\eqref{eq:classical_power_coefficient}, and the next terms give the
exponent in \eqref{eq:first_corrected_power_coefficient}, as shown in
appendix~\ref{app:wkb_amplitudes}.  Since $a_-=n+1/2$ is not large for
the lowest levels, the errors in table~\ref{tab:wkb_qes_amplitudes} are
not controlled by a small parameter.

Finally, the expansion in $\hbar$ and the expansion in the number of
traversals are independent
\cite{ZinnJustin:2004,ZinnJustin:2004b,vanSpaendonck:2024}.  The
remaining corrections to $C_{\beta,n}$ come from higher orders in
$\hbar$ within the one-traversal sector.  The multi-traversal sectors
do not contribute to the coefficient at fixed $r>0$.  Terms with an
even number of traversals shift the two levels of a doublet equally,
and the three-traversal term contributes to the splitting at order
$\kappa^{3r}$ up to logarithms.  The differences between $C^{(1,0)}$ or
$C^{(1,2)}$ and the exact coefficient are therefore truncation errors
of the WKB expansion and cannot be removed by multi-traversal terms.  This separation of the two
expansions is not uniform as $r\to0^+$.
\FloatBarrier

\section{Quasi-exact solvability and resonances}
\label{sec:resonance_qes}

At integer $\beta/\hbar$ the Razavy Hamiltonian is quasi-exactly
solvable: it preserves a finite-dimensional space of functions, and the
corresponding levels are the eigenvalues of a finite matrix.  This
property has been used to compute the level splitting analytically
\cite{UlyanovZaslavskii:1992,Konwent:1998,Finkel:1999,Becker:2017}.
In sections~\ref{sec:qes_module} and \ref{sec:qes_splitting} we review
this construction and the splitting of the QES doublets at small
$\kappa$, and compare the result with those of
sections~\ref{sec:heund} and \ref{sec:exact_wkb}.

The integer values of $\beta/\hbar$ are also the resonances of the
Floquet problem, at which the connection formula cannot be expanded
term by term.  At noninteger exponent the first terms of the expansion
of the block in $\lambda=\kappa^2$ give the first terms of the
splitting, as in section~\ref{sec:heund_splitting}.  At integer
exponent the spectrum is regular, but the coefficients of the block
have Kac poles, and the most singular terms must be summed to all
orders in $\lambda$.  The sum cancels the pole of the gamma-function
factor in the connection formula.  We show this in two steps.  In
section~\ref{sec:resonant_floquet} we reduce the Floquet problem to its
two resonant modes and obtain a square-root law for the characteristic
values, with a coefficient that we determine for every QES doublet.  In
section~\ref{sec:resonant_block} we assume that the Matone relation can
be continued to the resonance and obtain from this result the sum of
the leading Kac poles and the quantization conditions for even and odd
states.  Finally, in section~\ref{sec:qes_subbarrier_count} we use the
position of the QES levels in the spectrum to prove a sharp bound on
the number of levels below the barrier.

\subsection{The algebraic sector}
\label{sec:qes_module}

We keep $\hbar$ explicit and consider the integer values
\begin{equation}
  \hat\beta=\frac{\beta}{\hbar}=M,
  \qquad M\in\mathbb{Z}^+,
  \qquad j=\frac{M-1}{2}.
  \label{eq:qes_locus}
\end{equation}
The transformation
\cite{Razavy:1980,UlyanovZaslavskii:1992,Konwent:1998,Finkel:1999,Becker:2017}
\begin{equation}
  \psi(x)=\exp[-\hat \kappa\cosh x]\,\varphi(x),
  \qquad \hat \kappa=\frac{\kappa}{\hbar},
  \label{eq:qes_gauge_rotation}
\end{equation}
turns the Schr\"odinger equation into
\begin{equation}
  \mathcal L_M\varphi=E\varphi,
  \qquad
  \mathcal L_M
  =-\hbar^2\frac{\dd^2}{\dd x^2}
   +2\hbar \kappa\sinh x\frac{\dd}{\dd x}
   +\frac{\kappa^2}{2}-2j\hbar \kappa\cosh x.
  \label{eq:qes_gauged_operator}
\end{equation}
The $M$-dimensional space
\begin{equation}
\mathcal V_M
  =\operatorname{span}\{\ee^{kx}:k=-j,-j+1,\ldots,j\}
  =\ee^{-jx}\,\mathbb C_{M-1}[\ee^x]
  \label{eq:qes_module}
\end{equation}
is the space of polynomials of degree at most $M-1$ in $\ee^x$ used in
\cite{Finkel:1999}, multiplied by $\ee^{-jx}$.  It is invariant under
$\mathcal L_M$, since
\begin{align}
  \mathcal L_M\ee^{kx}
  ={}&\left(\frac{\kappa^2}{2}-\hbar^2k^2\right)\ee^{kx}
   +\hbar \kappa(k-j)\ee^{(k+1)x}
   -\hbar \kappa(k+j)\ee^{(k-1)x}.
  \label{eq:qes_tridiagonal_action}
\end{align}
The coefficient of $\ee^{(k+1)x}$ vanishes at $k=j$ and that of
$\ee^{(k-1)x}$ at $k=-j$, so that \eqref{eq:qes_tridiagonal_action}
defines an $M\times M$ matrix.  For $\kappa>0$ every element of
$\mathcal V_M$, multiplied by the exponential factor in
\eqref{eq:qes_gauge_rotation}, is square integrable.  The $M$
eigenvalues of this matrix are therefore eigenvalues of the
Hamiltonian.  We call them the QES levels.

The reflection $x\to-x$ exchanges $\ee^{kx}$ and $\ee^{-kx}$ and
decomposes $\mathcal V_M$ into an even and an odd subspace,
\begin{equation}
  \mathcal V_M=\mathcal V_M^{(+)}\oplus\mathcal V_M^{(-)},
  \qquad
  \dim\mathcal V_M^{(+)}=\left\lceil\frac M2\right\rceil,
  \qquad
  \dim\mathcal V_M^{(-)}=\left\lfloor\frac M2\right\rfloor.
  \label{eq:qes_parity_dimensions}
\end{equation}
This decomposition agrees with \cite{Becker:2017}.  For odd $M$ the
even subspace contains the function with $k=0$, which is unpaired.  The
other basis functions form pairs $k=\pm\ell$, which give rise to the
doublets at weak coupling.  The construction holds for any
$\hat\kappa>0$.  The potential is a double well for
$0<\hat\kappa<M/2$, and the expansion below assumes $\hat\kappa\ll1$.

The matrix of $\mathcal L_M$ in the basis $\ee^{kx}$ is not symmetric.
The operator is, however, self-adjoint with respect to the inner
product inherited from $L^2(\RR)$,
\begin{equation}
  \langle\varphi_1,\varphi_2\rangle_{\hat \kappa}
  =\int_{-\infty}^{\infty}
  \ee^{-2\hat \kappa\cosh x}
  \overline{\varphi_1(x)}\varphi_2(x)\,\dd x.
  \label{eq:qes_weighted_inner_product}
\end{equation}
Accordingly, the matrix is symmetrized by a diagonal matrix
$\mathsf D=\operatorname{diag}(d_k)$ with
\begin{equation}
  \frac{d_{k+1}}{d_k}
  =\sqrt{\frac{j-k}{j+k+1}}.
  \label{eq:qes_similarity_weights}
\end{equation}
The matrix $\mathsf D^{-1}\mathcal L_M\mathsf D$ is real and symmetric,
with the off-diagonal elements
\begin{equation}
  \left(\mathsf D^{-1}\mathcal L_M\mathsf D\right)_{k,k+1}
  =-\hbar \kappa\sqrt{(j-k)(j+k+1)}.
  \label{eq:qes_symmetric_matrix}
\end{equation}
It is the matrix of a spin-$j$ Hamiltonian in the basis of eigenstates
of $J_z$,
\begin{equation}
  H_M^{\rm spin}
  =\mathsf D^{-1}\mathcal L_M\mathsf D
  =\frac{\kappa^2}{2}-\hbar^2J_z^2-2\hbar\kappa J_x.
  \label{eq:qes_spin_hamiltonian}
\end{equation}
Up to the constant $\kappa^2/2$, this is the Hamiltonian of an
easy-axis spin in a transverse field studied in
\cite{ZaslavskiiUlyanov:1984,UlyanovZaslavskii:1992,Konwent:1998}.
The construction is an instance of the $\mathfrak{sl}(2)$ description
of quasi-exactly solvable models
\cite{Turbiner:1988QES,GonzalezLopezKamranOlver:1993}.

The QES levels are the lowest $M$ levels of the Hamiltonian, as shown
in \cite[sec.~2.1]{UlyanovZaslavskii:1992} (see also
\cite{Finkel:1999}).  Indeed, a QES wavefunction multiplied by the
positive factor $\exp[jx+\hat\kappa\cosh x]$ is a polynomial in $\ee^x$
of degree at most $M-1$, and therefore has at most $M-1$ nodes.  Since
the spectrum is nondegenerate, the $M$ QES levels are distinct, and by
the oscillation theorem they are the levels with $0,1,\ldots,M-1$
nodes.

\subsection{Level splitting from the spin Hamiltonian}
\label{sec:qes_splitting}

The splitting of the QES doublets at small $\hat\kappa$ follows from
degenerate perturbation theory for the spin Hamiltonian
\eqref{eq:qes_spin_hamiltonian}.  At $\hat\kappa=0$ the states with
$J_z=\pm\ell$ are degenerate.  For the pair that corresponds to the
$n$-th Morse level we write
\begin{equation}
  \ell=j-n,
  \qquad
  r=2\ell=M-1-2n>0,
  \qquad
  n=0,1,\ldots,
  \left\lfloor\frac{M-2}{2}\right\rfloor.
  \label{eq:qes_doublet_indices}
\end{equation}
The unperturbed energy and the exponent of
section~\ref{sec:wkb_small_kappa} are then
\begin{equation}
  E_n^{(0)}=-\hbar^2\ell^2
  =-\frac{[\beta-\hbar(2n+1)]^2}{4},
  \qquad
  r=2\ell=M-1-2n
   =\frac{\beta-\hbar(2n+1)}{\hbar}.
  \label{eq:qes_exponent_dictionary}
\end{equation}
The integer $r$ is thus the number of steps of $J_z$ between the two
states, and it coincides with the exponent
\eqref{eq:morse_exponent_general_hbar} obtained from the barrier
period.

Since $J_x$ changes $k$ by one unit, the two states are first connected
at order $r$ of perturbation theory, through the sequence
\begin{equation}
  -\ell\longleftrightarrow-\ell+1
  \longleftrightarrow\cdots\longleftrightarrow
  \ell-1\longleftrightarrow\ell,
  \qquad \text{number of links}=r.
  \label{eq:qes_shortest_chain}
\end{equation}
With the matrix elements \eqref{eq:qes_symmetric_matrix}, the product
of the $r$ off-diagonal elements along this sequence has the absolute
value
\begin{equation}
  \prod_{k=-\ell}^{\ell-1}
  \hbar \kappa\sqrt{(j-k)(j+k+1)}
  =(\hbar \kappa)^r\frac{(n+r)!}{n!},
  \label{eq:qes_path_numerator}
\end{equation}
and the product of the $r-1$ energy denominators is
\begin{equation}
  \prod_{k=-\ell+1}^{\ell-1}
  \bigl[E_n^{(0)}-(-\hbar^2k^2)\bigr]
  =(-\hbar^2)^{r-1}[(r-1)!]^2.
  \label{eq:qes_path_denominator}
\end{equation}
The $r$ matrix elements and the $r-1$ denominators are all negative, so
that the effective coupling of the two states is negative.  By the
reflection symmetry the two diagonal elements are equal.  In the basis
$(\lvert\ell\rangle,\lvert-\ell\rangle)$ the effective Hamiltonian of
the doublet is therefore
\begin{equation}
  H_{\rm eff}
  =E_{c,M,n}(\hat \kappa)\,\mathbf 1
  -\tau_{r,n}(\hat \kappa)
  \begin{pmatrix}0&1\\[1mm]1&0\end{pmatrix},
  \qquad
  \tau_{r,n}
  =\hbar^2\mathcal A_{r,n}\hat \kappa^r
  \left[1+\OO(\hat \kappa^2)\right],
  \label{eq:qes_effective_matrix}
\end{equation}
where $E_{c,M,n}=\hbar^2\hat E_{c,M,n}(\hat\kappa)$ and
\begin{equation}
  \mathcal A_{r,n}
  =\frac{(n+r)!}{n![(r-1)!]^2}
  =\frac{(M-1-n)!}{n![(M-2n-2)!]^2}.
  \label{eq:qes_tunnelling_amplitude}
\end{equation}
The even and odd states are the symmetric and antisymmetric
combinations of $\lvert\ell\rangle$ and $\lvert-\ell\rangle$.  Since
$\tau_{r,n}>0$, the even state is the lower one,
\begin{equation}
  E_{n,\mathrm e}=E_{c,M,n}-\tau_{r,n},
  \qquad
  E_{n,\mathrm o}=E_{c,M,n}+\tau_{r,n}.
  \label{eq:qes_parity_levels}
\end{equation}
The splitting is therefore
\begin{equation}
  E_{n,\mathrm o}-E_{n,\mathrm e}
  =2\hbar^2
  \frac{(M-1-n)!}{n![(M-2n-2)!]^2}
  \left(\frac{\kappa}{\hbar}\right)^{M-1-2n}
  \left[1+\OO\!\left(\frac{\kappa^2}{\hbar^2}\right)\right].
  \label{eq:qes_leading_splitting}
\end{equation}
The relative correction is of order $\hat\kappa^2$, since any other
path between the two states has at least two additional steps.

Equation \eqref{eq:qes_leading_splitting} coincides with
\eqref{eq:wkb_to_morse_matching} at $\beta/\hbar=M$, since
\begin{equation}
  \Gamma(M-n)=\Gamma(n+r+1)=(n+r)!,
  \qquad
  \Gamma(M-1-2n)=\Gamma(r)=(r-1)!.
  \label{eq:qes_morse_match}
\end{equation}
With $\alpha=\hbar^2$, $B=2\hbar\kappa$, and $S=(M-1)/2$, it is the
spin-tunneling formula (4.33) of
\cite[sec.~4.3]{UlyanovZaslavskii:1992}.  The
splitting of the ground doublet was first obtained in
\cite{KorenblitShender:1978} and that of an arbitrary doublet in
\cite{Garanin:1991} (see also \cite{Ulyanov:1999}).  Note that the
splitting is $2\tau_{r,n}$, twice the off-diagonal element of
\eqref{eq:qes_effective_matrix}.  This factor of two is relevant for
the normalization in section~\ref{sec:resonant_block}, and the matrix
\eqref{eq:qes_effective_matrix} determines the parity of the two
branches found in section~\ref{sec:resonant_floquet}.

The first correction is also known.  For $r>1$, equation~(12) of
\cite{Ulyanov:1999} gives, in our normalization,
\begin{equation}
  \Delta E_{M,n}
  =2\hbar^2\mathcal A_{r,n}\hat \kappa^r
  \left[
  1-\frac{M^2(r+2)}{(r^2-1)^2}\hat \kappa^2
  +\OO(\hat \kappa^4)
  \right].
  \label{eq:qes_subleading_splitting}
\end{equation}
This expression has a pole at $r=1$.  In that case equation~(13) of
\cite{Ulyanov:1999} gives
\begin{equation}
  \Delta E_{M,(M-2)/2}
  =M\hbar^2\hat \kappa 
  \left[1-\frac{M^2-4}{16}\hat \kappa^2
  +\OO(\hat \kappa^4)\right],
  \qquad M\ \text{even}.
  \label{eq:qes_r1_splitting}
\end{equation}
The leading term is reproduced in section~\ref{sec:resonant_block}, and
the corrections in \eqref{eq:qes_subleading_splitting} and
\eqref{eq:qes_r1_splitting} provide checks beyond leading order.

For odd $M$ the state that evolves from $k=0$ is even and has no
partner among the QES levels.  Its index is $n=j$, and its odd partner
$E_{j,\mathrm o}=\mathcal E_M$ is not a QES level.  The corresponding
exponent is $r=0$, the threshold at which the Morse level disappears,
and \eqref{eq:qes_leading_splitting} does not apply to this pair.

The spectra for $M=1,\ldots,5$ are given in
appendix~\ref{app:qes_examples}, where
\eqref{eq:qes_tunnelling_amplitude} and
\eqref{eq:qes_subleading_splitting} are verified explicitly.

\subsection{Two-mode reduction of the Floquet problem}
\label{sec:resonant_floquet}

We return to the Floquet problem of section~\ref{sec:irregular_block}
and set $\hbar=1$.  For the $n$-th QES doublet, with positive integer
exponent $r$, we introduce the detuning $\delta$ by
\begin{equation}
  \beta=M=r+1+2n,\qquad
  \nu=\frac r2+\delta,\qquad
  \mathcal A_{r,n}=\frac{(n+r)!}{n![(r-1)!]^2},
  \qquad \lambda=\kappa^2.
  \label{eq:resonant_variables}
\end{equation}
After the transformation \eqref{eq:qes_gauge_rotation}, a Floquet
solution is a series in the functions $\ee^{(k+\delta)x}$ with
$k\in r/2+\mathbb Z$, on which the operator
\eqref{eq:qes_gauged_operator} acts as
\begin{align}
  \mathcal L_M\ee^{(k+\delta)x}={}&
  \left[\frac{\kappa^2}{2}-(k+\delta)^2\right]\ee^{(k+\delta)x}
  +\kappa(k+\delta-j)\ee^{(k+\delta+1)x}
  \nonumber\\
  &-\kappa(k+\delta+j)\ee^{(k+\delta-1)x},
  \qquad j=n+\frac r2.
  \label{eq:off_shell_floquet_action}
\end{align}
Under $x\mapsto x+2\pi\ii$ these functions are multiplied by
$\exp(2\pi\ii\nu)$, so that $\nu$ is the Floquet exponent of the Matone
relation \eqref{eq:matone_energy_relation}.  At $\kappa=\delta=0$ the
diagonal elements with $k=\pm r/2$ are equal, and they differ from all
other diagonal elements by a finite amount.

Let $P$ be the projection onto the two functions with $k=\pm r/2$,
which we call the resonant modes, and $Q=1-P$.  The Feshbach--Schur
map \cite{DussonSigalStamm:2021} eliminates the other
modes and gives the $2\times2$ matrix
\begin{equation}
  H_{\rm eff}(E)=P\mathcal L_MP
  +P\mathcal L_MQ(E-Q\mathcal L_MQ)^{-1}Q\mathcal L_MP.
  \label{eq:resonant_feshbach_matrix}
\end{equation}
The characteristic values near the resonance are the solutions of
$\det[E I_P-H_{\rm eff}(E)]=0$, where $I_P$ is the identity on the
range of $P$.  No approximation has been made, since the other modes
enter through the resolvent.  In appendix~\ref{app:floquet_feshbach} we
show that for fixed $(r,n)$ and sufficiently small $\kappa$ and
$\delta$ the resolvent has a convergent expansion.  As in
section~\ref{sec:qes_splitting}, the off-diagonal element of the
symmetrized matrix is $-\mathcal A_{r,n}\kappa^r$ at leading order,
with corrections of relative order $\kappa^2$.  For real $\delta$ with
$|\delta|\leq D\kappa^r$ the two characteristic values are
\begin{equation}
  E_\pm=E_c(\kappa,\delta)\pm
  \sqrt{r^2\delta^2+\mathcal A_{r,n}^2\kappa^{2r}}
  +\OO_{r,n,D}(\kappa^{r+2}).
  \label{eq:resonant_energy_sheets}
\end{equation}
Here $E_c(\kappa,\delta)$ is a smooth function defined by
\eqref{eq:implicit_centroid}, and the error term is uniform in
$\delta$.  At $\delta=0$ the square root reduces to the coefficient
\eqref{eq:qes_tunnelling_amplitude} of the spin calculation.  The
derivation of \eqref{eq:resonant_energy_sheets} makes no use of the
conformal block.

\subsection{Resummation of the Kac poles}
\label{sec:resonant_block}

At order $\lambda^{kr}$ the coefficient of the block has a pole of
order $2k-1$ at $\nu=r/2$.  These leading poles are summed by a
function $\mathcal F_{\rm sing}$ with
\begin{equation}
  \lambda\partial_\lambda\mathcal F_{\rm sing}
  =r\delta-
  \sqrt{r^2\delta^2+\mathcal A_{r,n}^2\lambda^r},
  \label{eq:resonant_matone}
\end{equation}
where the square root is the branch that tends to $r\delta$ as
$\lambda\to0$.  Indeed, the expansion of \eqref{eq:resonant_matone} at
fixed $\delta\neq0$ gives the coefficients
\begin{equation}
  f_{kr}^{\rm lead}
  =-\frac{\binom{1/2}{k}}{k r^{2k}}
  \frac{\mathcal A_{r,n}^{2k}}
  {(\nu-r/2)^{2k-1}},
  \qquad k\geq1.
  \label{eq:kac_pole_tower}
\end{equation}
In appendix~\ref{app:kac_poles} we derive \eqref{eq:kac_pole_tower}
from the recurrence for the Floquet coefficients, using
\eqref{eq:floquet_to_block_coefficients}.  This derivation concerns the
individual coefficients of the block and involves no assumption.  To
use the sum \eqref{eq:resonant_matone} in the connection formula we
assume, in addition, that the Matone relation
\eqref{eq:matone_energy_relation} continues to hold near the resonance,
on the branch on which $E\to-\nu^2$ as $\kappa\to0$ for $\delta>0$.
The estimates of appendix~\ref{app:floquet_feshbach} then imply that
the remainder
$\mathcal F_{\rm reg}=\mathcal F-\mathcal F_{\rm sing}$ satisfies
\begin{equation}
  \partial_\nu\mathcal F_{\rm reg}=\OO_{r,n,D}(\kappa^2),
  \qquad 0<|\delta|\leq D\kappa^r,
  \label{eq:regular_block_power_counting}
\end{equation}
and that $\partial_\nu\mathcal F_{\rm reg}$ has finite limits as
$\delta\to0^\pm$.

We now evaluate the function $X$ of \eqref{eq:heun_exact_quantisation}
near the resonance.  The singular part of the block contributes the
factor
\begin{equation}
  \exp\!\left[-\frac12\partial_\nu\mathcal F_{\rm sing}\right]
  =\frac{2}{1+\sqrt{1+
  \left(\frac{\mathcal A_{r,n}\kappa^r}{r\delta}\right)^2}},
  \label{eq:resummed_singular_block_factor}
\end{equation}
and the remaining factors behave as
\begin{equation}
  \kappa^{2\nu}
  \frac{\Gamma(-2\nu)
  \Gamma\!\left(\frac{1-M}{2}+\nu\right)}
  {\Gamma(2\nu)
  \Gamma\!\left(\frac{1-M}{2}-\nu\right)}
  =\frac{\mathcal A_{r,n}\kappa^r}{2r\delta}[1+o(1)].
  \label{eq:correlated_gamma_asymptotic}
\end{equation}
For $\kappa\to0^+$ with $|\delta|/\kappa^r\to0$, the product of
\eqref{eq:resummed_singular_block_factor} and
\eqref{eq:correlated_gamma_asymptotic} tends to
$\operatorname{sgn}\delta$, and the regular part of the block changes
it by a factor $1+\OO(\kappa^2)$.  On the branch chosen above we
therefore find
\begin{equation}
  \delta\to0^+:\quad X\to+1,\quad E\to E_-,
  \qquad
  \delta\to0^-:\quad X\to-1,\quad E\to E_+.
  \label{eq:resonant_sheet_gluing}
\end{equation}
By the parity assignment of section~\ref{sec:irregular_block}, the
lower branch is even and the upper branch is odd.  The splitting at
integer exponent is therefore
\begin{equation}
  E_{n,\mathrm o}-E_{n,\mathrm e}
  =2\mathcal A_{r,n}\kappa^r[1+o(1)].
  \label{eq:resonant_block_splitting}
\end{equation}
Without the factor \eqref{eq:resummed_singular_block_factor}, the
condition $X=\pm1$ applied to \eqref{eq:correlated_gamma_asymptotic}
would give only half of this splitting.

The spin calculation of section~\ref{sec:qes_splitting}, the two-mode
reduction, and the connection formula thus give the same coefficient
and the same ordering of the two parities.  The third derivation relies
on the assumed continuation of the Matone relation.  It determines the
leading behavior of the two branches at the resonance, but it does not
provide an expression for the connection matrix in terms of conformal
blocks that is uniform in $\delta$.  We also note that the local
solutions at $z=0$ and at $z=\infty$ remain linearly independent at the
resonance, by \eqref{eq:local_wronskian_normalisations}.  Logarithms
can appear only in the Floquet solutions, when the monodromy matrix is
not diagonalizable \cite[section~4.1]{AbadGomezSesma:2008}.  We leave
the continuation of the conformal-block representation through the
resonance for future work.

\subsection{A sharp bound on the number of levels below the barrier}
\label{sec:qes_subbarrier_count}

The position of the QES levels in the spectrum implies a bound on the
number of levels below the barrier top, for all values of
$\beta/\hbar$.  With the eigenvalues ordered as
$\mathcal E_0(\kappa,\beta;\hbar)<\mathcal E_1(\kappa,\beta;\hbar)<\cdots$,
as in section~\ref{sec:double_well_chamber}, we define
\begin{equation}
  N_{\rm sub}(\kappa,\beta;\hbar)
  =\#\{k\geq0:\mathcal E_k(\kappa,\beta;\hbar)<\Vtop(\kappa,\beta)\}.
  \label{eq:subbarrier_count_definition}
\end{equation}
A level that coincides with the barrier top is not counted.

For all $\hbar>0$ and $0<\kappa<\beta/2$ we have
\begin{equation}
  \displaystyle
  N_{\rm sub}(\kappa,\beta;\hbar)
  \leq B(\hat\beta)
  :=2\left\lceil\frac{\hat\beta-1}{2}\right\rceil,
  \qquad \hat\beta=\frac{\beta}{\hbar}.
  \label{eq:sharp_subbarrier_bound}
\end{equation}
The bound is sharp for every $\hat\beta>0$,
\begin{equation}
  \sup_{\,0<\hat \kappa<\hat\beta/2}
  N_{\rm sub}(\hat \kappa,\hat\beta;1)
  =B(\hat\beta),
  \qquad \hat \kappa=\frac{\kappa}{\hbar},
  \label{eq:sharp_subbarrier_supremum}
\end{equation}
and it is attained for all sufficiently small $\hat\kappa>0$.

To prove the bound, we compare the Hamiltonian with the one at the next
odd integer value of $\hat\beta$.  By
\eqref{eq:hbar_rescaling_dictionary} and \eqref{eq:well_geometry}, the
Hamiltonian measured from the barrier top and divided by $\hbar^2$ is
\begin{equation}
  \mathcal K_{\hat\beta,\hat \kappa}
  =\frac{\Ham(\hbar;\kappa,\beta)-\Vtop}{\hbar^2}
  =-\partial_x^2+\hat \kappa^{\,2}\sinh^2x
   -\hat\beta\,\hat \kappa(\cosh x-1).
  \label{eq:barrier_shifted_operator}
\end{equation}
The number of its negative eigenvalues is $N_{\rm sub}$.  At fixed
$\hat\kappa$ these operators have the same form domain for all
$\hat\beta$,
\begin{equation}
  \mathcal Q
  =H^1(\RR)\cap
   L^2(\RR,\sinh^2x\,\dd x),
  \label{eq:barrier_shifted_form_domain}
\end{equation}
since $\cosh x-1$ is infinitesimally form bounded with respect to
$\sinh^2x$.  For $\hat\beta_2\geq\hat\beta_1$ we have, in the sense of
quadratic forms on $\mathcal Q$,
\begin{equation}
  \mathcal K_{\hat\beta_1,\hat \kappa}
  -\mathcal K_{\hat\beta_2,\hat \kappa}
  =(\hat\beta_2-\hat\beta_1)\hat \kappa 
   (\cosh x-1)\geq0.
  \label{eq:subbarrier_form_monotonicity}
\end{equation}
By the min--max principle, $N_{\rm sub}$ is therefore a nondecreasing
function of $\hat\beta$ at fixed $\hat\kappa$.

Let
\begin{equation}
  M_*(\hat\beta)
  =2\left\lceil\frac{\hat\beta-1}{2}\right\rceil+1
  \label{eq:next_odd_qes_value}
\end{equation}
be the smallest odd integer that is not smaller than $\hat\beta$.
Since $\hat\kappa<\hat\beta/2\leq M_*/2$, the potential with
$\hat\beta=M_*$ is also a double well, and
\begin{equation}
  N_{\rm sub}(\hat \kappa,\hat\beta;1)
  \leq N_{\rm sub}(\hat \kappa,M_*;1).
  \label{eq:subbarrier_qes_comparison}
\end{equation}
For odd $M=2j+1$ the spin $j$ is an integer, and the state
$\lvert j,0\rangle$ with $J_z=0$ exists.  Subtracting the barrier top
from \eqref{eq:qes_spin_hamiltonian} and dividing by $\hbar^2$, we
obtain
\begin{equation}
  \mathcal K_M^{\rm spin}
  =M\hat \kappa-J_z^2-2\hat\kappa J_x,
  \qquad
  \langle j,0|\mathcal K_M^{\rm spin}|j,0\rangle
  =M\hat \kappa>0.
  \label{eq:odd_qes_barrier_rayleigh}
\end{equation}
Since the QES levels are the lowest $M$ levels of the Hamiltonian, the
largest eigenvalue of $\mathcal K_M^{\rm spin}$ is the eigenvalue
$\lambda_{M-1}$ of $\mathcal K_{M,\hat\kappa}$, and
\begin{equation}
  \lambda_{M-1}(\mathcal K_{M,\hat \kappa})
  =\lambda_{\max}(\mathcal K_M^{\rm spin})
  \geq M\hat \kappa>0.
  \label{eq:odd_qes_highest_above_barrier}
\end{equation}
At most $M-1$ levels therefore lie below the barrier top, and
\eqref{eq:subbarrier_qes_comparison} gives the bound
\eqref{eq:sharp_subbarrier_bound}.  With the units restored,
\eqref{eq:odd_qes_highest_above_barrier} reads
$\mathcal E_{M-1}-\Vtop\geq M\hbar\kappa$.  The existence of a QES
level above the barrier top was noted in
\cite[sec.~2.3]{UlyanovZaslavskii:1992} and
\cite[p.~7549]{Konwent:1998}.

The bound is attained in the weak-coupling limit.  Let
$m=\lceil(\hat\beta-1)/2\rceil$.  For $m>0$ each of the two Morse wells
has the $m$ levels $-r_n^2/4<0$, $n=0,\ldots,m-1$.  We approximate the
corresponding Morse wavefunctions by smooth functions with compact
support in a fixed interval of the variable $z$.  In the variables
$z_R=\hat\kappa\ee^x$ and $z_L=\hat\kappa\ee^{-x}$ the supports of the
functions in the right and in the left well are disjoint for
sufficiently small $\hat\kappa$, and on the $2m$-dimensional space that
they span the quadratic form of the Hamiltonian converges to two copies
of the Morse form.  The form of $\mathcal K_{\hat\beta,\hat\kappa}$ is
therefore negative on this space for all sufficiently small
$\hat\kappa$.  The min--max principle then gives $N_{\rm sub}\geq2m$,
and with \eqref{eq:sharp_subbarrier_bound} we conclude that
$N_{\rm sub}=2m$.  For $m=0$ the bound itself gives $N_{\rm sub}=0$.
This proves \eqref{eq:sharp_subbarrier_supremum}.

Explicitly, the bound reads
\begin{equation}
  0<\hat\beta\leq1:\quad N_{\rm sub}=0,
  \qquad
  2m-1<\hat\beta\leq2m+1:\quad
  N_{\rm sub}\leq2m\quad(m\geq1).
  \label{eq:subbarrier_staircase}
\end{equation}
At the QES values $\hat\beta=M$ it becomes
\begin{equation}
  N_{\rm sub}\leq
  \begin{cases}
    M,&M\ \text{even},\\
    M-1,&M\ \text{odd}.
  \end{cases}
  \label{eq:qes_subbarrier_count}
\end{equation}
Since the QES levels are the lowest $M$ levels, all levels below the
barrier top are QES levels in this case.  For noninteger $\hat\beta$
we have
\begin{equation}
  B(\hat\beta)=
  \begin{cases}
    \lfloor\hat\beta\rfloor,
      &\lfloor\hat\beta\rfloor\ \text{even},\\
    \lfloor\hat\beta\rfloor+1,
      &\lfloor\hat\beta\rfloor\ \text{odd}.
  \end{cases}
  \label{eq:subbarrier_floor_comparison}
\end{equation}
The simpler expression $\lfloor\hat\beta\rfloor$ is therefore not a
bound when $\lfloor\hat\beta\rfloor$ is odd.  At a given $\hat\kappa$
the number of levels below the barrier top can be smaller than
$B(\hat\beta)$, and it is odd when the barrier top lies between the two
levels of a doublet.

\section{Conclusions and outlook}
\label{sec:sum}

We have studied the level splitting of the Razavy double well in the
weak-coupling limit $\kappa\to0^+$, in which the two wells separate
logarithmically and become Morse potentials.  The splitting of the
$n$-th doublet is proportional to $\kappa^{r_n}$ with
$r_n=\beta/\hbar-1-2n$, which is in general not an integer.  We have
derived this power law in three ways.  In the connection formula for
the double-confluent Heun equation it is the factor $\kappa^{2\nu}$,
evaluated at a pole of the gamma functions where the Floquet exponent
is $\nu=r_n/2$.  In exact WKB it is the exponential of a barrier action
that grows as $\hbar r_n\log(1/\kappa)$.  At integer $\beta/\hbar$ it
follows from the $r_n$ steps of perturbation theory that connect the
states with $J_z=\pm r_n/2$ of the spin Hamiltonian.

For noninteger $r_n$ we used the reflection symmetry to factorize the
spectral determinant and obtained the quantization conditions
\eqref{eq:parity_resolved_heun_quantisation} for even and odd states.
Their solution is the splitting
\eqref{eq:generic_splitting_subleading}, which consists of the
gamma-function coefficient, two corrections in powers of $\kappa^2$,
and the leading three-traversal term.  The coefficient extends the
known spin-tunneling result to noninteger $\beta/\hbar$.  In exact WKB
we expressed the classical periods in terms of complete elliptic
integrals and obtained the differential operator that generates their
first quantum correction, together with the finite part of the barrier
action as $\kappa\to0^+$.  The classical periods reproduce the leading
term of the Stirling expansion of the exact coefficient, and the first
quantum correction reproduces the next term.  A WKB approximation of
finite order thus gives the coefficient with an error that does not
decrease as $\kappa\to0^+$.  With the first correction this error is
about one percent for the lowest QES doublets.

At integer $r_n$, which includes all QES doublets, the spectrum is
regular, but the coefficients of the conformal block have Kac poles of
increasing order.  We showed that a two-mode reduction of the Floquet
problem gives the square-root law \eqref{eq:resonant_energy_sheets},
with an explicit coefficient for every QES doublet.  At zero detuning
its two branches are the even and the odd state, and their separation
is the known spin-tunneling splitting
\cite{Garanin:1991,UlyanovZaslavskii:1992}.  The same reduction
determines the leading Kac pole \eqref{eq:kac_pole_tower} of the block
at each order.  The sum of these poles cancels the pole of the gamma
functions in the connection formula, and without it the gamma functions
would give only half of the splitting.  This last statement assumes the
continuation of the Matone relation described in
section~\ref{sec:resonant_block}.  Finally,
we used the position of the QES levels in the spectrum to show that at
most $2\lceil(\beta/\hbar-1)/2\rceil$ levels lie below the barrier top,
for all $\beta/\hbar>0$.  The bound is attained at sufficiently weak
coupling.

Several questions remain open.  The first is to find a representation
of the connection matrix in terms of conformal blocks that is uniform
near the resonances.  This would remove the assumption on the Matone
relation and would make it possible to follow the even and odd levels
as $\beta/\hbar$ passes through an integer.  The second is an analysis
of the quantum periods to all orders, which should give the finite part
of the barrier period beyond the first quantum correction and reproduce
the full gamma-function coefficient.  The TBA equations for the quantum
Seiberg--Witten curve of the $N_f=2$ theory \cite{Imaizumi:2021} are a
natural starting point.  The third is the limit $r_n\to0^+$, in which a
Morse level reaches the threshold.  This limit has to be taken together
with $\kappa\to0^+$, and a uniform description would relate the
weak-coupling expansion of the doublets to the spectrum near the
threshold.  A related problem is the continuation through the barrier
top, where the cycles used in section~\ref{sec:exact_wkb} change
\cite{MisumiPazarbasi:2025,MisumiPazarbasi:2026}.

Our results can also be stated in the language of gauge theory.  The
Razavy operator is the quantum Seiberg--Witten curve of the
$\mathcal N=2$, $SU(2)$ gauge theory with two flavors of equal mass
\cite{Ashok:2016,Aminov:2022,Imaizumi:2021}.  Under the AGT
correspondence, and up to normalization conventions, the parameters of
section~\ref{sec:irregular_block} have the following interpretation in
units of $\hbar$.  The coupling $\kappa^2$ is the instanton counting
parameter, the Floquet exponent $\nu$ is the Coulomb parameter,
$\beta/2$ is the mass of the two flavors, and $\mathcal F$ is the
instanton part of the Nekrasov--Shatashvili free energy.  The
weak-coupling limit of the double well is the weak-coupling limit of
the gauge theory, and the series in $\kappa^2$ that multiplies each
traversal term is an instanton expansion.  The factor $\kappa^{r_n}$ is
in general not analytic in the instanton counting parameter.  It
originates from the power $\kappa^{2\nu}$ and the gamma functions in
the connection formula, and not from the block.  The poles
\eqref{eq:kac_pole_tower} are the leading singularities of the free
energy at $2\nu=r$ for $\beta=r+1+2n$, and their sum into a square root
is the two-flavor counterpart of the resummations of poles studied in
\cite{Gorsky:2017ndg,Beccaria:2016wop,Alekseev:2018kcn,Alekseev:2019gkl}.

\acknowledgments
J.R. thanks Edwan Pr\'eau and Hongfei Shu for helpful communications.

\appendix

\section{Coefficients of the conformal block and their poles}
\label{app:floquet_formal}

In this appendix we use the recurrence for the Floquet coefficients to
compute the coefficients of the classical block and their leading
poles.  We also derive the leading multi-traversal correction to the
splitting.  The estimates needed at the resonances are given in
appendix~\ref{app:floquet_feshbach}.

\subsection{Recursion for the coefficients}
\label{app:block_recursion}

Let $2\nu\notin\mathbb Z$ and $j=(\beta-1)/2$.  On the coefficients
$c_m$ of the Floquet series $\phi_\nu=\sum_mc_m\ee^{(\nu+m)x}$ we define
the difference operator
\begin{equation}
  H_1c_m:=(\nu+m-1-j)c_{m-1}
  -(\nu+m+1+j)c_{m+1}.
  \label{eq:floquet_H1_definition}
\end{equation}
Here $H_1c_m$ denotes the $m$-th component of $H_1$ applied to the
sequence $(c_m)$.  We expand in powers of $\kappa$ with the
normalization $c_0=1$,
\begin{equation}
  c_m=\sum_{q\geq0}c_m^{(q)}\kappa^q,\qquad
  E=-\nu^2+\sum_{q\geq1}e_q\kappa^q,\qquad
  c_m^{(0)}=\delta_{m0},\quad c_0^{(q>0)}=0.
  \label{eq:floquet_coefficient_expansion}
\end{equation}
Inserting these expansions in \eqref{eq:off_shell_floquet_action},
which holds for arbitrary $\beta$ with $j=(\beta-1)/2$ and with
$k+\delta$ replaced by $\nu+m$, we obtain
\begin{align}
  e_q&=H_1c_0^{(q-1)}+\frac12\delta_{q2},
  \label{eq:floquet_energy_recursion}\\
  c_m^{(q)}
  &=\frac{H_1c_m^{(q-1)}+\frac12c_m^{(q-2)}
  -\sum_{p=1}^{q}e_pc_m^{(q-p)}}
  {m(2\nu+m)},\qquad m\ne0,
  \label{eq:floquet_vector_recursion}
\end{align}
where coefficients with a negative superscript vanish.  At order $q$
only the coefficients with $|m|\leq q$ are nonzero, so that each step
of the recursion involves finitely many terms.  The recurrence is
invariant under $\kappa\to-\kappa$, $c_m\to(-1)^mc_m$, which implies
$e_{2q+1}=0$.  Comparing with the Matone relation
\eqref{eq:matone_energy_relation}, we find
\begin{equation}
  f_q=\frac{e_{2q}}{q}.
  \label{eq:floquet_to_block_coefficients}
\end{equation}

At the lowest order,
\begin{equation}
  c_1^{(1)}=\frac{\nu-j}{2\nu+1},\qquad
  c_{-1}^{(1)}=\frac{\nu+j}{2\nu-1},
  \label{eq:first_floquet_vectors}
\end{equation}
and hence
\begin{equation}
  e_2=\frac12+(\nu-1-j)c_{-1}^{(1)}
  -(\nu+1+j)c_1^{(1)}
  =-\frac{\beta^2}{2(4\nu^2-1)}=f_1.
  \label{eq:analytic_f1_derivation}
\end{equation}
At the next order we need
\begin{align}
  c_2^{(2)}
  &=\frac{(\nu+1-j)(\nu-j)}
  {4(\nu+1)(2\nu+1)},&
  c_{-2}^{(2)}
  &=\frac{(\nu-1+j)(\nu+j)}
  {4(\nu-1)(2\nu-1)},\nonumber\\
  c_1^{(3)}
  &=\frac{-(\nu+2+j)c_2^{(2)}
  +(\frac12-e_2)c_1^{(1)}}{2\nu+1},&
  c_{-1}^{(3)}
  &=\frac{(\nu-2-j)c_{-2}^{(2)}
  +(\frac12-e_2)c_{-1}^{(1)}}{1-2\nu}.
  \label{eq:third_order_floquet_vectors}
\end{align}
These give
\begin{equation}
  e_4=(\nu-1-j)c_{-1}^{(3)}
  -(\nu+1+j)c_1^{(3)}=2f_2,
  \label{eq:analytic_f2_derivation}
\end{equation}
which simplifies to
\begin{equation}
  f_2(\nu,\beta)=
  -\frac{
  (4\nu^2-1)^3
  -6\beta^2(4\nu^2-1)^2
  +\beta^4(20\nu^2+7)}
  {64(\nu^2-1)(4\nu^2-1)^3}.
  \label{eq:irregular_block_second_term}
\end{equation}
To write the coefficient \eqref{eq:generic_splitting_c2} explicitly, we
define
\begin{equation}
  \mathcal N_2(r,\beta)
  =(r^2-1)^4
  -6\beta^2(r^2-1)^2(2r^2-5)
  +3\beta^4(5r^4-4r^2-37).
  \label{eq:generic_splitting_N2}
\end{equation}
Inserting $f_2$, we find
\begin{equation}
  \displaystyle
  c_{2,n}
  =\frac{(r_n+4)
  \left[4\beta^4r_n(r_n^2-4)^2-\mathcal N_2(r_n,\beta)\right]}
  {8(r_n^2-4)^2(r_n^2-1)^4}.
  \label{eq:generic_splitting_c2_explicit}
\end{equation}

\subsection{Multi-traversal terms}
\label{app:multitraversal}

The leading multi-traversal correction is determined by the expansion
of the gamma functions in the connection formula around their pole.
For noninteger $r>0$ the residue is $\mathcal A_{r,n}/r$, with the
coefficient defined in \eqref{eq:block_series_and_morse_residue}.  We
use the notation
\begin{equation}
  \frac{\mathcal A_{r,n}}{r}
  =\frac{\Gamma(n+r+1)}{r\,n!\,\Gamma(r)^2},
  \qquad
  H_n^{(p)}=\sum_{k=1}^{n}\frac1{k^p},
  \qquad H_0^{(p)}=0.
  \label{eq:multitraversal_residue}
\end{equation}
We write $H_n=H_n^{(1)}$ and denote the digamma and trigamma functions
by $\psi$ and $\psi_1$.  We also define the constants
\begin{align}
  g_{1;r,n}
  ={}&H_n-\gamma_{\rm E}+\psi(n+r+1)-4\psi(r)
  -\frac2r-\pi\cot(\pi r),
  \label{eq:multitraversal_g1}\\
  g_{2;r,n}
  ={}&H_n^{(2)}+\frac{\pi^2}{6}+\psi_1(n+r+1)-8\psi_1(r)
  +\frac4{r^2}+3\pi^2\csc^2(\pi r).
  \label{eq:multitraversal_g2}
\end{align}
They are the coefficients in the expansion at $\nu=r/2+\delta$,
\begin{equation}
  \frac{\Gamma(-2\nu)\Gamma(-n+\delta)}
  {\Gamma(2\nu)\Gamma(-n-r-\delta)}
  =\frac{\mathcal A_{r,n}}{r\delta}
  \exp\!\left[
    g_{1;r,n}\delta+\frac12g_{2;r,n}\delta^2+\OO(\delta^3)
  \right].
  \label{eq:regularised_gamma_quotient}
\end{equation}
To derive these expressions, we consider the function
\begin{equation}
  \mathcal G_{r,n}(\delta)
  :=\frac{r\delta}{\mathcal A_{r,n}}
    \frac{\Gamma(-r-2\delta)\Gamma(-n+\delta)}
         {\Gamma(r+2\delta)\Gamma(-n-r-\delta)}.
\end{equation}
For noninteger $r>0$ it is holomorphic near $\delta=0$ with
$\mathcal G_{r,n}(0)=1$, and
$g_{j;r,n}=\left.\partial_\delta^j\log\mathcal G_{r,n}(\delta)
\right|_{\delta=0}$ for $j=1,2$.  Using
$\delta\Gamma(-n+\delta)=\Gamma(1+\delta)/
\prod_{k=1}^{n}(\delta-k)$, we find
\begin{equation}
  g_{2;r,n}=H_n^{(2)}+\frac{\pi^2}{6}
  +4\psi_1(-r)-4\psi_1(r)-\psi_1(-n-r).
\end{equation}
The identities
\begin{equation}
  \psi_1(z+1)=\psi_1(z)-\frac1{z^2},
  \qquad \psi_1(z)+\psi_1(1-z)=\pi^2\csc^2(\pi z),
\end{equation}
reduce this expression to \eqref{eq:multitraversal_g2}.  The expression
\eqref{eq:multitraversal_g1} follows in the same way from the
corresponding identities for $\psi$.  Note that $g_{2;r,n}$ is not the
derivative of $g_{1;r,n}$ with respect to $r$.

With $\ell_{r,n}(\kappa)=2\log\kappa+g_{1;r,n}$, the solution of the
quantization condition \eqref{eq:parity_resolved_heun_quantisation} at
zeroth order in the block is
\begin{equation}
  \delta\nu_\epsilon
  =\epsilon\frac{\mathcal A_{r,n}}{r}\kappa^r
   +\ell_{r,n}\frac{\mathcal A_{r,n}^2}{r^2}\kappa^{2r}
   +\frac{\epsilon\mathcal A_{r,n}^3}{2r^3}
      \left[3\ell_{r,n}^2+g_{2;r,n}\right]\kappa^{3r}
   +\OO\!\left(\kappa^{4r}|\log\kappa|^3\right).
  \label{eq:multitraversal_floquet_expansion}
\end{equation}
The block contributes further powers of $\kappa^2$ to each term.  The
term of order $\kappa^{2r}$ does not depend on the parity and shifts
both levels equally.  Converting to the energy with the Matone
relation, we find that the three-traversal term \eqref{eq:generic_three_traversal_gap} has the
coefficient
\begin{equation}
  \displaystyle
  d_{{\rm mt};r,n}(\kappa)
  =\frac{\mathcal A_{r,n}^2}{r^2}
  \left[
    \frac32\ell_{r,n}(\kappa)^2+\frac12g_{2;r,n}
    +\frac{2}{r}\ell_{r,n}(\kappa)
  \right].
  \label{eq:generic_multitraversal_coefficient}
\end{equation}
The square of the logarithm is characteristic of the three-instanton
contribution in a symmetric double well
\cite{ZinnJustin:2004,ZinnJustin:2004b}.  The result
\eqref{eq:generic_multitraversal_coefficient} holds for noninteger $r$,
and the integer case is treated in section~\ref{sec:resonant_block}.

\subsection{Leading Kac poles}
\label{app:kac_poles}

The poles of $c_{2,n}$ at $r_n=1$ and $r_n=2$ are examples of the Kac
poles of the block.  In general, let $\beta=r+1+2n$ and
$\nu=r/2+\delta$ with $r\in\mathbb Z^+$.  The denominator in
\eqref{eq:floquet_vector_recursion} at $m=-r$ is then $-2r\delta$.
Keeping $\delta\neq0$ and iterating the recursion to order
$\kappa^{2kr}$, one obtains any given coefficient $f_{kr}=e_{2kr}/(kr)$
as a Laurent series in $\delta$, and in particular its leading pole
\eqref{eq:kac_pole_tower}.

The leading poles for all $k$ can be obtained at once by eliminating
the modes with $m\neq0,-r$ from the recurrence that underlies
\eqref{eq:floquet_energy_recursion} and
\eqref{eq:floquet_vector_recursion}.  The two remaining modes have the
unperturbed energies $-\nu^2$ and $-(\nu-r)^2$, which differ by
$2r\delta$, and they are first connected at order $\kappa^r$.  At
$\delta=0$ the products of the matrix elements and of the energy
denominators along the shortest path are
\begin{equation}
  (-1)^r\kappa^r\frac{(n+r)!}{n!},
  \qquad
  \prod_{p=1}^{r-1}[-p(r-p)]=(-1)^{r-1}[(r-1)!]^2,
\end{equation}
so that the effective coupling is $-\mathcal A_{r,n}\kappa^r$ in both
directions (for $r=1$ the empty product is equal to one).  The
denominators of the eliminated modes do not vanish at $\delta=0$.  By
the reflection symmetry the correction to the difference of the
diagonal elements is $\OO(\delta\kappa^2)$, and the corrections to the
coupling are of order $\kappa^{r+2}$, as shown in
appendix~\ref{app:floquet_feshbach}.

We keep only the pole of highest order in $\delta$ at each order
$\lambda^{kr}$.  We denote the corresponding part of the shift of the
energy from $-\nu^2$, after subtraction of the regular shift of the
diagonal elements, by
$\Delta_{\rm lead}=\sum_{k\ge1}d_k(\delta)\lambda^{kr}$.  The
$2\times2$ determinant gives
\begin{equation}
  \Delta_{\rm lead}(\Delta_{\rm lead}-2r\delta)
       =\mathcal A_{r,n}^2\lambda^r,
  \qquad \Delta_{\rm lead}\longrightarrow0\quad(\lambda\to0).
\end{equation}
The corrections to the coupling, the energy dependence of the effective
matrix, and the terms that are analytic in $\delta$ contribute only
poles of lower order.  Comparing powers of $\lambda^r$, we find
\begin{equation}
 d_1=-\frac{\mathcal A_{r,n}^2}{2r\delta},\qquad
 d_k=\frac1{2r\delta}\sum_{a=1}^{k-1}d_a d_{k-a}\quad(k\ge2).
\end{equation}
The solution is
\begin{equation}
 d_k=(-1)^k\operatorname{Cat}_{k-1}
       \frac{\mathcal A_{r,n}^{2k}}{(2r\delta)^{2k-1}},
 \qquad
 \operatorname{Cat}_{k-1}=\frac1k\binom{2k-2}{k-1}.
\end{equation}
With $f_{kr}^{\rm lead}=d_k/(kr)$ and
$\binom{1/2}{k}=(-1)^{k-1}\operatorname{Cat}_{k-1}/2^{2k-1}$ this gives
\eqref{eq:kac_pole_tower}.

As a check, for $\beta=4$ and $\delta=\nu-3/2$ the recursion of
appendix~\ref{app:block_recursion} gives
\begin{align}
  f_3&=-\frac1{8\delta}+\frac{19}{16}-\frac{1759}{288}\delta+\OO(\delta^2),
  \label{eq:r3_f3_check}\\
  f_6&=\frac1{256\delta^3}-\frac{4001}{4608\delta}
  +\frac{25651}{2048}+\OO(\delta),
  \label{eq:r3_f6_check}
\end{align}
in agreement with \eqref{eq:kac_pole_tower} for
$\mathcal A_{3,0}=3/2$.  The term of order $\delta^{-2}$ in $f_6$ is
absent, which does not follow from \eqref{eq:kac_pole_tower}.
Similarly, for $\beta=5$ the residue of $f_4$ at $\nu=2$ agrees with
\eqref{eq:kac_pole_tower},
\begin{equation}
  \lim_{\nu\to2}(\nu-2)f_4(\nu,5)=-\frac1{72}
  =-\frac{\mathcal A_{4,0}^2}{2\cdot4^2},
  \qquad \mathcal A_{4,0}=\frac23.
  \label{eq:r4_f4_check}
\end{equation}

\section{Two-mode Floquet reduction and error estimates}
\label{app:floquet_feshbach}

In this appendix we derive the square-root law
\eqref{eq:resonant_energy_sheets} with bounds on the error terms.  The
two resonant modes are treated exactly, and all other modes, which are
separated from them by a finite gap, are eliminated.  The Matone
relation is used only in the last step, where the result is expressed
in terms of the block.  We set $\hbar=1$ and fix
\begin{equation}
  r\in\mathbb{Z}^+,\qquad n\in\mathbb Z_{\geq0},\qquad
  \ell=\frac r2,\qquad j=n+\ell,\qquad M=2j+1.
  \label{eq:feshbach_fixed_indices}
\end{equation}

\paragraph{The resonant pair.}
We write $\nu=\ell+\delta$.  The transformation of
section~\ref{sec:qes_module} leads to the expansion
\begin{equation}
  \psi(x)=\ee^{-\kappa\cosh x}\phi(x),\qquad
  \phi(x)=\sum_{k\in\mathbb Z+r/2}c_k\ee^{(k+\delta)x}.
\end{equation}
For $x=\ii\theta$ this is a Fourier series with the multiplier
$\ee^{2\pi\ii(\ell+\delta)}$.  We consider the operator on the space of
square-summable coefficients, with the domain
\begin{equation}
  \begin{aligned}
    \mathcal H_r^{\rm F}
      &=\left\{c=(c_k)_{k\in\mathbb Z+r/2}:\sum_k|c_k|^2<\infty\right\},\\
    \mathcal D^{\rm F}
      &=\left\{c\in\mathcal H_r^{\rm F}:\sum_k(1+k^4)|c_k|^2<\infty\right\}.
  \end{aligned}
  \label{eq:floquet_operator_domain}
\end{equation}
The weight $k^4$ accounts for the second derivative.  On the unit
vectors $e_k$ the operator of \eqref{eq:off_shell_floquet_action} acts
as
\begin{align}
  \mathbf L(\kappa,\delta)e_k={}&
  \left[\frac{\kappa^2}{2}-(k+\delta)^2\right]e_k
  +\kappa(k+\delta-j)e_{k+1}
  -\kappa(k+\delta+j)e_{k-1}.
  \label{eq:sequence_floquet_operator}
\end{align}

The off-diagonal terms change $k$ by one unit.  Their coefficients grow
as $|k|$, whereas the diagonal terms grow as $k^2$, so that $\mathbf L$
is a closed operator with compact resolvent.  Its eigenvalues are the
characteristic values of the Floquet problem at the given $\delta$.  At
$\kappa=\delta=0$ the eigenvalues are $-k^2$, and only those with
$k=\pm\ell$ coincide.  We define the projection
$Pc=c_\ell e_\ell+c_{-\ell}e_{-\ell}$ and $Q=1-P$.  The other modes are
separated from the resonant pair by the gap
\begin{equation}
  g_r=\inf_{\substack{k\in\ell+\mathbb Z\\k\neq\pm\ell}}
  |k^2-\ell^2|
  =
  \begin{cases}
    2,&r=1,\\
    r-1,&r\geq2.
  \end{cases}
  \label{eq:floquet_complement_gap}
\end{equation}

For $r\geq2$ the nearest mode is $k=\ell-1$, with
$\ell^2-(\ell-1)^2=r-1$.  For $r=1$ the resonant modes are $k=\pm1/2$
and the nearest other modes are $k=\pm3/2$, at a distance $2$.

\paragraph{Eliminating the complementary modes.}
We split $\mathbf L=\mathbf D+\mathbf T$ into its diagonal and
off-diagonal parts.  On the range of $Q$ the resolvent of the diagonal
part is
\begin{equation}
 R_D(E)e_k=\frac{e_k}{E-\kappa^2/2+(k+\delta)^2},\qquad k\ne\pm\ell.
\end{equation}
For sufficiently small $\kappa$ and $\delta$, and for $E$ in a fixed
disk inside $|E+\ell^2|<g_r/2$, the denominators do not vanish, and
\begin{equation}
  R_D(E)=[E-Q\mathbf DQ]^{-1},\qquad
  \|R_D(E)Q\mathbf TQ\|\leq C_{r,n}|\kappa|<1.
  \label{eq:diagonal_resolvent_bound}
\end{equation}

Indeed, for large $|k|$ the diagonal resolvent is $\OO(k^{-2})$ and the
off-diagonal elements are $\OO(\kappa|k|)$, so that their product is
bounded, and the finitely many remaining indices are controlled by the
gap.  The Neumann series then gives
\begin{equation}
  [E-Q\mathbf LQ]^{-1}
  =\sum_{m=0}^{\infty}
  [R_D(E)Q\mathbf TQ]^mR_D(E).
  \label{eq:complementary_neumann_series}
\end{equation}

Each factor of $\mathbf T$ contributes one step and a power of
$\kappa$, and each factor of $R_D$ an energy denominator.  Projecting
the eigenvalue equation $\mathbf Lc=Ec$ with $P$ and $Q$, we obtain
\begin{equation}
 (E-P\mathbf LP)Pc=P\mathbf LQ\,Qc,\qquad
 (E-Q\mathbf LQ)Qc=Q\mathbf LP\,Pc.
\end{equation}
Solving the second equation for $Qc$ and inserting the result in the
first, we find $H_{\rm eff}(E)Pc=EPc$ with
\begin{equation}
  H_{\rm eff}(E)
  =P\mathbf LP+
  P\mathbf LQ[E-Q\mathbf LQ]^{-1}Q\mathbf LP.
  \label{eq:exact_feshbach_matrix}
\end{equation}

This is the Feshbach--Schur map.  It involves no approximation, and the
characteristic values in the disk are the solutions of
\begin{equation}
  \det[E-H_{\rm eff}(E)]=0.
  \label{eq:feshbach_characteristic_equation}
\end{equation}

The self-energy $\Sigma=H_{\rm eff}-P\mathbf LP$ is the sum over the
paths that leave the range of $P$ and return to it, each of which has
at least two steps.  Since
$\partial_E(E-Q\mathbf LQ)^{-1}=-(E-Q\mathbf LQ)^{-2}$ is bounded, we
have
\begin{equation}
  \|\Sigma(E)\|\leq C_{r,n}\kappa^2,\qquad
  \|\partial_E\Sigma(E)\|\leq C_{r,n}\kappa^2.
  \label{eq:self_energy_bounds}
\end{equation}

\paragraph{Common shift and detuning.}
The diagonal elements of $P\mathbf LP$ are
$\kappa^2/2-(\pm\ell+\delta)^2
=-\ell^2-\delta^2+\kappa^2/2\mp r\delta$.
They differ by $2r\delta$ and have the common part
$-\ell^2-\delta^2+\kappa^2/2$.  To include the common part of the
self-energy we define $E_c$ by
\begin{equation}
  E_c=\frac12\operatorname{tr}H_{\rm eff}(E_c).
  \label{eq:implicit_centroid}
\end{equation}

By \eqref{eq:self_energy_bounds} the derivative of the right-hand side
with respect to $E_c$ is $\OO(\kappa^2)$, so that this equation has a
unique solution near $-\ell^2$.  Since $H_{\rm eff}$ depends on $E$,
the solution $E_c$ is in general not the mean of the two characteristic
values.  The reflection $(\mathsf Jc)_k=c_{-k}$ exchanges the two
resonant modes and satisfies
\begin{equation}
  \mathsf J\mathbf L(\kappa,\delta)\mathsf J=\mathbf L(\kappa,-\delta).
  \label{eq:floquet_reflection_identity}
\end{equation}

It follows that $E_c$ is even in $\delta$.  In the basis
$(e_\ell,e_{-\ell})$ we also find
\begin{equation}
  \Sigma_{++}(E;\kappa,\delta)=\Sigma_{--}(E;\kappa,-\delta),\qquad
  \Sigma_{++}-\Sigma_{--}=\OO_{r,n}(\delta \kappa^2).
  \label{eq:diagonal_self_energy_symmetry}
\end{equation}
The second relation holds because the difference of the diagonal
elements of $\Sigma$ is analytic and odd in $\delta$, and
$\Sigma=\OO(\kappa^2)$.  The common shift of order $\kappa^2$ can be
larger than the splitting of order $\kappa^r$, but it does not
contribute to the splitting.

At order $\kappa^2$ the self-energy is the sum of the two paths
$\ell\to\ell\pm1\to\ell$.  For $r>1$, with the denominators evaluated
at $E=-\ell^2-\delta^2$, we find
\begin{equation}
  \Sigma_{++}^{(2)}=\kappa^2\bigg[ \frac{(n-\delta)(n+r+1+\delta)} {r+1+(r+2)\delta}
  +\frac{(n+r+\delta)(n+1-\delta)} {-(r-1)+(r-2)\delta} \bigg]+\OO(\kappa^4).
  \label{eq:first_diagonal_self_energy}
\end{equation}

At $\delta=0$ the two denominators are $r+1$ and $-(r-1)$.  For $r=1$
the mode $\ell-1$ is the other resonant mode, and the corresponding
path is already contained in $P\mathbf LP$.  Taking the mean of the two
diagonal elements and adding the term $\kappa^2/2$, we obtain
\begin{equation}
  E_c=-\ell^2-\delta^2+c_{r,n}\kappa^2
  +\OO_{r,n}(\kappa^4+\delta^2\kappa^2),
  \label{eq:centroid_first_correction}
\end{equation}
\begin{equation}
  c_{r,n}=
  \begin{cases}
    \displaystyle\frac{(n+1)^2}{2},&r=1,\\[2mm]
    \displaystyle\frac12+
    \frac{n(n+r+1)}{r+1}
    -\frac{(n+1)(n+r)}{r-1},&r>1.
  \end{cases}
  \label{eq:centroid_coefficient}
\end{equation}

\paragraph{Coupling and eigenvalues.}
The modes $e_{-\ell}$ and $e_\ell$ are connected by paths along which
the index changes by $2\ell=r$.  The shortest path,
$-\ell\to-\ell+1\to\cdots\to\ell$, has $r$ steps and $r-1$ intermediate
denominators.  Its contributions to the two off-diagonal elements are
\begin{align}
  B_{\uparrow}^{(r)}(E)
  &=
  \frac{\kappa^r\prod_{p=0}^{r-1}(-\ell+p+\delta-j)}
  {\prod_{p=1}^{r-1}
  [E-\kappa^2/2+(-\ell+p+\delta)^2]},
  \label{eq:oriented_path_up}\\
  B_{\downarrow}^{(r)}(E)
  &=
  \frac{(-\kappa)^r\prod_{p=0}^{r-1}(\ell-p+\delta+j)}
  {\prod_{p=1}^{r-1}
  [E-\kappa^2/2+(-\ell+p+\delta)^2]}.
  \label{eq:oriented_path_down}
\end{align}

For $r=1$ the products in the denominators are empty and equal to one.
At $\delta=0$ and $E=-\ell^2$ the denominators are
$-\ell^2+(-\ell+p)^2=-p(r-p)$, with the product
$(-1)^{r-1}[(r-1)!]^2$, and both numerators are equal to
$(-1)^r\kappa^r(n+r)!/n!$.  The ratio is negative.

For $\delta\neq0$ the two off-diagonal elements $B_\uparrow$ and
$B_\downarrow$ of $H_{\rm eff}$ are not equal, but the characteristic
polynomial depends only on their product.  They are made equal by the
similarity transformation $S=\operatorname{diag}(\varsigma,1)$ with
$\varsigma^2=B_\uparrow/B_\downarrow$, which is bounded because the
ratio tends to one.  At $E=E_c$ the off-diagonal element of the
symmetrized matrix is, at leading order,
\begin{equation}
  -\sqrt{B_{\uparrow}^{(r)}B_{\downarrow}^{(r)}} =-\mathcal A_{r,n}\kappa^r
  +\OO_{r,n}(\kappa^{r+2}+\kappa^r\delta^2),\qquad \mathcal A_{r,n}
  =\frac{(n+r)!}{n![(r-1)!]^2}.
  \label{eq:symmetrised_shortest_path}
\end{equation}

The square root is positive at $\delta=0$.  The product of the
numerators is even in $\delta$,
\begin{equation}
  \prod_{m=n+1}^{n+r}\sqrt{(m-\delta)(m+\delta)}
  =\prod_{m=n+1}^{n+r}\sqrt{m^2-\delta^2}.
  \label{eq:even_delta_path_product}
\end{equation}

Pairing the factors $p$ and $r-p$ shows that the denominator at $E=E_c$
is also even, so that the coupling has no term linear in $\delta$.
Longer paths have at least two additional steps and contribute at order
$\kappa^{r+2}$.

We now take $\delta$ real with $|\delta|\leq D\kappa^r$, for fixed $D$
and sufficiently small $\kappa>0$.  In this range the detuning is
comparable to the coupling, and the resonance $\delta=0$ is included.
The effective matrix takes the form
\begin{equation}
  S^{-1}H_{\rm eff}(E_c)S
  =E_c\mathbf1+
  \begin{pmatrix}
    -r\delta&-\mathcal A_{r,n}\kappa^r\\
    -\mathcal A_{r,n}\kappa^r&r\delta
  \end{pmatrix}
  +\mathcal R_c,\qquad
  \|\mathcal R_c\|\leq C\kappa^{r+2}.
  \label{eq:controlled_effective_matrix}
\end{equation}

The error of the diagonal elements is $\OO(\delta\kappa^2)$ and that of
the off-diagonal elements is $\OO(\kappa^{r+2}+\kappa^r\delta^2)$.  For
$|\delta|\leq D\kappa^r$ both are $\OO(\kappa^{r+2})$, since
$3r\geq r+2$ for $r\geq1$.  The characteristic equation
\eqref{eq:feshbach_characteristic_equation} involves $H_{\rm eff}(E)$
instead of $H_{\rm eff}(E_c)$.  Its solutions satisfy
$|E-E_c|=\OO(\kappa^r)$, and by \eqref{eq:self_energy_bounds} the
difference of the two matrices is
$\OO(\kappa^2|E-E_c|)=\OO(\kappa^{r+2})$.  The two characteristic
values are therefore
\begin{equation}
  E_\pm=E_c(\kappa,\delta)\pm
  \sqrt{r^2\delta^2+\mathcal A_{r,n}^2\kappa^{2r}}
  +\OO_{r,n,D}(\kappa^{r+2}).
  \label{eq:rigorous_local_energy_sheets}
\end{equation}

The constants in the error term depend on $r$, $n$, and $D$, but not on
$\kappa$ and $\delta$.  At $\delta=0$ the splitting is
$2\mathcal A_{r,n}\kappa^r+\OO(\kappa^{r+2})$.  As an example, for
$r=3$ and $n=0$ we have $M=4$ and the path
$-3/2\to-1/2\to1/2\to3/2$.  The numerator is $-6\kappa^3$ and the
denominator is $(-2)(-2)=4$, which gives the coupling $-3\kappa^3/2$
and the splitting $3\kappa^3+\OO(\kappa^5)$.  The common shift at
$\delta=0$ is $-\kappa^2+\OO(\kappa^4)$, which is much larger than the
splitting.

\paragraph{Estimate for the block.}
We now assume the continuation of the Matone relation described in
section~\ref{sec:resonant_block}.  Since the connection formula
involves $\partial_\nu\mathcal F$, we need an estimate for the
derivative of the energy with respect to $\delta$.  At fixed $E$ the
resolvent $R_Q=(E-Q\mathbf LQ)^{-1}$ satisfies
$\partial_\delta R_Q=R_Q(Q\partial_\delta\mathbf LQ)R_Q$.  The
coefficients of $\partial_\delta\mathbf L$ grow as $|k|$ and are
controlled as before.  Including the derivatives of the factors
$P\mathbf LQ$ and $Q\mathbf LP$, we find
\begin{equation}
  \|\partial_\delta\Sigma(E;\kappa,\delta)\|
  \leq C_{r,n}\kappa^2.
  \label{eq:self_energy_detuning_bound}
\end{equation}

With $\lambda=\kappa^2$ we subtract \eqref{eq:resonant_matone} from
$E+\nu^2=\lambda\partial_\lambda\mathcal F$, which defines
$\lambda\partial_\lambda\mathcal F_{\rm reg}$.  The estimates for the
derivatives of the diagonal and off-diagonal elements then give
$\partial_\delta[\lambda\partial_\lambda\mathcal F_{\rm reg}]=\OO(\lambda)$
for real $\delta$.  This bound cannot be obtained by differentiating
the error term in \eqref{eq:rigorous_local_energy_sheets}.

To integrate over the coupling at fixed $\delta\neq0$, we note that at
coupling $t$ the off-diagonal element is of order $t^{r/2}$, which is
comparable to $|\delta|$ at $t_*=|\delta|^{2/r}$.  We split the
integral at $\min\{\lambda,t_*\}$ and use nondegenerate perturbation
theory below this value and the two-mode estimate above it.  With the
normalization $\mathcal F_{\rm reg}=0$ at $\lambda=0$ we have
\begin{equation}
  \partial_\delta\mathcal F_{\rm reg}(\lambda,\delta)
  =\int_0^\lambda\frac{\dd t}{t}\,
  \partial_\delta[t\partial_t\mathcal F_{\rm reg}(t,\delta)],
  \qquad \mathcal F_{\rm reg}(0,\delta)=0.
  \label{eq:regular_block_normalised_integral}
\end{equation}

The numerator is $\OO(t)$, so that the integrand is bounded and the
integral is $\OO(\lambda)$.  Since $\partial_\nu=\partial_\delta$ at
fixed $r$, this is \eqref{eq:regular_block_power_counting}.  At
$\delta=0$ the estimate holds for the limits $\delta\to0^\pm$.

\section{Classical and quantum periods}
\label{app:period_calculations}

In this appendix we derive the expressions for the periods used in
section~\ref{sec:exact_wkb}.  We reduce the classical periods to
complete elliptic integrals, compute the finite part of the barrier
action in the limit $\kappa\to0^+$, and express the first quantum
correction through derivatives of the classical periods.

\subsection{Classical periods in terms of elliptic integrals}
\label{app:elliptic_reduction}

We use the parameter $m=k^2$ and the complete elliptic integrals
\begin{align}
  \K(m)&=\int_0^{\pi/2}\frac{\dd\theta}
  {\sqrt{1-m\sin^2\theta}},
  &
  \Eell(m)&=\int_0^{\pi/2}
  \sqrt{1-m\sin^2\theta}\,\dd\theta,
  \nonumber\\
  \Pell(n|m)&=\int_0^{\pi/2}
  \frac{\dd\theta}
  {(1-n\sin^2\theta)\sqrt{1-m\sin^2\theta}}.
  \label{eq:elliptic_integral_definitions}
\end{align}
The periods can be expressed in terms of the integrals
\begin{align}
  \cJ(n,m)
  &=\int_0^{\pi/2}
  \frac{\sin^2\theta\cos^2\theta\,\dd\theta}
  {(1-n\sin^2\theta)^2\sqrt{1-m\sin^2\theta}},
  \label{eq:J_integral_definition}\\
  \cH(n,m)
  &=\int_0^{\pi/2}
  \frac{\cos^2\theta\sqrt{1-m\sin^2\theta}}
  {(1-n\sin^2\theta)^2}\,\dd\theta.
  \label{eq:H_integral_definition}
\end{align}
The identity
\begin{equation}
  \partial_n\Pell(n|m)
  =\frac{n\Eell(m)+(m-n)\K(m)+(n^2-m)\Pell(n|m)}
  {2(m-n)(n-1)n}
  \label{eq:elliptic_Pi_derivative}
\end{equation}
and a decomposition into partial fractions give
\begin{align}
  \cJ(n,m)
  &=
  \frac{n\Eell(m)+(n-m)\K(m)
  +(n^2-2n+m)\Pell(n|m)}
  {2n^2(m-n)},
  \label{eq:J_elliptic_reduction}\\
  \cH(n,m)
  &=
  \frac{-n\Eell(m)+(m+n)\K(m)
  +(n^2-m)\Pell(n|m)}
  {2n^2}.
  \label{eq:H_elliptic_reduction}
\end{align}

For the $A$-cycle we set
\begin{equation}
  n_A=\frac{u_+-u_-}{u_+-1},
  \qquad
  m_A=\frac{2(u_+-u_-)}
  {(u_+-1)(u_-+1)}
  \label{eq:A_characteristic_and_parameter}
\end{equation}
and substitute
\begin{equation}
  u=\frac{u_--n_A\sin^2\theta}
  {1-n_A\sin^2\theta}.
  \label{eq:A_period_substitution}
\end{equation}
This substitution maps the interval $0\leq\theta\leq\pi/2$ onto
$u_-\leq u\leq u_+$, and we obtain
\begin{equation}
  t_0(E)=\frac{2\kappa}{\pi}
  \frac{(u_--1)(u_+-u_-)^2}
  {(u_+-1)\sqrt{(u_+-1)(u_-+1)}}
  \cJ(n_A,m_A).
  \label{eq:A_period_complete_elliptics}
\end{equation}
This is $t_0=\Pi_{A,0}/(2\pi)$ in the normalization of
\eqref{eq:action_conventions}.

For the barrier cycle we define
\begin{equation}
  n_B=\frac{u_--1}{u_-+1},
  \qquad
  m_B=\frac{(u_--1)(u_++1)}
  {(u_-+1)(u_+-1)}
  \label{eq:B_characteristic_and_parameter}
\end{equation}
and substitute
\begin{equation}
  u=\frac{1+n_B\sin^2\theta}
  {1-n_B\sin^2\theta}.
  \label{eq:B_period_substitution}
\end{equation}
This gives
\begin{equation}
  t_{D,0}(E)=
  8\kappa(u_--1)\sqrt{\frac{u_+-1}{u_-+1}}
  \cH(n_B,m_B).
  \label{eq:B_period_complete_elliptics}
\end{equation}
Below the barrier top we have
\begin{equation}
  0<m_A<n_A<1,\qquad
  0<n_B<m_B<1,\qquad
  m_A+m_B=1,
  \label{eq:elliptic_parameter_ordering}
\end{equation}
so that all functions in \eqref{eq:A_period_complete_elliptics} and
\eqref{eq:B_period_complete_elliptics} are real and are evaluated on
their principal branches.

The derivatives of the periods with respect to the energy are most
easily obtained from the integral representations.  The contributions
of the endpoints vanish, and we need the integrals
\begin{align}
  \int_{u_-}^{u_+}
  \frac{\dd u}
  {\sqrt{(u_+-u)(u-u_-)(u^2-1)}}
  &=
  \frac{2\K(m_A)}
  {\sqrt{(u_+-1)(u_-+1)}},
  \label{eq:A_holomorphic_period}\\
  \int_1^{u_-}
  \frac{\dd u}
  {\sqrt{(u_--u)(u_+-u)(u^2-1)}}
  &=
  \frac{2\K(m_B)}
  {\sqrt{(u_-+1)(u_+-1)}}.
  \label{eq:B_holomorphic_period}
\end{align}
They give \eqref{eq:A_period_derivative} and
\eqref{eq:B_period_derivative}.  At the bottom of the well,
\begin{equation}
  \frac{\partial t_0}{\partial E}
  \longrightarrow
  \frac1{\sqrt{\beta^2-4\kappa^2}}
  =\frac1{\Omega_0}
  \qquad (E\to\Vmin),
  \label{eq:period_harmonic_check}
\end{equation}
as required for a harmonic oscillator of frequency $\Omega_0$ with the
kinetic term of \eqref{eq:razavy_hamiltonian}.

\textbf{Finite part of the barrier action.} For fixed $-\beta^2/4<E<0$ we set
\begin{equation}
  d=\sqrt{\beta^2+4E},\qquad
  R=2\sqrt{-E}=\hbar r,\qquad
  \lambda_\pm(\kappa)=\kappa u_\pm
  =\frac{\beta\pm\sqrt{d^2+2\kappa^2}}{2}.
  \label{eq:scaled_barrier_endpoints}
\end{equation}
In the variable $y=\kappa u$ the barrier action is
\begin{equation}
  S_B(E;\kappa)=2\int_\kappa^{\lambda_-(\kappa)}
  \frac{\sqrt{[\lambda_-(\kappa)-y][\lambda_+(\kappa)-y]}}
  {\sqrt{y^2-\kappa^2}}\,\dd y.
  \label{eq:scaled_barrier_action}
\end{equation}
Let $A=(\beta+d)/2$, $B=(\beta-d)/2$, and $c=\sqrt{AB}=R/2$.
Subtracting from the numerator its value at $y=0$, we are led to the
convergent integral
\begin{equation}
  I(A,B)=\int_0^B
  \frac{\sqrt{(A-y)(B-y)}-\sqrt{AB}}{y}\,\dd y.
  \label{eq:barrier_finite_integral}
\end{equation}
The substitution $t^2=(B-y)/(A-y)$ gives
\begin{equation}
  I(A,B)=c\log\frac{4A}{A-B}
  +\frac{A+B}{2}
  \log\frac{1-\sqrt{B/A}}{1+\sqrt{B/A}}-c.
  \label{eq:barrier_finite_integral_result}
\end{equation}
At $E=-\hbar^2r^2/4$, with $d=\sqrt{\beta^2-\hbar^2r^2}$, we therefore
obtain
\begin{equation}
  S_B(E;\kappa)=\hbar r\log\frac1\kappa
  +\hbar r\left[\log\frac{2\hbar^2r^2}{d}-1\right]
  +\frac{\beta}{2}\log\frac{\beta-\hbar r}{\beta+\hbar r}
  +\OO\!\left(\kappa^2\log\frac1\kappa\right),
  \label{eq:barrier_finite_part_appendix}
\end{equation}
which is \eqref{eq:barrier_action_degeneration} with the constant
\eqref{eq:barrier_finite_part}.

\subsection{The first quantum correction}
\label{app:quantum_periods}

To express the first quantum correction \eqref{eq:first_quantum_period}
through derivatives of the classical periods, we write $w=u-\xi$ and
$p=\kappa\sqrt{\eta^2-w^2}$.  At fixed $u$ we have
\begin{equation}
  \partial_\eta^2p=-\frac{\kappa^4w^2}{p^3},
  \qquad
  \partial_\eta\partial_\xi p
  =-\frac{\kappa^4\eta w}{p^3},
  \qquad
  \partial_\xi^2p=-\frac{\kappa^4\eta^2}{p^3}.
  \label{eq:classical_differential_parameter_derivatives}
\end{equation}
With $V''(x)=2\kappa^2(2u^2-\xi u-1)$ it follows that
\begin{equation}
  \frac1{24\kappa^2} \left[ -2\partial_\eta^2
  -\frac{3\xi}{\eta}\partial_\eta\partial_\xi -\frac{\xi^2-1}{\eta^2}\partial_\xi^2
  \right]p =\frac{\kappa^2[2w^2+3\xi w+\xi^2-1]}{24p^3} =\frac{V''(x)}{48p^3}.
  \label{eq:quantum_period_operator_pointwise}
\end{equation}
Since the measure $\dd x=\dd u/\sqrt{u^2-1}$ does not depend on the
parameters, integration of \eqref{eq:quantum_period_operator_pointwise}
over a cycle and \eqref{eq:first_quantum_period} give
\begin{equation}
  \Pi_{\gamma,2}
  =\frac1{24\kappa^2}
  \left[
    -2\partial_\eta^2
    -\frac{3\xi}{\eta}\partial_\eta\partial_\xi
    -\frac{\xi^2-1}{\eta^2}\partial_\xi^2
  \right]\Pi_{\gamma,0},
  \qquad \gamma=A,B.
  \label{eq:first_quantum_period_operator}
\end{equation}
The same relations hold for $t=\Pi_A/(2\pi)$ and $t_D=-\ii\Pi_B$.

For the limit $\kappa\to0^+$ it is convenient to use the variables
\begin{equation}
  D=2\kappa\eta,\qquad
  \xi=\frac{\beta}{2\kappa},\qquad
  E=-\frac{\beta^2}{4}-\frac{\kappa^2}{2}+\frac{D^2}{4}.
  \label{eq:quantum_period_scaled_variables}
\end{equation}
At fixed $\kappa$ we have $\partial_\xi=2\kappa\partial_\beta$ and
$\partial_\eta=2\kappa\partial_D$, and hence
\begin{equation}
  \Pi_{\gamma,2}=\mathcal L_\kappa \Pi_{\gamma,0},
  \qquad
  \mathcal L_\kappa
  =-\frac13\partial_D^2
   -\frac{\beta}{2D}\partial_D\partial_\beta
   -\frac{\beta^2-4\kappa^2}{6D^2}\partial_\beta^2.
  \label{eq:scaled_quantum_period_operator}
\end{equation}
\begin{samepage}
We also define
\begin{equation}
  R=\sqrt{\beta^2-D^2},\qquad
  F(\beta,D)
  =R\log\frac{2R^2}{D}
   +\frac{\beta}{2}\log\frac{\beta-R}{\beta+R}-R.
  \label{eq:general_barrier_finite_part}
\end{equation}
\end{samepage}
In terms of these functions the result of
appendix~\ref{app:elliptic_reduction} reads
\begin{equation}
  t_{D,0}=2R\log\frac1\kappa+2F(\beta,D)
  +\OO\!\left(\kappa^2\log\frac1\kappa\right).
  \label{eq:dual_period_general_degeneration}
\end{equation}
At $\kappa=0$ the operator
\begin{equation}
  \mathcal L_0
  =-\frac13\partial_D^2
   -\frac{\beta}{2D}\partial_D\partial_\beta
   -\frac{\beta^2}{6D^2}\partial_\beta^2
  \label{eq:limiting_quantum_period_operator}
\end{equation}
annihilates $R$, so that the first quantum correction does not change
the coefficient of $\log(1/\kappa)$.  Moreover,
\begin{equation}
  \mathcal L_0[2F(\beta,D)]
  =\frac{2D^2-\beta^2}{3D^2R}.
  \label{eq:finite_part_operator_result}
\end{equation}
At $E=-\hbar^2r^2/4$ we have $R\to\hbar r$ and
$D^2\to\beta^2-\hbar^2r^2$ as $\kappa\to0^+$, and therefore
\begin{equation}
  \lim_{\kappa\to0^+}t_{D,2}
  =\frac{\beta^2-2\hbar^2r^2}{3\hbar r(\beta^2-\hbar^2r^2)},
  \label{eq:first_quantum_period_limit_appendix}
\end{equation}
which is \eqref{eq:first_B_period_degeneration}.

\subsection{Coefficient of the power law and the first WKB correction}
\label{app:wkb_amplitudes}

Here we supply the details of the comparison in
section~\ref{sec:wkb_amplitude_comparison}, including the normalization
of the one-traversal term and the corrections to the perturbative
energy, and we compare the result with the Stirling expansion of the
exact coefficient.

\paragraph{Normalization.}
With the kinetic term $-\hbar^2\partial_x^2$ the one-traversal
splitting \eqref{eq:exact_wkb_leading_splitting} is
\begin{equation}
  \Delta E^{(1)}
  =\frac{\hbar}{\pi\,\partial_Et(E_{\rm p};\hbar)}
    \exp\!\left[-\frac{t_D(E_{\rm p};\hbar)}{2\hbar}\right],
  \qquad t(E_{\rm p};\hbar)=\hbar(N+1/2).
  \label{eq:amplitude_period_formula}
\end{equation}
With the classical periods the prefactor is
$\hbar\kappa\sqrt{(u_+-1)(u_-+1)}/\K(m_A)$.  For $\kappa\to0$ the
parameter behaves as $m_A=8d\kappa/(\hbar^2r^2)+\OO(\kappa^2)$ with
$d=\sqrt{\beta^2-\hbar^2r^2}$, so that $\K(m_A)\to\pi/2$.
The finite part in \eqref{eq:barrier_finite_part_appendix} and
$\partial_Et_0\to1/(\hbar r)$ then give
\eqref{eq:classical_power_coefficient} at $\hbar=1$.

At the next order in $\hbar$ all quantities in
\eqref{eq:amplitude_period_formula} are corrected.  At fixed $N$ we
write $E_{\rm p}=E_0+\hbar^2e_2+\OO(\hbar^4)$ with
$e_2=-t_2(E_0)/t_0'(E_0)$, where a prime denotes a derivative with
respect to the energy.  Then
\begin{align}
  t_D(E_{\rm p};\hbar)
    &=t_{D,0}(E_0)+\hbar^2
      [t_{D,2}(E_0)+e_2t_{D,0}'(E_0)]+\OO(\hbar^4),
      \label{eq:amplitude_shifted_dual_period}\\
  \partial_Et(E_{\rm p};\hbar)
    &=t_0'(E_0)+\hbar^2
      [t_2'(E_0)+e_2t_0''(E_0)]+\OO(\hbar^4).
      \label{eq:amplitude_shifted_A_derivative}
\end{align}
We now consider the coefficients of these expansions in the limit
$\kappa\to0^+$.

In the variables \eqref{eq:quantum_period_scaled_variables}, with
$a=(\beta-D)/2$, $b=(\beta+D)/2$, and $R=\sqrt{\beta^2-D^2}$, we have
\begin{equation}
  t_0=\frac1\pi\int_a^b
    \frac{\sqrt{(b-y)(y-a)}}{\sqrt{y^2-\kappa^2}}\,\dd y
    =\frac{\beta-R}{2}+\frac{D^2}{2R^3}\kappa^2
      +\OO(\kappa^4).
  \label{eq:morse_A_period_expansion}
\end{equation}
For fixed $0<D<\beta$ the lower limit $a$ stays away from zero, so that
the integrand can be expanded in $\kappa^2$, uniformly on compact sets
of the parameters and together with its derivatives.  Applying
\eqref{eq:scaled_quantum_period_operator}, we find
\begin{equation}
  t_2=\frac{\beta^2}{2R^5}\kappa^2+\OO(\kappa^4),
  \qquad t_2'=\OO(\kappa^2),\qquad e_2=\OO(\kappa^2).
  \label{eq:morse_A_quantum_correction}
\end{equation}
At fixed negative energy, $R\to\hbar r=2\sqrt{-E}$.  The energy
determined by the classical quantization condition differs from
$-\hbar^2r_n^2/4$ by $\OO(\kappa^2)$, which changes the barrier action
by $\OO(\kappa^2\log(1/\kappa))$.  Since
$t_{D,0}'=\OO(\log(1/\kappa))$, the term $e_2t_{D,0}'$ also vanishes in
the limit.  The shift of the energy and the correction to
$1/\partial_Et$ therefore do not contribute to the coefficient of the
power law, and only the correction $t_{D,2}$ to the barrier period
remains.  This simplification is special to the limit
$\kappa\to0^+$.

\paragraph{The correction to the coefficient.}
From appendix~\ref{app:quantum_periods} we have
\begin{equation}
  T_2(\beta,r;\hbar):=\lim_{\kappa\to0^+}t_{D,2}
   =\frac{\beta^2-2\hbar^2r^2}{3\hbar r(\beta^2-\hbar^2r^2)}.
  \label{eq:amplitude_T2_definition}
\end{equation}
The logarithmic term is absent because $\mathcal L_0R=0$, and the
constant is $\mathcal L_0(2F)$.  Here the expansion
\eqref{eq:dual_period_general_degeneration} is differentiated term by
term, which is justified on compact subsets of $0<R<\beta$.  With
$\mathsf L=\log(2R^2/D)$ and
$\mathsf H=\log[(\beta-R)/(\beta+R)]$, the derivatives of $F$ are
\begin{equation}
  F_D=-\frac D R\mathsf L,\qquad
  F_\beta=\frac\beta R\mathsf L+\frac12\mathsf H,
  \qquad \mathcal L_0(2F)=\frac{2D^2-\beta^2}{3D^2R}.
  \label{eq:amplitude_finite_part_check}
\end{equation}
Since the period is divided by $2\hbar$ in
\eqref{eq:amplitude_period_formula}, the term $\hbar^2t_{D,2}$
contributes $-\hbar T_2/2$ to the exponent.  Keeping it in the exponent
gives the factor $\exp(-\hbar T_2/2)$, whereas an expansion to first
order gives $1-\hbar T_2/2$.  The two expressions differ at higher
order.  In table~\ref{tab:wkb_qes_amplitudes} we use the exponential,
as in \eqref{eq:first_corrected_power_coefficient}.

\paragraph{Comparison with the exact coefficient.}
At $\hbar=1$ we write $T_2(\beta,r)$ for $T_2(\beta,r;1)$ and
\begin{equation}
  a_\pm=\frac{\beta\pm r}{2},\qquad a_-=n+\frac12,
  \qquad
  C_{\beta,n}
    =\frac{2\Gamma(a_++1/2)}{\Gamma(a_-+1/2)\Gamma(r)^2}.
  \label{eq:amplitude_gamma_shifted}
\end{equation}
For large positive $z$ the Stirling expansions
\cite[eqs.~(5.11.1), (5.11.8)]{DLMF:Gamma} are
\begin{align}
  \log\Gamma(z+1/2)
    &=z\log z-z+\tfrac12\log(2\pi)-\frac1{24z}+\OO(z^{-3}),
       \label{eq:amplitude_half_shift_stirling}\\
  \log\Gamma(z)
    &=(z-1/2)\log z-z+\tfrac12\log(2\pi)+\frac1{12z}+\OO(z^{-3}).
       \label{eq:amplitude_unshifted_stirling}
\end{align}
Their leading terms give
\begin{equation}
  \log C_{\beta,n}^{(1,0)}
  =\log\frac r\pi+r+a_+\log a_+-a_-\log a_--2r\log r,
  \label{eq:amplitude_stirling_classical}
\end{equation}
which is \eqref{eq:classical_power_coefficient}, since $a_+-a_-=r$ and
$4a_+a_-=\beta^2-r^2$.  The terms of order $1/z$ give
\begin{equation}
  \log\frac{C_{\beta,n}}{C_{\beta,n}^{(1,0)}}
  =-\frac1{24a_+}+\frac1{24a_-}-\frac1{6r}
    +\OO(a_+^{-3}+a_-^{-3}+r^{-3}).
  \label{eq:amplitude_stirling_correction}
\end{equation}
The three terms add up to
$-(\beta^2-2r^2)/[6r(\beta^2-r^2)]=-T_2(\beta,r)/2$.  The first quantum
correction to the barrier period thus reproduces the first correction
in the Stirling expansion of the exact coefficient.

The remainder in \eqref{eq:amplitude_stirling_correction} is small only
if $a_+$, $a_-$, and $r$ are all large, which is not the case for the
entries of table~\ref{tab:wkb_qes_amplitudes}.  For instance,
$a_-=1/2$ for the ground doublet at any $M$.  With $\hbar$ restored,
the arguments of the gamma functions are $(\beta/\hbar\pm r)/2$ and
$r$, which are of order $1/\hbar$ at fixed $\hbar r=2\sqrt{-E}$, and
the correction in \eqref{eq:amplitude_stirling_correction} becomes
$-\hbar T_2(\beta,r;\hbar)/2$.  This comparison identifies the first
correction to the classical coefficient.  We have not examined the
higher orders.

\section{Explicit QES spectra}
\label{app:qes_examples}

In this appendix we give the QES spectra for $M\leq5$ and use them to
check the results of section~\ref{sec:resonance_qes}.  The spectra
agree with \cite{Konwent:1998,Becker:2017}, except for a sign in the
case $M=4$ discussed below.  We set $\hbar=1$, so that $\beta=M$ and
the parameters of \eqref{eq:hbar_rescaling_dictionary} are written
without hats.  The matrices $H_+$ and $H_-$ are the even and odd blocks
of the spin Hamiltonian \eqref{eq:qes_spin_hamiltonian} without the
constant $\kappa^2/2$.

For $M=1$ the only QES level is even, with
\begin{equation}
  E=\frac{\kappa^2}{2}.
  \label{eq:qes_M1_appendix}
\end{equation}

For $M=2$ the QES levels form one doublet,
\begin{equation}
  E_{\mathrm e}
  =\frac{\kappa^2}{2}-\frac14-\kappa,
  \qquad
  E_{\mathrm o}
  =\frac{\kappa^2}{2}-\frac14+\kappa,
  \qquad
  \Delta E=2\kappa.
  \label{eq:qes_M2}
\end{equation}
Here $r=1$ and $\mathcal A_{1,0}=1$, and the splitting $2\kappa$ is
exact.

For $M=3$ the two blocks are
\begin{equation}
  H_+=
  \begin{pmatrix}
    -1&-2\kappa\\
    -2\kappa&0
  \end{pmatrix},
  \qquad
  H_-=(-1).
  \label{eq:qes_M3_blocks}
\end{equation}
Including the constant $\kappa^2/2$, we obtain one odd and two even
levels,
\begin{align}
  E_{\mathrm o}
  &=-1+\frac{\kappa^2}{2},
  \label{eq:qes_M3_odd}\\
  E_{\mathrm e,\pm}
  &=\frac{\kappa^2-1
  \pm\sqrt{1+16\kappa^2}}{2}.
  \label{eq:qes_M3_even}
\end{align}
In the notation of section~\ref{sec:double_well_chamber},
$E_{\mathrm e,-}=E_{0,\mathrm e}$, $E_{\mathrm o}=E_{0,\mathrm o}$, and
$E_{\mathrm e,+}=E_{1,\mathrm e}$.  The level $E_{\mathrm e,+}$ tends
to zero as $\kappa\to0$ and has no partner among the QES levels.  The
other two levels form the doublet with $r=2$, and their splitting is
\begin{equation}
  E_{\mathrm o}-E_{\mathrm e,-}
  =\frac{1}{2}
  \left(\sqrt{1+16\kappa^2}-1\right)
  =4\kappa^2
   -16\kappa^4
   +\OO(\kappa^6).
  \label{eq:qes_M3_gap}
\end{equation}
Since $\mathcal A_{2,0}=2$, the leading term agrees with
\eqref{eq:qes_leading_splitting} and with the result
\eqref{eq:resonant_block_splitting} of
section~\ref{sec:resonant_block}.

In the cases $M=4$ and $M=5$ the corrections to the leading splitting
can also be checked.  We write $\mathcal E=E-\kappa^2/2$.

For $M=4$ we have
\begin{equation}
  H_\pm=
  \begin{pmatrix}
    -\frac94&-\sqrt3\kappa\\
    -\sqrt3\kappa&-\frac14\mp2\kappa 
  \end{pmatrix},
  \label{eq:qes_M4_blocks}
\end{equation}
and the four levels are
\begin{align}
  E_{\mathrm e,\pm}
  &=\frac{\kappa^2}{2}-\frac54-\kappa 
    \pm\sqrt{1-2\kappa+4\kappa^2},
  \label{eq:qes_M4_even_energies}\\
  E_{\mathrm o,\pm}
  &=\frac{\kappa^2}{2}-\frac54+\kappa 
    \pm\sqrt{1+2\kappa+4\kappa^2}.
  \label{eq:qes_M4_odd_energies}
\end{align}
The splittings of the lower and of the upper doublet are
\begin{equation}
  \Delta E_{4,0} =2\kappa+\sqrt{1-2\kappa+4\kappa^2}-\sqrt{1+2\kappa+4\kappa^2}
  =3\kappa^3\left(1-\frac54\kappa^2-\frac78\kappa^4+\OO(\kappa^6)\right),
  \label{eq:qes_M4_lower_gap}
\end{equation}
\begin{equation}
  \Delta E_{4,1} =2\kappa+\sqrt{1+2\kappa+4\kappa^2}-\sqrt{1-2\kappa+4\kappa^2}
  =4\kappa\left(1-\frac34\kappa^2+\OO(\kappa^4)\right). \label{eq:qes_M4_upper_gap}
\end{equation}
The leading terms agree with equations~(52)--(54) of
\cite{Becker:2017}.  The term of order $\kappa^3$ in
\eqref{eq:qes_M4_upper_gap} is negative.  With the identifications
$\beta_{\rm B}=-\kappa$, $\kappa_{\rm B}=M=4$, and
$E=E_h-\kappa^2/2$, the exact energies in table~IV of
\cite{Becker:2017} give the same expansion,
$\Delta E_{4,1}=4\kappa-3\kappa^3+\OO(\kappa^5)$.  The plus sign of
this term in their equation~(54) is therefore not consistent with their
table.

For $M=5$ the even and odd blocks are
\begin{equation}
  H_+=
  \begin{pmatrix}
    -4&-2\kappa&0\\
    -2\kappa&-1&-2\sqrt3\kappa\\
    0&-2\sqrt3\kappa&0
  \end{pmatrix},
  \qquad
  H_-=
  \begin{pmatrix}
    -4&-2\kappa\\
    -2\kappa&-1
  \end{pmatrix}.
  \label{eq:qes_M5_blocks}
\end{equation}
Their characteristic polynomials are
\begin{align}
  P_+(\mathcal E)
  &=\mathcal E(\mathcal E+1)(\mathcal E+4)
    -16\kappa^2\mathcal E-48\kappa^2,
  \label{eq:qes_M5_even_polynomial}\\
  P_-(\mathcal E)
  &=(\mathcal E+1)(\mathcal E+4)-4\kappa^2.
  \label{eq:qes_M5_odd_polynomial}
\end{align}
From the roots near $\mathcal E=-4$ and $\mathcal E=-1$ we find
\begin{align}
  \Delta E_{5,0}
  &=\frac43\kappa^4\left(1-\frac23\kappa^2+\frac1{81}\kappa^4+\OO(\kappa^6)\right),
  \label{eq:qes_M5_lower_gap}\\
  \Delta E_{5,1}
  &=12\kappa^2\left(1-\frac{100}{9}\kappa^2+\OO(\kappa^4)\right).
  \label{eq:qes_M5_upper_gap}
\end{align}
The leading terms and the corrections of relative order $\kappa^2$ of
all these splittings agree with \eqref{eq:qes_leading_splitting} and
\eqref{eq:qes_subleading_splitting}, or with
\eqref{eq:qes_r1_splitting} for $r=1$.  Moreover, the terms
$-7\kappa^4/8$ and $\kappa^4/81$ in \eqref{eq:qes_M4_lower_gap} and
\eqref{eq:qes_M5_lower_gap} agree with the coefficient
\eqref{eq:generic_splitting_c2_explicit} evaluated at
$(\beta,r)=(4,3)$ and $(5,4)$, where it is finite.

\section{Connection formula from a recurrence relation}
\label{app:recurrence_connection}

An alternative to the connection formula in terms of conformal blocks
is the construction of Wolf \cite{Wolf:1998}, in which the connection
coefficients are obtained as limits of the solutions of a recurrence
relation.  In this appendix we summarize the construction, correct the
prefactor of the published formula with the help of a result of
Sch\"afke \cite{Schaefke:1984}, and apply it to the Razavy equation.
The derivations can be found in \cite{Wolf:1998,Schaefke:1984}.  The
correction affects the normalization of the connection coefficient, but
not its zeros for $\alpha\neq0$.

\subsection{Canonical form and solutions}
\label{app:recurrence_symmetric_form}
\label{app:recurrence_bases}

With $D_\zeta=\zeta\,\dd/\dd\zeta$, the symmetric canonical form of the
DCHE is
\cite[eq.~(1.1)]{Wolf:1998}
\begin{equation}
  \left\{D_\zeta^2+\alpha(\zeta+\zeta^{-1})D_\zeta
  +\alpha\left(\beta_1+\frac12\right)\zeta
  +\frac{\alpha^2}{2}-\gamma
  +\alpha\left(\beta_{-1}-\frac12\right)\zeta^{-1}
  \right\}y=0,
  \qquad \alpha\ne0.
  \label{eq:recurrence_symmetric_dche}
\end{equation}
Here $\beta_{-1}$, $\beta_1$, and $\gamma$ are complex parameters, and
$\zeta=0$ and $\zeta=\infty$ are irregular singular points of rank
one.

The normal form \eqref{eq:canonical_dche} is obtained with the change
of variables
\begin{equation}
  \zeta=\sqrt{\frac{L_1}{L_2}}\,z,
  \qquad \alpha=\sqrt{L_1L_2},
  \qquad \alpha\zeta=L_1z,
  \qquad \frac{\alpha}{\zeta}=\frac{L_2}{z},
  \label{eq:recurrence_coordinate_dictionary}
\end{equation}
where $L_1L_2\neq0$, together with
\begin{equation}
  y(\zeta)=z^{-1/2}
  \exp\!\left(-\frac{L_1z}{2}+\frac{L_2}{2z}\right)\Phi(z),
  \qquad
  \beta_1=m_1,\quad \beta_{-1}=m_2,\quad
  \gamma=\frac14-u.
  \label{eq:recurrence_normal_form_dictionary}
\end{equation}

We use the normalization of the solutions of
\cite[sec.~2]{Wolf:1998}.  Let
$\Psi(\alpha,\beta_{-1},\beta_1,\gamma;\zeta)$ be the solution with the
asymptotic behavior
\begin{equation}
  \Psi(\alpha,\beta_{-1},\beta_1,\gamma;\zeta)
  \sim(\alpha\zeta)^{-\beta_1-1/2}
  \left[1+\OO(\zeta^{-1})\right],
  \quad -\frac\pi2<\arg(\alpha\zeta)<\frac{5\pi}{2}.
  \label{eq:recurrence_basic_solution}
\end{equation}
The sector and the powers are understood on the Riemann surface of the
logarithm.  The other solutions are defined by
\begin{align}
  y_{\infty1}(\zeta)
  & =\Psi(\alpha,\beta_{-1},\beta_1,\gamma;\zeta),
  \nonumber\\
  y_{\infty2}(\zeta)
  & =\ii\ee^{-\ii\pi\beta_1}\ee^{-\alpha\zeta}
    \Psi(\ee^{\ii\pi/2}\alpha,\beta_{-1},-\beta_1,\gamma;
    \ee^{\ii\pi/2}\zeta),
  \nonumber\\
  y_{01}(\zeta)
  & =\ee^{-\alpha\zeta+\alpha/\zeta}
    \Psi(\alpha,\beta_1,\beta_{-1},\gamma;\zeta^{-1}),
  \nonumber\\
  y_{02}(\zeta)
  & =\ii\ee^{-\ii\pi\beta_{-1}}\ee^{-\alpha\zeta}
    \Psi(\ee^{\ii\pi/2}\alpha,\beta_1,-\beta_{-1},\gamma;
    \ee^{\ii\pi/2}\zeta^{-1}).
  \label{eq:recurrence_canonical_bases}
\end{align}
Their leading behavior is
\begin{align}
  y_{\infty1}&\sim(\alpha\zeta)^{-\beta_1-1/2},
  & y_{\infty2}&\sim\ee^{-\alpha\zeta}
    (\alpha\zeta)^{\beta_1-1/2},
  &&\zeta\to\infty,
  \nonumber\\
  y_{01}&\sim\ee^{\alpha/\zeta}
    (\alpha/\zeta)^{-\beta_{-1}-1/2},
  & y_{02}&\sim(\alpha/\zeta)^{\beta_{-1}-1/2},
  &&\zeta\to0.
  \label{eq:recurrence_endpoint_asymptotics}
\end{align}
The solutions themselves are defined by
\eqref{eq:recurrence_basic_solution} and
\eqref{eq:recurrence_canonical_bases}.

With $S=\beta_{-1}+\beta_1$ we write the connection relation as
\begin{equation}
  (y_{01},y_{02})=(y_{\infty1},y_{\infty2})\,\mathsf C.
  \label{eq:recurrence_connection_relation}
\end{equation}
The matrix $\mathsf C$ is expressed through a single function $c$,
which is symmetric under $\beta_{-1}\leftrightarrow\beta_1$
\cite[eqs.~(2.5)--(2.6)]{Wolf:1998},
\begin{equation}
  \mathsf C=\ee^{-\ii\pi S/2}
  \begin{pmatrix}
    -\ii c(\ee^{\ii\pi/2}\alpha,\beta_{-1},-\beta_1,\gamma)
      & c(\ee^{\ii\pi}\alpha,-\beta_{-1},-\beta_1,\gamma)\\[2pt]
    c(\alpha,\beta_{-1},\beta_1,\gamma)
      & \ii c(\ee^{\ii\pi/2}\alpha,-\beta_{-1},\beta_1,\gamma)
  \end{pmatrix},
  \qquad \det\mathsf C=-1.
  \label{eq:recurrence_connection_matrix}
\end{equation}
Inverting \eqref{eq:recurrence_connection_relation}, we find
\begin{equation}
  y_{\infty2}=\mathsf C_{12}y_{01}-\mathsf C_{11}y_{02}.
  \label{eq:recurrence_inverse_relation}
\end{equation}
Hence $\mathsf C_{12}$ is the coefficient of the growing solution
$y_{01}$ in the continuation of $y_{\infty2}$ to $\zeta=0$.  In the
comparison with the solutions $\Phi$ of
section~\ref{sec:local-solutions}, the powers of $\alpha$ in
\eqref{eq:recurrence_endpoint_asymptotics} have to be taken into
account.

\subsection{Recurrence and connection coefficient}
\label{app:recurrence_construction}
\label{app:recurrence_corrected_limit}

By a Laplace transformation and changes of variables
\cite[secs.~3--4]{Wolf:1998}, the DCHE is transformed into an equation
with a regular singular point at $\xi=0$, an irregular singular point
of rank one at $\xi=1$, and a further singular point at $\xi=2$.  Its
solution at $\xi=0$ is written as
\begin{equation}
  w(\xi)=\sum_{n=0}^{\infty}
    \frac{d_n(\alpha,\beta_{-1},\beta_1,\gamma)}
         {\Gamma(n+1-S)}\,\xi^n,
  \qquad d_0=1,\qquad d_{-1}=d_{-2}=0.
  \label{eq:recurrence_w_series}
\end{equation}
The coefficients satisfy, for $n\geq0$, the four-term recurrence
\cite[eq.~(4.11)]{Wolf:1998}
\begin{equation}
  2(n+1)d_{n+1}=\mathsf a_n d_n -2(n-S)\mathsf b_n d_{n-1} +(n-S)(n-1-S)\mathsf c_n
  d_{n-2},
  \label{eq:recurrence_general}
\end{equation}
with
\begin{align}
  \mathsf a_n={}&
    \left(n+\frac32-2\beta_{-1}\right)
    \left(n+\frac32-2\beta_1\right)
    +2(n+\alpha)\left[\alpha+2(n+1-S)\right]
    \nonumber\\[-2pt]
    &+2n(\alpha-1)+4\gamma-\frac14,
  \nonumber\\
  \mathsf b_n={}&
    \left(n+\frac12-2\beta_{-1}\right)
    \left(n+\frac12-2\beta_1\right)
    +(n-1+2\alpha)(n-S),
  \nonumber\\
  \mathsf c_n={}&
    \left(n-\frac12-2\beta_{-1}\right)
    \left(n-\frac12-2\beta_1\right).
  \label{eq:recurrence_general_coefficients}
\end{align}
The recurrence is symmetric under $\beta_{-1}\leftrightarrow\beta_1$
and gives $d_1=\mathsf a_0/2$.  It is the same as in \cite{Wolf:1998},
and only the prefactor of the connection formula below is corrected.

For $\alpha\neq0$ with $|\arg\alpha|<\pi$ we define
\begin{equation}
  \mathcal L(\alpha,\beta_{-1},\beta_1,\gamma)
  =\lim_{n\to\infty}
    \ee^{-4\sqrt{\alpha n}}n^{3/4}
    \frac{d_n(\alpha,\beta_{-1},\beta_1,\gamma)}
         {\Gamma(n+1-S)}.
  \label{eq:recurrence_limit_definition}
\end{equation}
All powers of $\alpha$ are defined with $-\pi<\arg\alpha<\pi$.  The
corrected connection formula is
\begin{equation}
  c(\ee^{\ii\pi}\alpha,-\beta_{-1},-\beta_1,\gamma)
  =2\sqrt2\,\pi\,\ee^{\ii\pi S/2}\alpha^{1/4}
    \mathcal L(\alpha,\beta_{-1},\beta_1,\gamma).
  \label{eq:recurrence_corrected_c}
\end{equation}
Equivalently, the matrix element in
\eqref{eq:recurrence_connection_relation} is
\begin{equation}
  \mathsf C_{12}
    =2\sqrt2\,\pi\,\alpha^{1/4}
      \mathcal L(\alpha,\beta_{-1},\beta_1,\gamma).
  \label{eq:recurrence_corrected_matrix_entry}
\end{equation}
The argument $\ee^{\ii\pi}\alpha$ in \eqref{eq:recurrence_corrected_c}
denotes the continuation of $\alpha$ by a counterclockwise rotation
through $\pi$, without a return to the principal value of the
argument.

To derive the prefactor, we note that the growing solution at $\xi=1$
behaves as $\exp[4\alpha/(1-\xi)]\,[1+\OO(1-\xi)]$
\cite[eqs.~(4.5)--(4.7)]{Wolf:1998}.  Applying
\cite[Corollary~3.3]{Schaefke:1984} with the exponential parameter
$\Lambda=4\alpha$ and vanishing algebraic exponent, we find for its
coefficient
\begin{equation}
  \widetilde\gamma_2
  =2\sqrt\pi\,\ee^{-2\alpha}(4\alpha)^{-1/4}
     \mathcal L(\alpha,\beta_{-1},\beta_1,\gamma).
  \label{eq:recurrence_corrected_gamma}
\end{equation}
The saddle-point evaluation of the Laplace integral in
\cite[sec.~5, eq.~(5.9)]{Wolf:1998} then gives
\begin{equation}
  \mathsf C_{12}
  =2\ee^{2\alpha}\sqrt{\pi\alpha}\,\widetilde\gamma_2,
  \label{eq:recurrence_laplace_transfer}
\end{equation}
which proves \eqref{eq:recurrence_corrected_matrix_entry}.

The difference from the published formula originates from a misprint
in \cite[Theorem~4.15]{Schaefke:1984}.  In the notation of that theorem
the correct power is
\begin{equation}
  (-\alpha_{\rm S})^{(\mu_{\infty2}-\mu_{\infty1})/2-1/4},
  \quad\hbox{not}\quad
  (-\alpha_{\rm S})^{(\mu_{\infty2}-\mu_{\infty1}-2)/2}.
  \label{eq:recurrence_schaefke_correction}
\end{equation}
Here $\alpha_{\rm S}$ is Sch\"afke's parameter, which is distinct from
$\alpha$ in \eqref{eq:recurrence_symmetric_dche}.  With
$\alpha_{\rm S}=-4\alpha$ and equal algebraic exponents, the misprint
affects \cite[Propositions~4.1 and~5.1]{Wolf:1998}, whose right-hand
sides have to be multiplied by $(4\alpha)^{3/4}$.  In particular, the
prefactor $\pi\ee^{\ii\pi S/2}/\sqrt\alpha$ of Proposition~5.1 becomes
$2\sqrt2\,\pi\ee^{\ii\pi S/2}\alpha^{1/4}$.  The exponential, the power
of $n$, and the gamma function in the limit
\eqref{eq:recurrence_limit_definition} are not affected.

The other elements of \eqref{eq:recurrence_connection_matrix} follow
from the transformations of the parameters given there and from
analytic continuation \cite[sec.~6]{Wolf:1998}.  The rotated arguments
do not all lie in $|\arg\alpha|<\pi$.  The remaining element can be
obtained from $\det\mathsf C=-1$, provided that the element by which
one divides does not vanish.  The sequence in
\eqref{eq:recurrence_limit_definition} has an asymptotic expansion in
powers of $n^{-1/2}$ \cite{Schaefke:1984,Wolf:1998}, so that the limit
is best computed by extrapolation in $n^{-1/2}$.  The expansion holds
at fixed $\alpha\neq0$ and is not uniform in the limit $\alpha\to0$.

\subsection{Application to the Razavy equation}
\label{app:recurrence_razavy}
\label{app:recurrence_integer_parameters}

For the Hamiltonian \eqref{eq:razavy_hamiltonian}, with $\hbar$ kept
explicit, we set
\begin{equation}
  \zeta=\ee^x,\qquad
  y(\zeta)=\exp\!\left[-\frac{\kappa}{\hbar}\sinh x\right]\psi(x),
  \qquad
  \alpha=\frac{\kappa}{\hbar},\quad
  \beta_{-1}=\beta_1=\frac{\beta}{2\hbar},\quad
  \gamma=-\frac{E}{\hbar^2}.
  \label{eq:recurrence_razavy_dictionary_hbar}
\end{equation}
With these identifications \eqref{eq:recurrence_symmetric_dche} is the
Schr\"odinger equation divided by $\hbar^2$.  For the potential written
as $V=a\cosh2x-b\cosh x$ we have $\alpha=\sqrt{2a}/\hbar$ and
$\beta_{-1}=\beta_1=b/(2\hbar\sqrt{2a})$.

We now set $\hbar=1$, as in section~\ref{sec:heund}, so that
$\alpha=\kappa$, $S=\beta$, $\gamma=-E$, and the variable of
\eqref{eq:liouville_transformation} is $z=\kappa\zeta$.  We define
\begin{equation}
  d_n^{\rm R}(E;\kappa,\beta)
  =d_n\left(\kappa,\frac\beta2,\frac\beta2,-E\right).
  \label{eq:recurrence_razavy_coefficients_definition}
\end{equation}
The recurrence becomes
\begin{equation}
  2(n+1)d_{n+1}^{\rm R}=\mathsf a_n^{\rm R}d_n^{\rm R} -2(n-\beta)\mathsf b_n^{\rm
  R}d_{n-1}^{\rm R} +(n-\beta)(n-1-\beta)\mathsf c_n^{\rm R}d_{n-2}^{\rm R},
  \label{eq:recurrence_razavy}
\end{equation}
with $d_0^{\rm R}=1$, $d_{-1}^{\rm R}=d_{-2}^{\rm R}=0$, and the
coefficients
\begin{align}
  \mathsf a_n^{\rm R}&= \left(n+\frac32-\beta\right)^2
  +2(n+\kappa)\left[\kappa+2(n+1-\beta)\right] +2n(\kappa-1)-4E-\frac14,\nonumber\\
  \mathsf b_n^{\rm R}&=
    \left(n+\frac12-\beta\right)^2
  +(n-1+2\kappa)(n-\beta),
  \nonumber\\
  \mathsf c_n^{\rm R}&=
    \left(n-\frac12-\beta\right)^2.
  \label{eq:recurrence_razavy_coefficients}
\end{align}
The connection coefficient is
\begin{equation}
  \begin{split}
  \mathcal L_{\rm R}(E;\kappa,\beta)
    &=\lim_{n\to\infty}\ee^{-4\sqrt{\kappa n}}n^{3/4}
       \frac{d_n^{\rm R}(E;\kappa,\beta)}{\Gamma(n+1-\beta)},\\
  \mathsf C_{12}(E;\kappa,\beta)
    &=2\sqrt2\,\pi\,\kappa^{1/4}
       \mathcal L_{\rm R}(E;\kappa,\beta).
  \end{split}
  \label{eq:recurrence_razavy_limit}
\end{equation}
Since $\kappa>0$, no continuation in $\alpha$ is needed, and the result
is real for real $E$ and $\beta$.

To relate this coefficient to the spectral determinant
\eqref{eq:heun_spectral_determinant}, we set
\begin{equation}
  G(z)=z^{1/2}\exp\!\left(\frac z2-\frac{\kappa^2}{2z}\right).
  \label{eq:recurrence_razavy_gauge_factor}
\end{equation}
The asymptotic behavior \eqref{eq:recurrence_endpoint_asymptotics}
implies
\begin{equation}
  G\,y_{\infty1}=\Phi_{\infty,+}, \quad G\,y_{\infty2}=\Phi_{\infty,-},\quad
  G\,y_{01}=\kappa^{-1-\beta}\Phi_{0,+}, \quad G\,y_{02}=\kappa^{\beta-1}\Phi_{0,-}.
  \label{eq:recurrence_razavy_basis_normalisations}
\end{equation}
Here the solutions $y$ are evaluated at $\zeta=z/\kappa$, and the
sectors are chosen consistently.  The coefficient of the growing
solution in \eqref{eq:dche_connection_matrix} is therefore
$C_{-,+}=\kappa^{1-\beta}\mathsf C_{12}$.  With
$W[\Phi_{\infty,+},\Phi_{\infty,-}]=-1$ we obtain the spectral
determinant in the form
\begin{equation}
  \mathcal D(E;\kappa,\beta)
    =-2\sqrt2\,\pi\,\kappa^{5/4-\beta}
      \lim_{n\to\infty}\ee^{-4\sqrt{\kappa n}}n^{3/4}
      \frac{d_n^{\rm R}(E;\kappa,\beta)}{\Gamma(n+1-\beta)}.
  \label{eq:recurrence_razavy_determinant}
\end{equation}
The eigenvalues are thus the solutions of
$\mathcal L_{\rm R}(E;\kappa,\beta)=0$.  For $\hbar\neq1$ the
parameters are replaced by those of
\eqref{eq:hbar_rescaling_dictionary}.

The zeros of the determinant include the levels of both parities.  To
distinguish them, we note that the wavefunction that decays for
$x\to+\infty$ is
$\psi_R(x)=\exp[\kappa\sinh x]y_{\infty2}(\ee^x)$.  Since
$\beta_{-1}=\beta_1$, we have
$\exp[\kappa\sinh x]y_{02}(\ee^x)=\psi_R(-x)$, so that at an
eigenvalue
\begin{equation}
  \mathsf C_{22}=\epsilon,
  \qquad
  \epsilon=+1\quad\hbox{for even states},\qquad
  \epsilon=-1\quad\hbox{for odd states}.
  \label{eq:recurrence_razavy_parity}
\end{equation}
Equivalently, the even and odd levels are determined by
$\psi_R'(0)=0$ and $\psi_R(0)=0$.  The matrix element is
\begin{equation}
  \mathsf C_{22}
    =\ii\ee^{-\ii\pi\beta/2}
      c\left(\ee^{\ii\pi/2}\kappa,-\frac\beta2,
               \frac\beta2,-E\right).
  \label{eq:recurrence_razavy_parity_entry}
\end{equation}
It can be evaluated with \eqref{eq:recurrence_corrected_c} for
$(\alpha,\beta_{-1},\beta_1,\gamma)
=(-\ii\kappa,\beta/2,-\beta/2,-E)$, which satisfies
$|\arg\alpha|<\pi$.

The construction also applies to integer $S$.  The recurrence
\eqref{eq:recurrence_general} involves a division by $2(n+1)$ only and
is therefore well defined, and $1/\Gamma(n+1-S)$ is an entire function
of $S$.  A recurrence for the ratios $d_n/\Gamma(n+1-S)$ would instead
involve a division by $n+1-S$, which is avoided by computing $d_n$
first.

At the QES values $\beta=M\in\mathbb{Z}^+$ we have
$1/\Gamma(n+1-M)=0$ for $0\leq n<M$, so that the series
\eqref{eq:recurrence_w_series} begins with the term proportional to
$d_M^{\rm R}$.  Because of the factors $n-\beta$ in
\eqref{eq:recurrence_razavy}, the coefficients $d_n^{\rm R}$ with
$n>M$ are proportional to $d_M^{\rm R}$.  If $d_M^{\rm R}(E)=0$, they
all vanish, and the limit \eqref{eq:recurrence_razavy_limit} is zero.
For $M=1$ and $M=2$ we find
\begin{align}
  \beta=1:\qquad
  d_1^{\rm R}&=\kappa^2-2E,
  & E&=\frac{\kappa^2}{2},
  \nonumber\\
  \beta=2:\qquad
  d_2^{\rm R}&=2\left[
       \left(E-\frac{\kappa^2}{2}+\frac14\right)^2-\kappa^2
       \right],
  & E&=\frac{\kappa^2}{2}-\frac14\pm\kappa.
  \label{eq:recurrence_qes_checks}
\end{align}
These are the eigenvalues of \eqref{eq:qes_spin_hamiltonian} given in
\eqref{eq:qes_M1_appendix} and \eqref{eq:qes_M2}, and the lower sign in
the second line corresponds to the even state.  The recurrence can thus
be used without modification at integer $\beta$.

\providecommand{\href}[2]{#2}\begingroup\raggedright\endgroup
\end{document}